\documentclass[reprint,amsmath,amssymb,aps,prb,superscriptaddress,floatfix]{revtex4-2}
\usepackage{array}
\usepackage{bm}
\usepackage{color}
\usepackage{float}
\usepackage{graphicx}
\usepackage{hyperref}
\hypersetup{colorlinks=true,allcolors=blue}
\usepackage{natbib}
\usepackage{newtxtext}
\usepackage{newtxmath}
\usepackage{enumitem}
\usepackage[caption=false]{subfig}

\usepackage{lipsum}
\usepackage{amsmath}
\usepackage{amssymb}

\begin{document}

\title{Random Transverse Field Effects on Magnetic Noise in Spin Systems}

\author{Yufei Pei}
\affiliation{T.C.M. Group, Cavendish Laboratory, JJ Thomson Avenue, Cambridge CB3 0HE, United Kingdom}
\address{Max Planck Institute for the Physics of Complex Systems, 01187 Dresden, Germany}

\author{Claudio Castelnovo}
\affiliation{T.C.M. Group, Cavendish Laboratory, JJ Thomson Avenue, Cambridge CB3 0HE, United Kingdom}

\author{Roderich Moessner}
\address{Max Planck Institute for the Physics of Complex Systems, 01187 Dresden, Germany}

\begin{abstract}

Motivated by experimental developments in non-Kramers spin ice materials and the unclear role of disorder therein, we study the impact of random transverse fields on the dynamics of correlated magnetic systems. We model the effect of dilute, randomly placed transverse fields on quantities such as magnetic noise/susceptibility and the diffusivity of topological excitations.  We consider a random ferromagnetic Ising chain (RTFIC) as well as three-dimensional spin ice. At low temperatures, both exhibit (sub-)diffusive defect dynamics, i.e., of domain walls and  magnetic monopoles, respectively. Introducing sparse transverse fields leads to the emergence of an additional timescale on the order of the single-spin flip time. We develop a Lindbladian framework that combines Monte Carlo simulations and exact diagonalization which allows us to  characterize the dynamics and develop an analytical understanding of the phenomenon. This framework can be benchmarked in detail for the RTFIC. Our findings provide insights into the magnetization dynamics of disordered non-Kramers oxides, such as oxygen-diluted Ho$_2$Ti$_2$O$_7$, and offer a framework for interpreting experimental observations in these systems.
\end{abstract}


\maketitle
%
%


\section{\label{section:1:introduction}
Introduction}
Dynamics in spin liquids \cite{savary2016quantum, Knolle_2019, broholm2020quantum, knolle2014dynamics, knolle2016dynamics, udagawa2016out, gohlke2018dynamical} are highly intertwined with the properties and behavior of their characteristic fractionalized quasiparticles. This interplay offers a window to probe spin liquid behavior as well as to uncover new and striking topological phenomena. Key quantities, such as dynamical correlators \cite{knolle2014dynamics, knolle2016dynamics, udagawa2016out, gohlke2018dynamical} and magnetic noise \cite{revell2013evidence, samarakoon2022anomalous, hallen2022dynamical, hsu2024dichotomous}, can be modeled and understood within an effective quasiparticle framework, linking macroscopic observables to the underlying microscopic excitations. This is exemplified by some of the most prominent spin liquid states, such as those found in Kitaev systems \cite{knolle2014dynamics, knolle2016dynamics, udagawa2016out, gohlke2018dynamical}, dimer \cite{misguich2008quantum, yan2021topological} and vertex models \cite{ritort2003glassy}. 

Disorder plays a dual role in this framework: it is not only an unavoidable feature of experimental systems but also a catalyst for novel phenomena. Usually manifesting as site dilution or randomized couplings, it can stabilize zero modes \cite{willans2011site}, bound states \cite{petrova2015hydrogenic}, order-by-disorder \cite{andreanov2015order}, and even trigger transitions into spin liquid states \cite{furukawa2015quantum, zhu2017disorder}. Despite its rich consequences, modeling the interplay of disorder with strong correlations and dynamics remains a formidable challenge, as these competing effects often resist conventional analytical or numerical approaches.

This work bridges the gap between disorder, strong correlations, and dynamics by investigating their interplay in the quasiparticle-dominated regime  of spin liquids. We focus on classical spin ice (CSI) \cite{udagawa2021spin}, a paradigmatic system much pursued both experimentally and theoretically as the main 3D quantum spin liquid candidate. There, frustration in the classical antiferromagnetic Ising model on the pyrochlore lattice enforces the `ice rules': at low temperatures, spins on each tetrahedron adopt a two-in/two-out configuration. In pristine CSI, dynamics under ideal conditions is governed by emergent magnetic monopoles \cite{castelnovo2008magnetic, castelnovo2012spin}, mobile point defects that propagate via stochastic spin flips. These monopoles unify the interpretation of diverse macroscopic phenomena --- from transport properties like thermal \cite{kolland2012thermal} and magnetic conductivity \cite{bramwell2009measurement} to response functions such as power spectral density \cite{samarakoon2022anomalous}, susceptibility \cite{hsu2024dichotomous}, and magnetization dynamics under thermal/field quenches \cite{castelnovo2010thermal, mostame2014tunable}. 

Our paper aims to investigate how disorder perturbs this well-established phenomenology. A concrete motivation are the advances in rare-earth oxide materials, the primary experimental realizations of classical spin ice (CSI). In these systems, the lattice distortions \cite{savary2016disorder, benton2018instabilities} modify spin interactions and enable new dynamical phenomena. A key example is $\mathrm{Ho_2Ti_2O_7}$, where non-Kramers $\mathrm{Ho}^{3+}$ ions exhibit a ground-state doublet susceptible to distortion-induced splitting. This splitting effectively introduces local transverse field terms, inducing quantum effects onto the classical CSI framework. Investigating the dynamics of such a modified spin ice model, and the interplay between monopolar excitations and `quantum clusters' of spins where the transverse field is active, offers a rich and unexplored research direction, and potentially a stepping stone between classical and quantum spin ice whose dynamics is far less understood to date.

To unravel the interplay of disorder and quasiparticle dynamics, we proceed in two steps. First, we analyze a simplified 1D analog of locally distorted spin ice: a ferromagnetic Ising chain with sparse random transverse fields. This system hosts domain walls --- topological defects that act as 1D counterparts to magnetic monopoles, exhibiting two distinct charge flavors. Such a one-dimensional model, with its simplicity, allows for the establishment of foundational insights into how transverse field disorder perturbs quasiparticle dynamics, both analytically and numerically.

We propose a generalization of the classical stochastic single spin-flip dynamics to incorporate the random transverse fields. This is done by formulating a quantum dissipative dynamics in Lindblad form~\cite{lindblad1976generators, gorini1976completely}, using jump operators between eigenstates of the quantum clusters of spins acted upon by the transverse field. Our modified Monte-Carlo then samples the diagonal part of the density matrix in the eigenbasis of the Hamiltonian.

We use the modified MC to study the behavior of the model in thermal equilibrium. Inspired by recent advancements in magnetic noise probes of monopole dynamics in spin ice \cite{dusad2019magnetic, samarakoon2022anomalous, morineau2025satisfaction}, we compute the power spectral density $\mathrm{PSD}(\omega)$ of the magnetic noise generated by the 1D model. We discover the emergence of two diffusive regimes that differ from the behavior of the system in the limit where the transverse field is absent. Namely, we find that the PSD behaves diffusively in the limits $\omega\to 0$ and $\omega\to\infty$, transitioning between the two around $\omega\sim1\left(\text{MC time}\right)^{-1}$. 

We are able to relate such structure to the random walk behavior of the excitations in presence of random transverse fields. Taking advantage of the fluctuation-dissipation theorem in equilibrium, one can relate the PSD to the mean-square displacement (MSD) of single excitations, and to the magnetic susceptibility of the system, $\chi(\omega)$. 
We correspondingly find that the MSD behaves diffusively in the limits $t\to 0$ and $t\to \infty$, with a transition between the two around $t\sim 1\text{ MC time}$. 
We also investigate the AC-susceptibility of the system by coupling it to an oscillatory longitudinal external driving field. We find an enhancement of the imaginary part $\chi''(\omega)$ at high frequencies, and a pronounced deviation of the phase shift $\phi=\tan^{-1}(\chi''/\chi')$ from the classical limit $\pi/2$, when $\omega$ is in the transition region of the PSD.

Having acquired ample insights in 1D, we use these results in the three-dimensional setting of spin ice, where the monopoles execute random walks subject to constraints from the spin configuration on one hand, and to non-uniformity from disorders on the other. There, the methodology developed in Sec.~\ref{section:2:model}, namely the modified Monte-Carlo algorithm, can still be used. We find that the phenomenology largely carries over to $d=3$ --- namely, the PSD behaves diffusively at $\omega\to 0$ and $\omega\to \infty$ while transitioning at intermediate frequencies. We show that, although this is already present in classical spin ice, such an effect is augmented by the random transverse fields. This proves the generality of the effects of the transverse fields irrespective of the dimensionality. 

The rest of this paper is organized as follows. Secs.~\ref{section:2:model} to~\ref{section:4:driven} are dedicated to the 1D model. In Sec.~\ref{section:2:model}, we introduce the model, the proposed dissipative dynamics, and our modified Monte-Carlo method. In Sec.~\ref{section:3:equilibrium_dynamics}, we present our results for the dynamics in thermal equilibrium, namely our study of the PSD and related MSD. In Sec.~\ref{section:4:driven}, we consider the driven model coupled to an external oscillatory longitudinal field and we study the behavior of the magnetic AC susceptibility, $\chi(\omega)$. Then, Sec.~\ref{section:5:rtfsi} is dedicated to presenting results in three dimensions. The final section, Sec.~\ref{section:6:discussion}, discusses the results and their relevance with recent experimental and theoretical progresses in spin ice models and materials. 
In the appendices, we discuss the thermodynamics of the one-dimensional model and fill in the technical details omitted in the main text. 
%
%

\section{\label{section:2:model}Model}
In the first part of this work, we are interested in studying how transverse fields applied at sparse random sites of an Ising chain affect its dynamical properties. 
We therefore need to develop a description that encompasses both the Markovian stochastic dynamics of a classical Ising chain, as well as the quantum dynamics due to the off-diagonal matrix elements introduced by the local applied fields. 
A study of the thermodynamic properties of our model is presented for completeness in App.~\ref{app:A}. 
%
%

\subsection{\label{section:2:dim}Ising chain and stochastic dynamics}
Our starting point is the classical, ferromagnetic nearest-neighbor Ising chain, with Hamiltonian 
\begin{equation}\label{eq:2:dim}
    \mathcal{H}_\text{C}=-\sum_{i} s_i s_{i+1}
    \, ,
\end{equation} 
where the $s_{i} = \pm 1$ are classical variables, and we have set the exchange coupling strength $J=1$ as the reference energy scale (we shall further assume $k_B=1$ for the Boltzmann constant). 
We endow this model with single-spin-flip Markovian stochastic dynamics, governed by Metropolis-Hastings acceptance probabilities~\cite{hastings1970monte}. {Under such dynamics, each spin attempts to flip on a characteristic time scale which is the microscopic single-spin-flip time, conventionally set to $1$ in Metropolis-Hastings dynamics. The spin-flip events are accepted with probability $P(\bm{\sigma}\to\bm{\sigma}')=\min\left(1,\mathrm{e}^{-\beta \Delta E}\right)$, where $\beta=1/T$ is the inverse temperature and $\Delta E$ denotes the energy difference between the two spin configurations $\bm{\sigma}$ and $\bm{\sigma}'$. }
We shall refer to this system as the kinetic Ising chain \cite{augusiak2010quantum, krapivsky2010kinetic}. 
Our choice of stochastic discrete-time dynamics is equivalent to the one generated by the classical Master equation.

A configuration of the Ising chain can be thought of as a sequence of ferromagnetic domains separated by domain walls. The latter in 1D take the form of topological point-like defects forming an alternating sequence of kinks and anti-kinks (similarly to the magnetic monopoles in spin ice, they live on the dual lattice and carry charge $\pm1$). Flipping a spin can either create or annihilate a kink-anti-kink pair, or make them hop by one lattice spacing; the latter does not change the energy of the system, whereas the former changes it by $\pm 4$. 
In the low-temperature limit, defects become sparse 
(their density is suppressed by the corresponding Boltzmann factor $e^{-2/T}$) 
and the magnetization dynamics of the system is effectively governed by the motion of sparse topological defects. 
In turn, their motion (at least between occasional creation and annihilation events, that become rarer the lower the temperature) can be described by a standard random walk in 1D.
%
%

\subsection{\label{section:2:rtfic}Random transverse field}
We then consider a one-dimensional random transverse field Ising chain (RTFIC), where the classical Hamiltonian $\mathcal{H}_\mathrm{C}$ is supplemented by transverse fields in the $x$-direction, with fixed magnitude $h$, applied at sparse randomly selected sites $\alpha$ across the system with uniform probability $r$: 
\begin{equation}\label{eq:2:rtfic}
    \mathcal{H} = -\sum_{i}S_i^zS_{i+1}^z-h\sum_{\alpha\in\{\alpha\}}S_\alpha^x
    \, .
\end{equation}

This model can also be understood as a version of the random transverse Ising system~\cite{Suzuki2013} with a two-point distribution of transverse fields. 
Our RTFIC can be regarded as an interpolation between the classical and the quantum limits that obtain for $r=0$ and $r=1$, respectively. 

In the dilute limit of interest here, $r \ll 1$, most of the spins are unaffected and behave classically (hereafter referred to as classical spins). They appear in the eigenstates of the Hamiltonian as single-spin eigenstates of the corresponding $S^z_i$ operators in a tensor product with the rest of the system. 
The spins affected by the transverse field (hereafter referred to as quantum spins) behave differently. 
It is convenient to identify each group of $n \geq 1$ consecutive quantum spins, labeled $1,2,\dots n$, as a `quantum cluster' of size $n$, or an `$n$-cluster' for short, with local $2^n$-dimensional Hamiltonian 
\begin{equation}\label{eq:2:Hn}
H^{(n)}=-\left(s_0S_1^z+S_{n}^zs_{n+1}+h\sum_{i=1}^{n}S_i^x+\sum_{i=1}^{n-1}S_i^zS_{i+1}^z\right)
\, ,
\end{equation}
where $s_0, s_{n+1}=\pm1$ represent the orientations of the two classical adjacent (boundary) spins. The density of $n$-clusters (per site) in a RTFIC system with transverse field density $r$ is given by $p_n=r^n(1-r)^2$, which decays exponentially with $n$. In the dilute limit, it is reasonable to approximate the system by neglecting the rare clusters larger than some cutoff size. We can then use exact diagonalisation (ED) to study the four Hamiltonians $H^{(n)}$ corresponding to different combinations of the two boundary spin states for each $n$ below the cutoff, build the eigenstates of the system Hamiltonian $\mathcal{H}$ as tensor products of single classical spins and quantum clusters, and investigate the behavior of our RTFIC model. (For details on the ED results, see App.~\ref{app:A}.)

In order to develop a dynamical version of the RTFIC, we need to generalize the classical { Metropolis-Hastings} dynamics to account for the presence of the quantum clusters. This is done by coupling the system described by Eq.~\eqref{eq:2:rtfic} to an external heat bath at a given temperature and inducing dissipative dynamics. Correspondingly, the reduced density matrix of the system, $\rho$, follows the quantum master equation~\cite{breuer2002theory, landi2022nonequilibrium}
\begin{equation}\label{eq:2:lind}
    \frac{\mathrm{d}}{\mathrm{d}t}\rho=\mathcal{L}[\rho]=-i[\mathcal{H}, \rho]+\mathcal{D}[\rho]
    \, ,
\end{equation}
where $\mathcal{L}$ is the Liouvillian superoperator. The first term on the right hand side is the Hamiltonian-time evolution, whereas the second dissipative term $\mathcal{D}[\rho]$ describes the interaction between the system and the heat bath, and has the usual Lindblad form~\cite{weisbrich2018decoherence}: 
\begin{equation}\label{eq:2:dissp}
    \mathcal{D}[\rho]=\sum_{i, \omega} \gamma_i(\omega) \left( A_{i}(\omega) \rho A_{i}^{\dagger}(\omega) - \frac{1}{2} \left\{ A_{i}^{\dagger}(\omega) A_{i}(\omega), \rho\right\} \right)
    \, .
\end{equation}
The $A_i(\omega)$ are the jump operators of the form
\begin{equation} \label{eq:2:jumpA}
    A_i(\omega)=\sum_{a, b}\delta_{E_b - E_a, \omega}|a\rangle\langle a|S_i^x|b\rangle\langle b|
    \, ,
\end{equation}
where the $|a\rangle$ and $|b\rangle$ are eigenstates of the system Hamiltonian and $\delta_{\omega', \omega}$ is a Kronecker delta.
The heat bath attempts to flip a spin at site $i$ by applying the Pauli operator $S_i^x$. { To maintain detailed balance~\cite{rajagopal1998principle}, the jump operators that complete such flips, Eq.~\eqref{eq:2:jumpA}, decompose the $S_i^x$ operators in the basis of system eigenstates and select the part that connects two states with a fixed energy difference, $\omega$ (with $\hbar=1$)~\footnote{This $\omega$ is not to be confused with the angular frequency used in the power spectral density study later on in the manuscript.}. 
These transition rates are then scaled by factors of the system-bath coupling strength $\gamma_i(\omega)$.} 
Detailed balance is obeyed for $\gamma_i(\omega)=e^{-\beta\omega}\gamma_i(-\omega)$.

\begin{figure*}[htb!]
    \includegraphics[width=0.88\columnwidth]{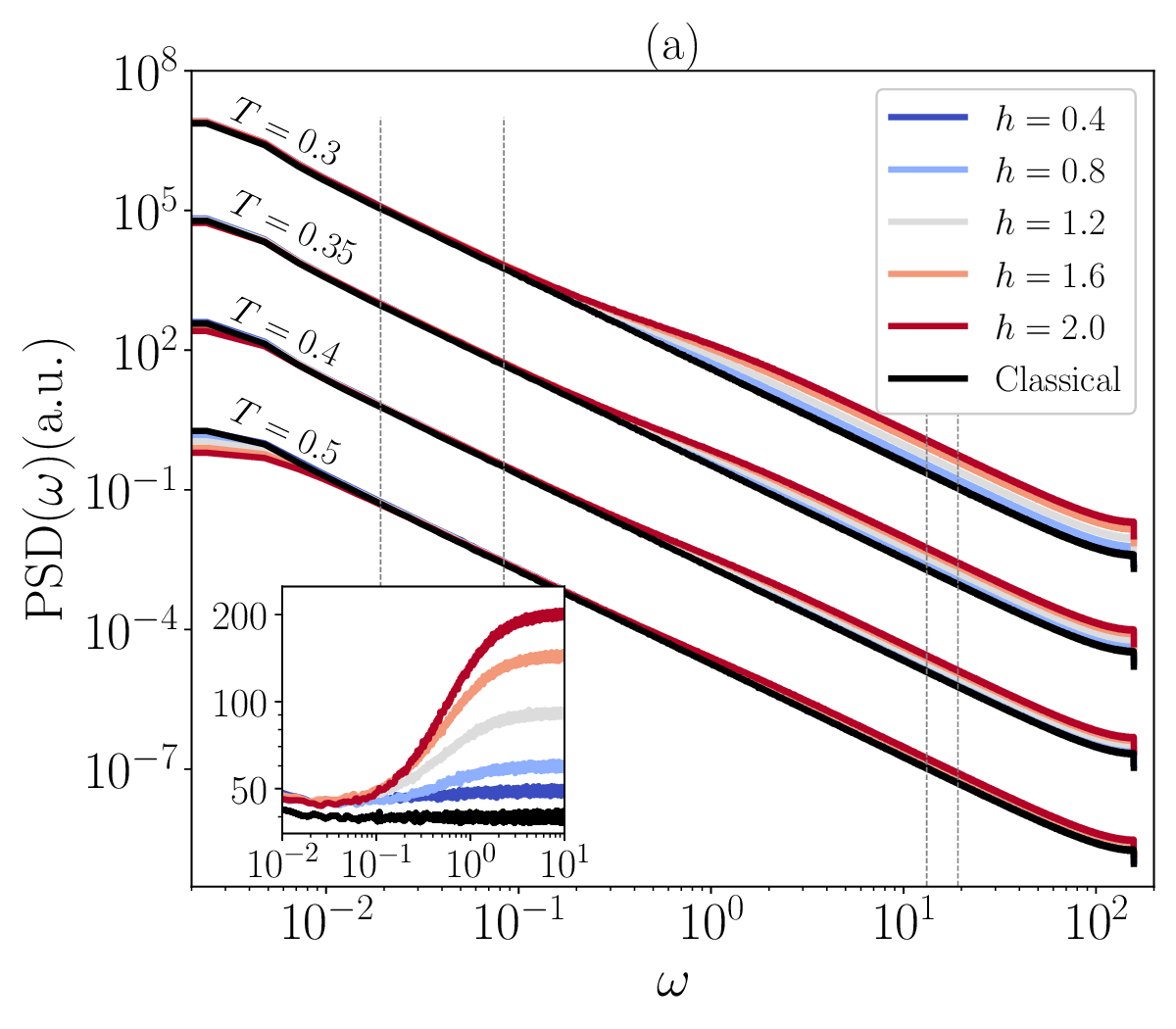}
    \hspace{3em}
    \includegraphics[width=0.88\columnwidth]{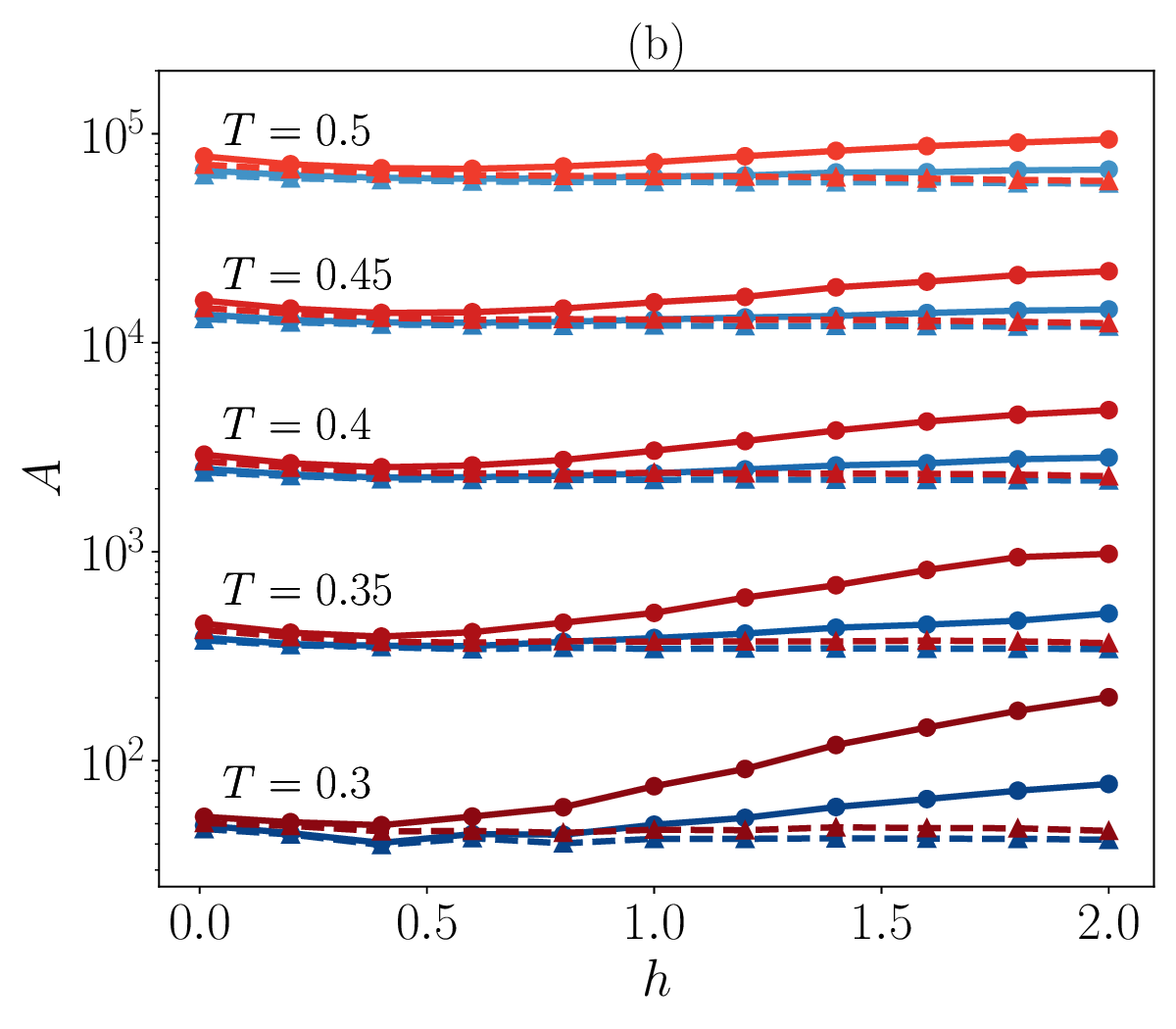}
    \caption{\label{fig:PSD}
    Magnetic noise spectrum of the RTFIC: (a) PSDs of the RTFIC extracted using Welch's method on MC simulations, with $r=0.2$ and varying values of $h$ and $T$. PSDs for the {classical kinetic Ising model} are also shown for reference. Inset: $\omega^2\,\mathrm{PSD}(\omega)$ as a function of $\omega$, with $r=0.2$ and varying values of $h$ and $T=0.3$, focusing on the intermediate region connecting the two inverse-square scaling regions. (b) {Estimates for the values of the parameters $A_\infty$ and $A_0$, as defined in the main text}, obtained by fitting the two regions of the PSD bounded by the dashed vertical lines in (a) to the form $\mathrm{PSD}(\omega)=A\omega^{-2}$, at $r=0.1$ (blue) or $0.2$ (red). While the low-frequency ({dotted}) and high-frequency (solid) results shall overlap {for the classical model}, they differ for the RTFIC, with the discrepancy becoming more pronounced with increasing $h, \, r$ and decreasing $T$. In both figures, results for different temperatures are shifted vertically for clarity.}
\end{figure*}

We now focus on the time evolution of the diagonal part of the density matrix, $\langle a | \rho | a \rangle = \rho_{aa}$, with respect to an eigenbasis of the system Hamiltonian, $\{|a\rangle\}$.
The Hamiltonian time-evolution term in the Liouvillian trivially vanishes for diagonal terms, and we are left with
\begin{equation}\label{eq:2:pme}
\begin{aligned}
    \frac{\mathrm{d}\rho_{aa}}{\mathrm{d}t}=\mathcal{L}&[\rho]_{aa} 
    =\sum_{i, b, c}\gamma(E_b-E_a)M_{i,bac}\rho_{bc} \\
    &-\gamma(E_a-E_b)\frac{M_{i,abc}+M_{i,cba}}{2}\rho_{ac}
    \, ,
\end{aligned}
\end{equation}
where $|b\rangle$, $|c\rangle$ are states drawn from the Hamiltonian eigenbasis, and we have introduced the short-hand notation $M_{i, abc}\equiv\langle a|S_i^x|b\rangle\langle b|S_i^x|c\rangle\delta_{E_a,E_c}$.
For the { classical kinetic Ising chain}, we select the eigenbasis $\{|\alpha\rangle\}$ to be identical to the spin-configuration basis $\{\bm{\sigma}\}$, under which we have $M_{i, abc}=\delta_{a,c}\delta_{b, S_i^xa}$. In this case, Eq.~\eqref{eq:2:pme} reduces to the classical Master equation,
by identifying $\langle \bm{\sigma}|\rho|\bm{\sigma}\rangle=P(\bm{\sigma})$ { as the statistical probability of the spin configuration $\bm{\sigma}$} and under the selection $\gamma_i(\omega)=\min(1,\mathrm{e}^{-\beta\omega})$ where the single-spin-flip time $\tau_0$ has been set to $1$. For the RTFIC, we select $\{|\alpha\rangle\}$ to be the set of the product states introduced above. Now, the question arises whether one still has $M_{i,abc}\propto\delta_{a,c}$ so that the RHS of Eq.~\eqref{eq:2:pme} involves only diagonal elements of the density matrix. In App.~\ref{app:B} we show that, although there exist some rare cases where the above proportionality is violated, they have low statistical weight at low temperature and can be safely neglected. 
Using the approximation $M_{i, abc}=\left|\langle a|S_i^x|b\rangle\right|^2\delta_{a, c}$, we arrive at the following Master equation for the RTFIC:
\begin{equation}\label{eq:2:rme}
    \frac{\mathrm{d}P(a)}{\mathrm{d}t}=\sum_{b}\tilde{w}_{ba}P(b)-\tilde{w}_{ab}P(a)\, ,
\end{equation}
with the transition rate between eigenstates given by:
\begin{equation}\label{eq:2:qcmc}
    \tilde{w}_{ab}=\sum_{i}\gamma(E_b-E_a)\left|\langle a | S_i^x | b \rangle \right|^2
    \, ,
\end{equation}
where we have once again selected $\gamma(\omega)=\min(1, e^{-\beta\omega})$ and set $\tau_0=1$. 

The above Master equation can be conveniently simulated by a modified version of the classical Monte Carlo algorithm operating between the eigenstates $\{|a\rangle\}$ introduced above. The matrix elements $\langle a | S_i^x | b \rangle$ involve limited computational effort due to the product nature of the eigenstates. In particular, if spin $i$ and its neighbors are classical, the spin-flip operation represented by $S_i^x$ maps an eigenstate to another eigenstate, and the corresponding matrix element is $1$; if spin $i$ belongs to a quantum cluster initially in state $|\psi\rangle$, the spin-flip operation maps it to $|\psi'\rangle$ with probability $|\langle\psi|S_i^x|\psi'\rangle|^2$ ($|\psi'\rangle$ may be identical to $|\psi\rangle$). If spin $i$ is classical but serves as the boundary spin of a quantum cluster initially in state $|\psi\rangle$, the spin is mapped to the opposite state and the adjacent cluster is mapped to the state $|\psi'\rangle$ with probability $|\langle\psi|\psi'\rangle|^2$ (note that the two states are eigenstates of different Hamiltonians and therefore not generally orthonormal to one another). Similarly for the case where spin $i$ is classical but serves as the boundary spin of two clusters, where the final state of the two clusters is drawn with probability given by the product of two such factors. After each spin-flip is performed, it is accepted with probability 
\begin{equation}\label{eq:2:qcmcp}
    p(a\to b)=\gamma(E_b-E_a) = \min\left(1, e^{-\beta (E_b-E_a)}\right)
    \, . 
\end{equation}
%
%

\section{\label{section:3:equilibrium_dynamics}Equilibrium dynamics}
We begin by studying the equilibrium dynamics of the RTFIC model introduced above, probing the magnetic fluctuations of the system at low temperature, where they are dominated by the motion of sparse defects (namely, randomly walking kinks and anti-kinks, as mentioned in Sec.~\ref{section:2:dim}). 
%
%

\subsection{\label{section:3:psd}Magnetic noise spectrum}
We start by studying the magnetic noise of the system as a function of (Monte-Carlo) time, which is done by measuring the time evolution of the magnetization of the system in the $z$-direction, $m(t)$, 
and by calculating its power spectral density, $\mathrm{PSD}(\omega)$, defined as 
\begin{equation}\label{eq:3:psd}
\begin{aligned}
    \mathrm{PSD}(\omega) &\equiv \left\langle\left|\int_{-\infty}^{\infty}\mathrm{d}t\,m(t)\mathrm{e}^{-i\omega t}\right|^2\right\rangle \\
    &=\int_{-\infty}^{\infty}\mathrm{d}t\left\langle  m(0)m(t)\right\rangle\mathrm{e}^{-i\omega t}\, ,
\end{aligned}
\end{equation}
where the angular brackets denote statistical averaging over Monte Carlo histories. The equivalence of the two definitions comes from the Wiener-Khinchin Theorem~\cite{wiener1930generalized}.

In the limit of isolated kink or anti-kink motion in the classical system ($r=0$), the mapping to a random walk allows to derive the known scaling result ${\mathrm{PSD}}(\omega)\propto \omega^{-2}$~\cite{cardiner1985handbook}. 
This provides a starting point to study the behavior of the RTFIC when the randomly walking domain walls are affected by the presence of quantum clusters. 

The results for the PSD, obtained using Welch's method~\cite{welch1967use} on MC simulations with different values of $r$, $h$ and $T$, are presented in Fig.~\ref{fig:PSD}a. (For details of the parameters used to run the MC simulations, see App.~\ref{app:F}.) A deviation from the { classical} $\omega^{-2}$ behavior at low frequency is observed around $\omega\sim 10^{-1}$ (in units of $2\pi/\text{MC time}$), when the PSD of the RTFIC exhibits a plateau-like upturn 
which terminates at $\omega \sim 10^1$, followed by another inverse square scaling regime, $\mathrm{PSD}(\omega)\propto\omega^{-2}$, with different proportionality factor depending on the value of $h$. 

We fit the function $\mathrm{PSD}(\omega)=A\omega^{-2}$ to the low- and high-frequency part of the curve {(see the dashed vertical lines in Fig.~\ref{fig:PSD}), and use the values obtained for $A$ as estimates of the proportionality factors} $A_0 = \lim_{\omega\to0}\omega^2\, \mathrm{PSD}(\omega)$ and $A_\infty = \lim_{\omega\to\infty}\omega^2\, \mathrm{PSD}(\omega)$. {We plot these values} as functions of the parameters $r$, $h$ and $T$. 
As seen in~Fig.~\ref{fig:PSD}b, { while these two values agree in the classical scenario, they differ in the RTFIC; the effect of the quantum clusters is most prominent in the high-frequency regime of the PSD, as $A_\infty$ has a more pronounced dependence on $h$ and $r$ comparing with $A_0$. The difference between these two quantities increases with $h, \, r$ and decreases with $T$. In addition, we note that the results obtained in the limit $h\to0$ (see App.~\ref{app:F}) give values of $A$ that do not converge to the classical result. We shall discuss the reason for this discrepancy later in the manuscript.}
%
%

\subsection{\label{section:3:rw}Quasiparticle random walk dynamics}
\begin{figure*}[htb!]
\begin{minipage}[]{0.43\linewidth}
    \includegraphics[width=\linewidth]{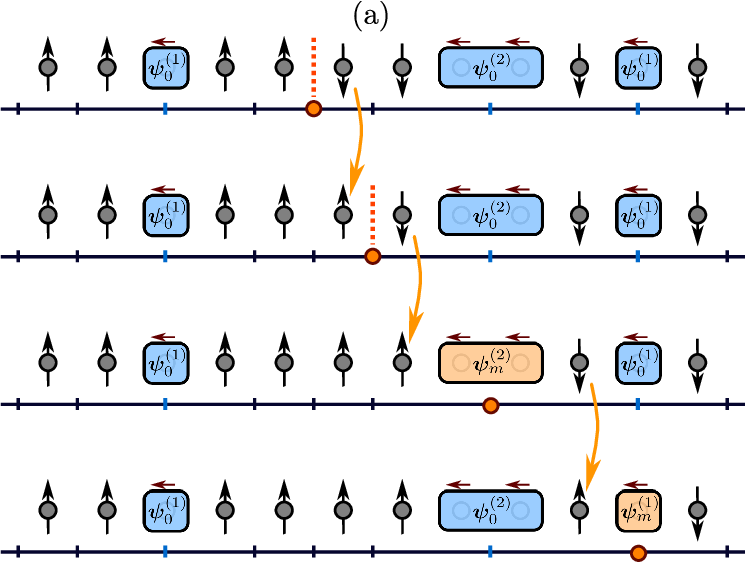}
    \vspace{1em}
\end{minipage}
\hspace{3em}
\begin{minipage}[]{0.90\columnwidth}
    \includegraphics[width=\linewidth]{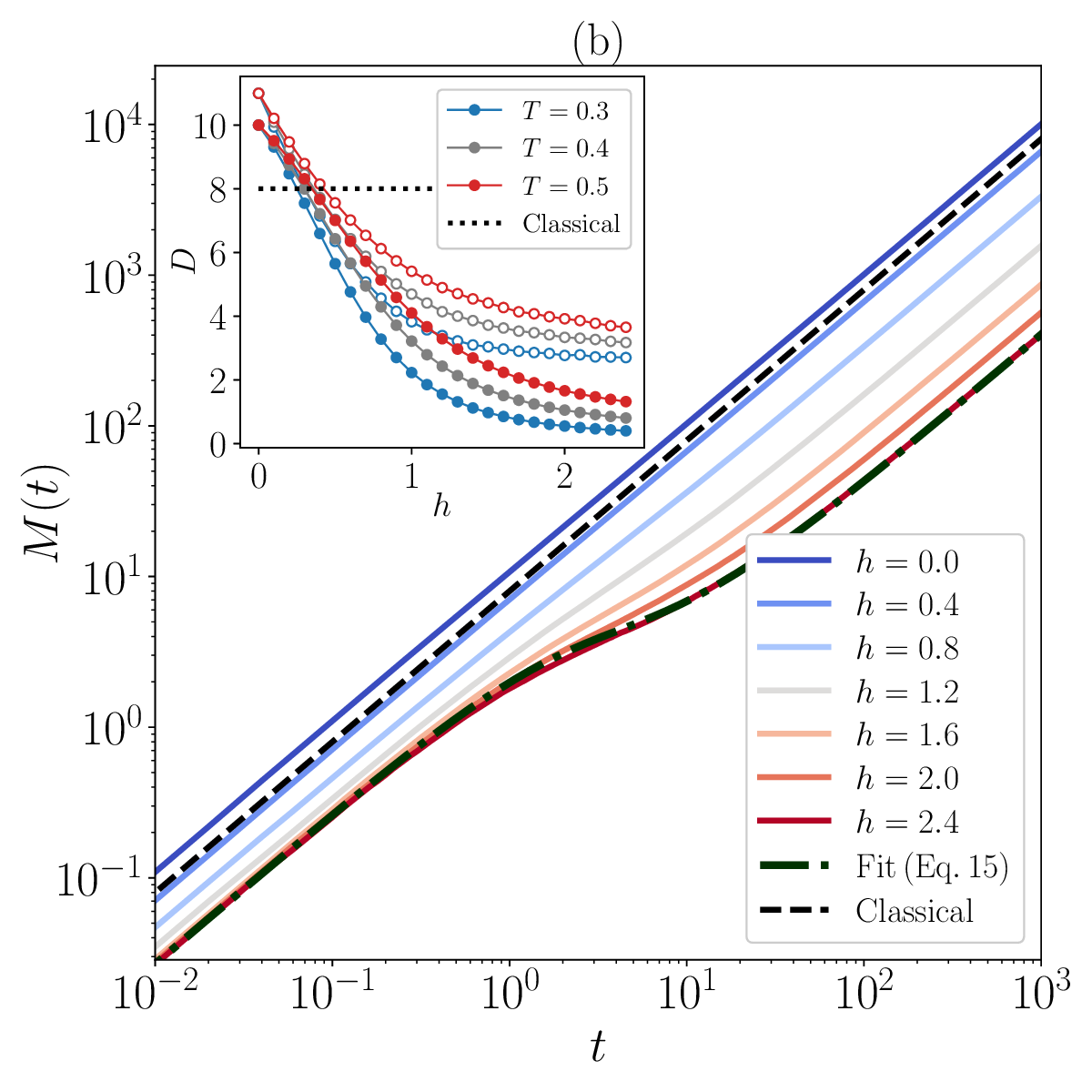}
\end{minipage}
    \caption{\label{fig:rw} Quasiparticle random walk dynamics: (a) Illustration of the random walk model. We associate each pair of nearest-neighbor classical spins and each cluster with a single site on the {random walk} lattice. An excitation can manifest itself either classically (as anti-aligned spin pairs) or as a cluster with anti-aligned boundary spins. These excitations move by flipping classical spins in the system. Labeling of the quantum clusters follow the convention given in App.~\ref{app:A}. (b) Results for the MSD, $M(t)$, with $T = 0.3,\, r = 0.2$. One sees the emergence of two distinct diffusive regimes for $t\to0$ and $t\to\infty$ as $h$ increases, deviating (in both cases) from the {\text{red} classical scenario}, where $M(t)$ exhibits the same linear scaling throughout. We also plot the fits to the data with $h=2.4$ according to Eq.~\eqref{eq:3:expfit}. Inset: estimates of the diffusion constants $D_0$ (empty markers) and $D_\infty$ (filled markers), plotted as a function of $h$ and $T$, for $r=0.2$. They both decrease as $h$ increases, yet their difference increases with $h$.} 
\end{figure*}
As mentioned above, the magnetization dynamics of the RTFIC is dominated by the motion of the sparse kinks and anti-kinks at low temperature. The sparseness allows us to neglect any interactions between them and focus on the random walks of single excitations for the rest of this section. 

It can be demonstrated \cite{chandler1987introduction, nishi2018symmetrical} that the PSD of the signal generated by a single random walker, $X(t)$, can be expressed as
\begin{equation}\label{eq:3:rwpsd}
    \mathrm{PSD}(\omega)=\frac{1}{\omega}\int_0^\infty\mathrm{d}t\sin(\omega t)\dot{M}(t)\, ,
\end{equation}
where $M(t)$ is the mean-square displacement of the random walk, defined as
\begin{equation}\label{eq:3:msd}
    M(t) \equiv \left\langle \left[X(t)-X(0)\right]^2\right\rangle \, .
\end{equation}
In our case, $X(t)$ is identified as the magnetization of the system. The modifications of the PSD seen in the previous section due to the quantum clusters can then be attributed to the interplay between the clusters and the mobile {random walk} excitations, which alters $M(t)$. 

The focus of the first part of this section, then, is to establish a random walk model. For convenience, we present here only the key steps and details can be found in App.~\ref{app:C}. 
We simulate a random walk taking place on a one-dimensional lattice where each quantum cluster is identified with a single lattice site --- see Fig.~\ref{fig:rw}a for a pictorial description. 

Each site of the {random walk} lattice corresponds to a possible position of a single excitation, which can be mapped to the magnetization of the system injectively (see App.~\ref{app:C}). In the classical scenario, excitations manifest themselves classically as domain walls, i.e., anti-aligned nearest-neighbor classical spin pairs. On the {random walk} lattice, this corresponds to a `classical' site in the middle of the corresponding bond. When quantum clusters are present, excitations may also manifest themselves as clusters with anti-aligned boundary spins, see Fig.~\ref{fig:rw}a. The cases $h=0$ and finite $h$ are drastically different for the clusters: at $h=0$, the ground state of such a cluster with anti-aligned boundary spins is $n+1$-fold degenerate, corresponding to the $n+1$ possible positions of a single domain wall; any infinitesimal but nonzero $h$ lifts this degeneracy, creating quantum superpositions that necessitate the cluster to be treated as a single site. Such modification of the {random walk} lattice for arbitrary $h>0$ explains the deviation between the RTFIC at $h\to 0$ to the classical scenario as mentioned above.

Motion on the sites of the {random walk} lattice takes place by flipping spins. In between classical sites, domain walls move by one lattice spacing when one of their spins is flipped, inducing a magnetization change $\Delta m = 2$. On the contrary, moving into, out of and in between quantum clusters is accomplished by flipping one of the boundary spins. Such moves can involve several sites of the original lattice, and non-trivial changes in magnetization; according to our mapping, however, these processes still correspond to nearest-neighbor motion on the {random walk} lattice. This subtlety will require careful modeling in the following. 

The nearest-neighbor nature of the random walk yields the following Master equation \cite{haus1987diffusion}:
\begin{equation}\label{eq:3:rwme}
    \frac{\mathrm{d}}{\mathrm{d}t}P_i = W_{i+1, i}P_{i+1}+W_{i-1, i}P_{i-1}-(W_{i,i+1}+W_{i,i-1})P_{i}
    \, ,
\end{equation}
where $P_i$ denotes the probability of finding the walker at site $i$ of the {random walk} lattice, and $W_{ij}$ is the transition rate from site $i$ to site $j$. The latter is related to the {Metropolis-Hastings} probability, Eq.~\eqref{eq:2:qcmcp}, of the spin flip that induces the hopping from site $i$ to site $j$. Such probability is $1$ when both sites are classical, giving $W_{ij} = 1$; in all other cases, $W_{ij}$ take a value determined by the complex energy landscapes of the relevant quantum clusters. These values are explicitly evaluated in App.~\ref{app:C}.

As explained above, not all nearest-neighbor hops on the {random walk} lattice result in the same change in magnetization of the system. It is therefore necessary to complement the Master equation, Eq.~\eqref{eq:3:rwme}, with a way to account for the size and energy level structure of the quantum clusters. 
This is equivalent to a heterogeneous {random walk} lattice with varying step lengths across the quantum clusters. In App.~\ref{app:C}, we present details of the evaluation of the magnetization changes and corresponding step lengths. 

Simulating this random walk and extracting the MSD, $M(t)$, yields Fig.~\ref{fig:rw}b. Similarly to the PSD, $M(t)$ also demonstrates two distinct diffusive regimes, and a crossover between them. For $t\to 0$ and $t\to \infty$, the MSD exhibits linear scaling ($M(t)\propto t$). These two regimes are separated by an intermediate shoulder with slower-than-linear scaling. 

By fitting the asymptotic linear behavior, $\lim_{t\to0}M(t)=D_0 t$ and $\lim_{t\to \infty}M(t)=D_\infty t$, we obtain the diffusion constants $D_0$ and $D_\infty$. In the classical scenario, where the excitations perform a conventional random walk, we have $M(t)=Dt$ with $D = 8$ at all times (where a factor of $2$ comes from coordination number, and another factor of $2^2=4$ comes from the unit step length $\Delta m=2$ induced by flipping a classical spin and the square scaling relationship between the diffusion constants and the unit bond length \cite{haus1987diffusion}). For the RTFIC, results for $D_0$ and $D_\infty$ for varying $h$ and $T$ are shown in the inset of Fig.~\ref{fig:rw}b, where we see the most prominent deviations between $D_0$ and $D_\infty$ at large $h$ and small $T$. 
Note that, as mentioned before, the results for $h=0^{+}$ for the RTFIC are not continuously connected with the { classical behavior}, due to the degeneracy-lifting effect of the transverse fields. 

\begin{figure}[t!]
    \includegraphics[width=\columnwidth]{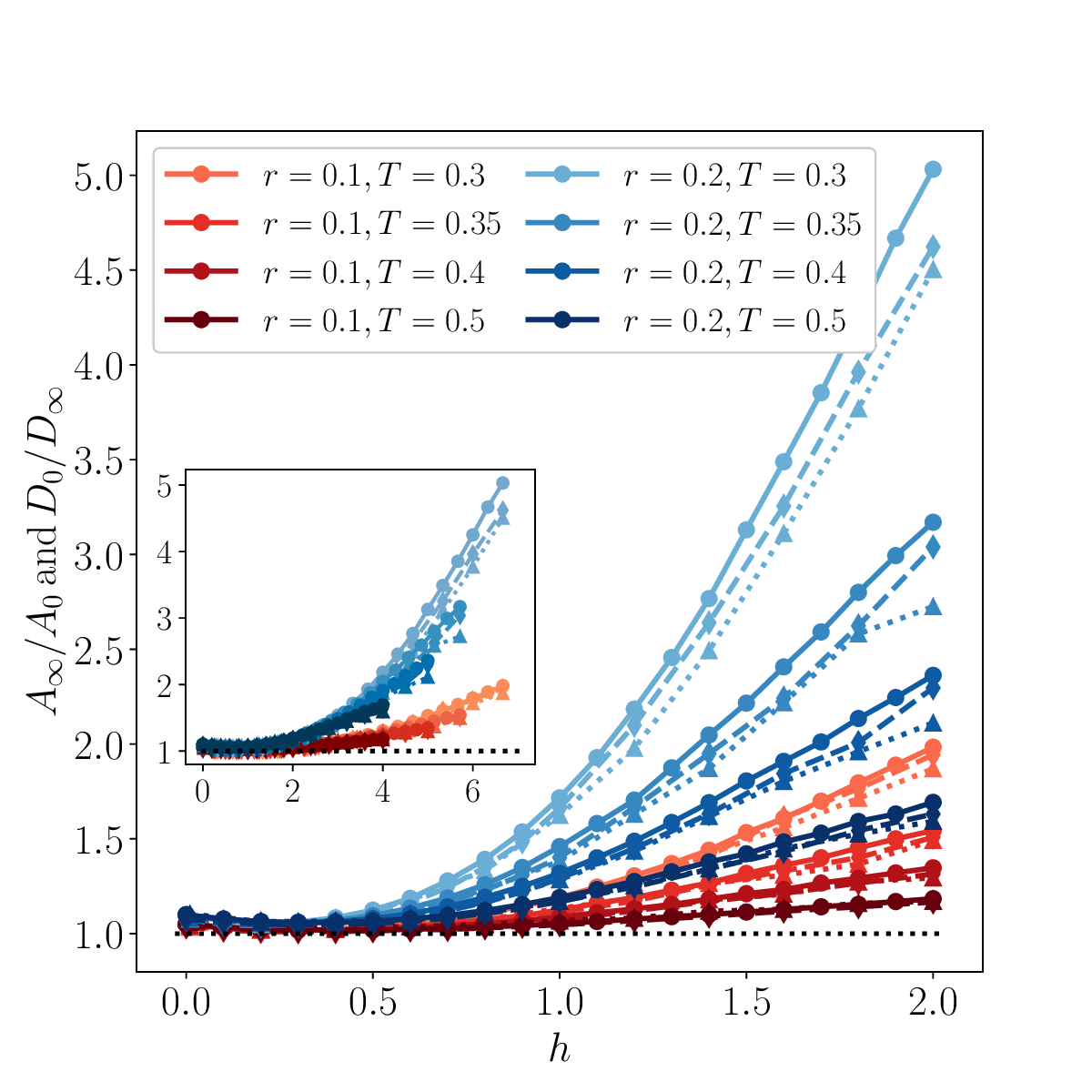}
    \caption{\label{fig:psdmsd} Results for $D_0/D_\infty$ from the random walk model (round points, solid lines) and $A_\infty / A_0$ extracted from computing the PSD of the random walk (rhombic points, dashed lines) and the full-scale spin simulations (triangular points, dotted lines) presented in Fig.~\ref{fig:PSD}. The horizontal dotted line represents the value of $A_\infty/A_0=D_0/D_\infty=1$ {for the classical case}. Consistency between the three groups of data validates the single-excitation picture and the approximation of the form Eq.~\eqref{eq:3:expfit} used in the main text. Inset: the same data plotted against $h/T$ rather than $h$; they scale with $h/T$ at small values, while they deviate at large values.}
\end{figure}

%
%

\subsection{\label{section:3:discussion}Discussion}
Similar behavior for the MSD has been observed in other contexts, such as studies of colloid diffusion \cite{choudhury2017active, su2017colloidal, hanes2012dynamics} and Brownian motion in nonuniform potentials \cite{evers2013particle, fulde1975problem, festa1978diffusion}, which can all be reduced to heterogeneous random walk problems. The appearance of the two diffusion constants in Fig.~\ref{fig:rw}b signals different behaviors in the two limits: for $t\to 0$ the random walker has not yet felt the heterogeneity of the system; for $t\to\infty$ the walker effectively moves on a coarse-grained landscape with a renormalized diffusion constant $D_\infty$. 

One can approximate the MSD, using a form conventionally adopted in the literature \cite{choudhury2017active}:
\begin{equation}\label{eq:3:expfit}
    M(t) = D_\infty t - (D_0-D_\infty)\tau (\mathrm{e}^{-t/\tau}-1)
    \, ,
\end{equation}
where the only fitting parameter $\tau$ can be regarded as some emergent, phenomenological crossover timescale; it is of the order of unity for all parameter combinations considered. This form agrees well with our data, as can be seen from the fit in Fig.~\ref{fig:rw}b (dash-dotted line). Note that the fit does deviate slightly in the crossover region, signaling the incompleteness of this single-timescale analysis. 

The usefulness of this approximation lies in the fact that it enables us to evaluate explicitly Eq.~\eqref{eq:3:rwpsd} using Eq.~\eqref{eq:3:expfit}. We can therefore check the consistency between the results obtained for the MSD discussed in Sec.~\ref{section:3:rw}, and those for the PSD discussed in Sec.~\ref{section:3:psd}. Importantly, Eq.~\eqref{eq:3:expfit} implies the relation
\begin{equation}\label{eq:3:psdmsd}
    \frac{A_\infty}{A_0} = \frac{D_0}{D_\infty}
    \, ,
\end{equation}
obtained by substituting Eq.~\eqref{eq:3:expfit} into Eq.~\eqref{eq:3:rwpsd}.

To verify this, we plot the quantities on both sides in Fig.~\ref{fig:psdmsd}. A good consistency is observed between the data for $D_0/D_\infty$, obtained from the random walk, and the data for $A_\infty/A_0$ obtained by simulating both the PSD of the random walk and the full spin system. This validates the single-excitation picture proposed at the beginning of Sec.~\ref{section:3:rw} and the approximation Eq.~\eqref{eq:3:expfit} just made (and the deviation between these data for large $h$ are also likely due to them being inaccurate in this region). Both quantities are seen to be increasing with $h, \, r$ and decreasing with $T$. Furthermore, an interesting data collapse of these quantities is observed at small $h$ when plotted against $h/T$. 

The above investigations highlight the usefulness of two key quantities --- $A_\infty/A_0$, derived from the magnetic noise of the system, and $D_0/D_\infty$, obtained from studying the single-excitation dynamics. Both take the trivial value of $1$ in the {classical scenario} and can serve to quantify deviations from diffusive behaviors. A full analytical solution for the diffusion constants $D_0$ and $D_\infty$ is provided in Appendix~\ref{app:C}.
%
%

\begin{figure*}[htb!]
    \includegraphics[width=0.3\textwidth]{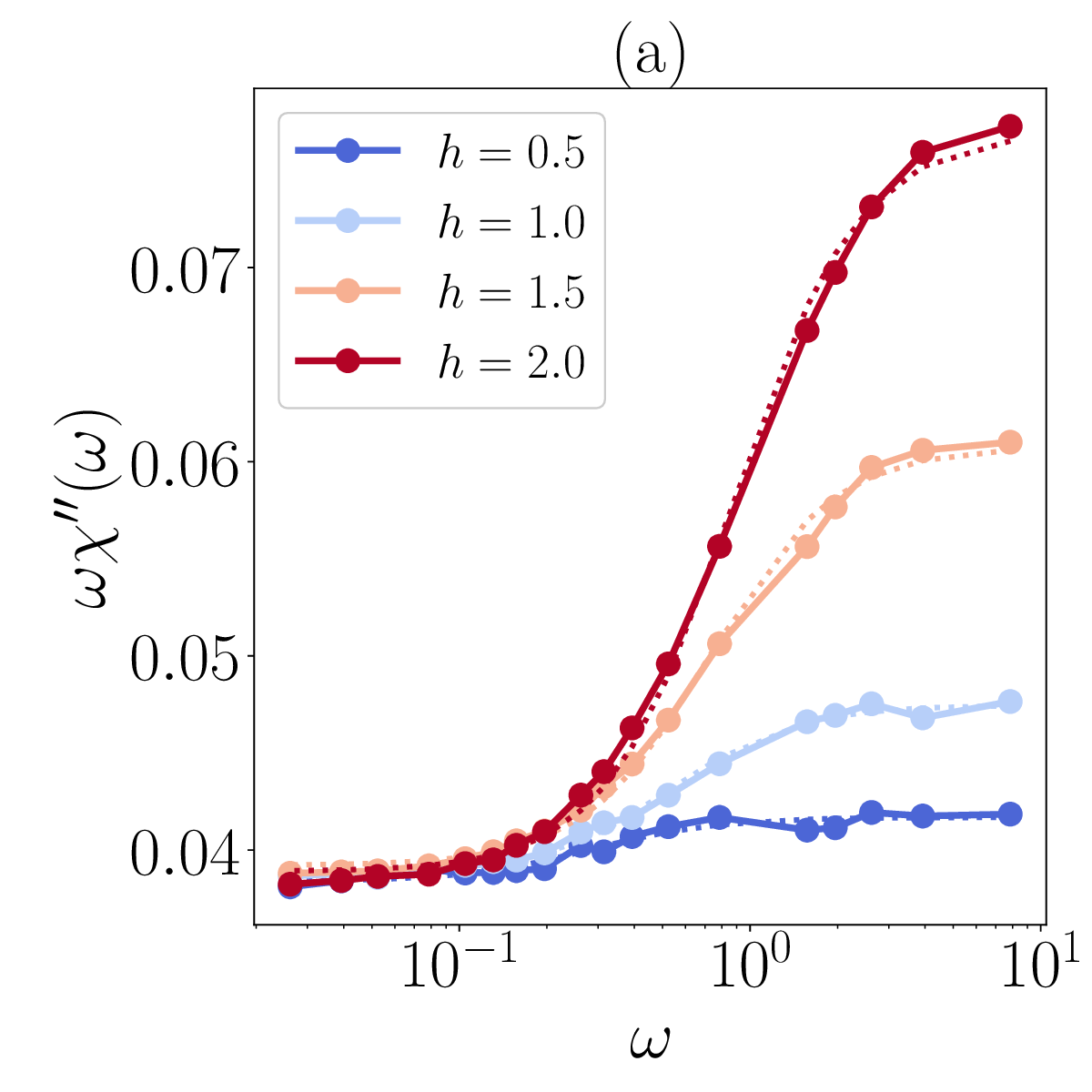}
    \includegraphics[width=0.3\textwidth]{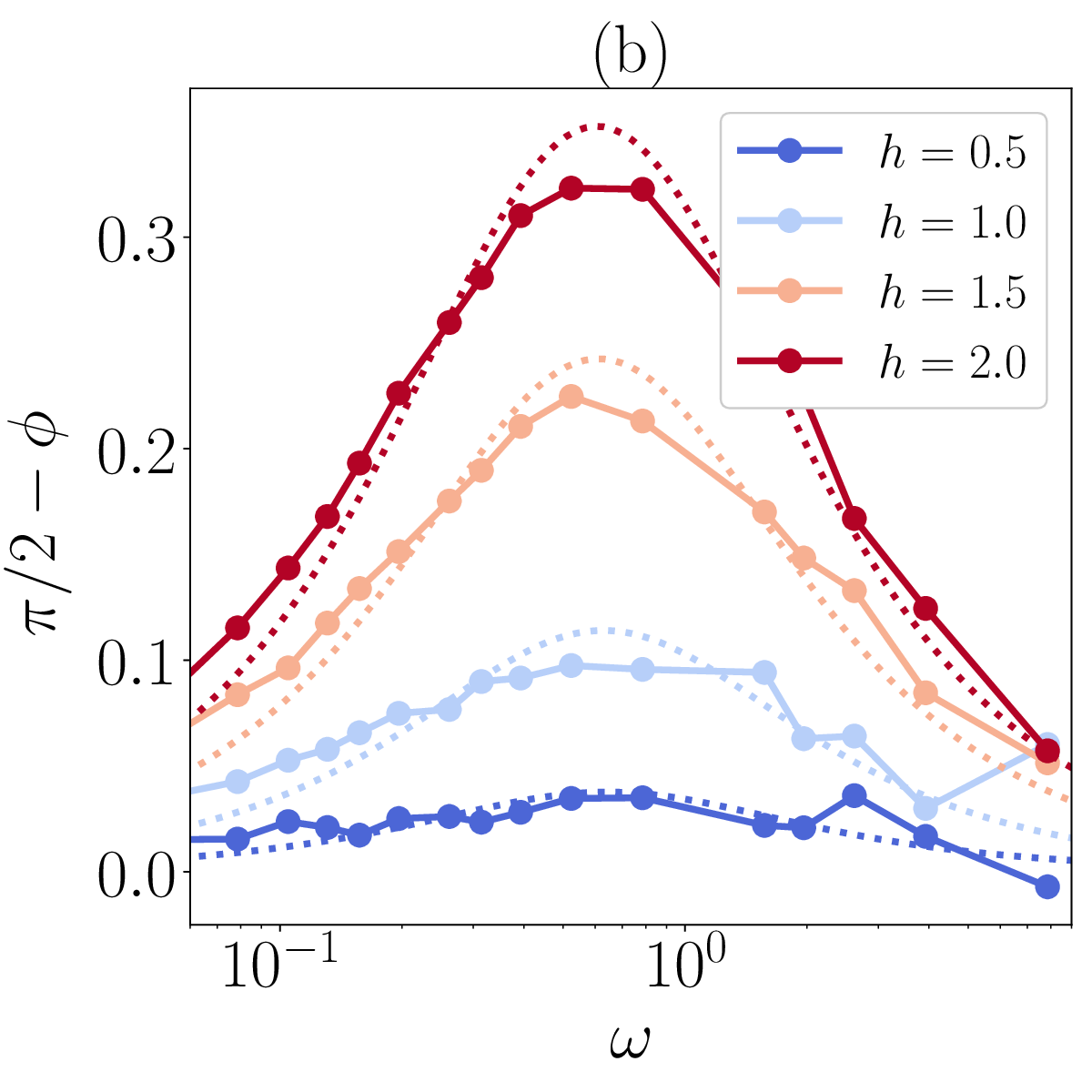}
    \includegraphics[width=0.3\textwidth]{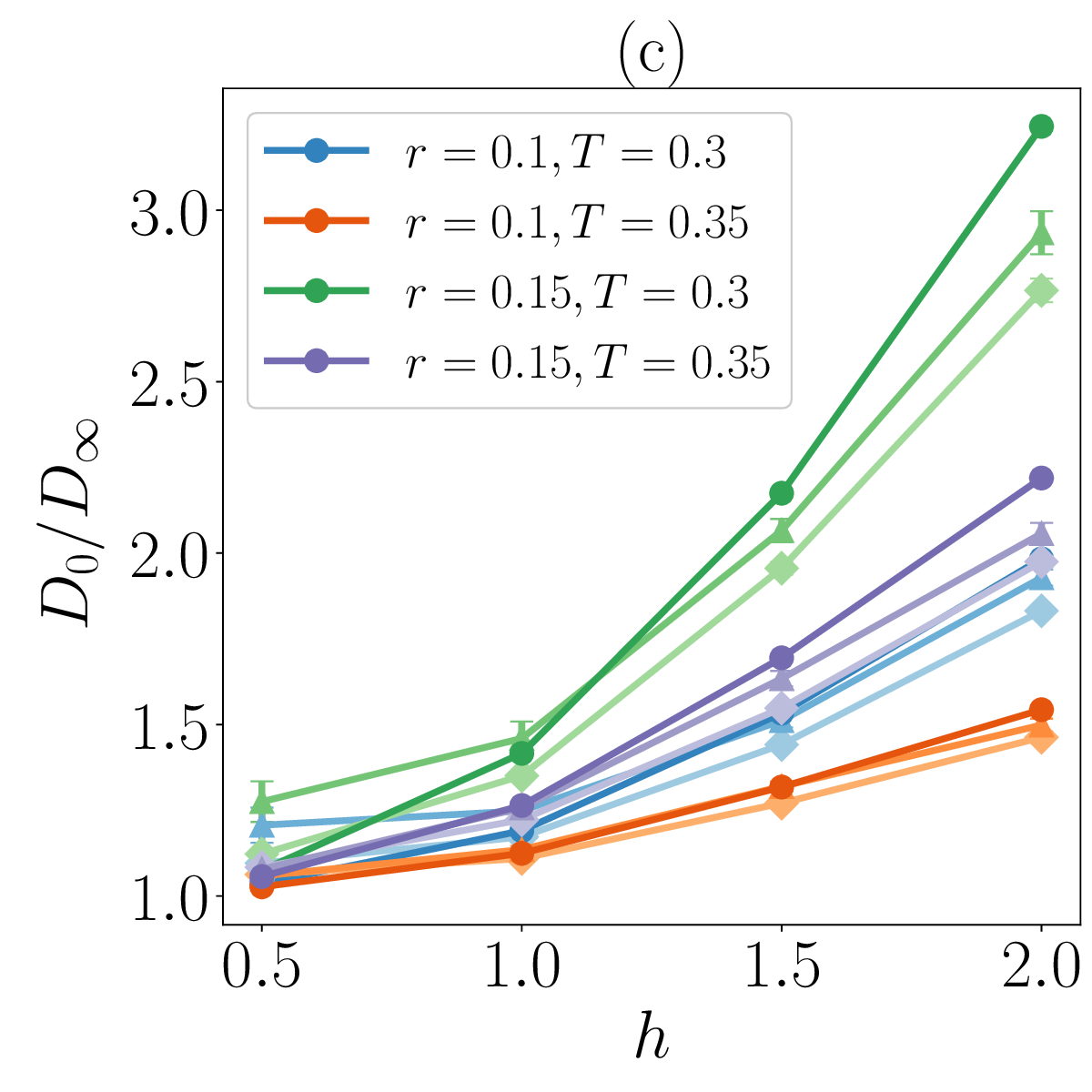}
    \caption{\label{fig:4:drtfic} 
    Monte Carlo simulations of the driven RTFIC: (a) $\omega\chi''(\omega)$ and (b) $\pi/2-\phi$, for $T=0.35$, $r=0.15$ and $B_0=0.02$. Fits to the data, discussed in the main text, are presented as dotted lines. $\omega\chi''(\omega)$ and $\omega^2\,\mathrm{PSD}(\omega)$ display similar behavior, as demonstrated in the inset of Fig.~\ref{fig:PSD}, and $\phi$ deviates from the classical value $\pi/2$ most prominently at large $h$ and $\omega\sim10^0$. (c) Fitted values for $D_0/D_\infty$ using $\chi''(\omega)$ data (rhombic points) and $\phi$ data (triangular points), are compared with those calculated in Sec.~\ref{section:3:rw} (circular points).}
\end{figure*}

\section{\label{section:4:driven}Driven dynamics}
In the previous section, we studied the properties of the magnetic fluctuations of the RTFIC in thermodynamic equilibrium, which we were able to model as a heterogeneous random walk process performed by emergent excitations. It would then be of interest to study the behavior of these excitations under an external driving field. Recent studies on spin ice systems \cite{nilsson2023dynamics, hsu2024dichotomous} have provided an example of how microscopic details of the dynamics can reflect in a major way in its response to external driving, which we intend to investigate here in our model.
%
%

\subsection{\label{section:4:model}Model}
We consider the RTFIC, Eq.~\eqref{eq:2:rtfic}, under a sinusoidally oscillating longitudinal driving field {\cite{chakrabarti1999dynamic}}
\begin{equation}\label{eq:4:Drtfic}
    \mathcal{H}' = -\sum_{i}S_i^zS_{i+1}^z-h\sum_{\alpha\in\{\alpha\}}S_\alpha^x - B(t)\sum_{i}S_i^z
    \, ,
\end{equation}
where $B(t)=B_0\sin(\omega t)$; $B_0$ and $\omega$ are the field strength and angular frequency parameters. Once again, we couple Eq.~\eqref{eq:4:Drtfic} to the dynamics of Eq.~\eqref{eq:2:qcmc}, assuming that the Lindbladian approach formulated in Sec.~\ref{section:2:rtfic} remains valid, and that the state of a quantum cluster changes adiabatically under driving, i.e., it stays on the same eigenstate (which now is time-dependent) unless impacted by a spin flip event.

In the driven model, the system responds to the oscillating field $B(t)$ with a time-dependent magnetization of the form $m(t) = m_0\sin(\omega t + \phi)$, where $\phi$ is the phase shift. We are interested in the magnetic susceptibility of the system, $\chi(\omega) = m_0\mathrm{e}^{i\phi}/B_0$ and its real and imaginary components, denoted by $\chi'(\omega)$ and $\chi''(\omega)$ respectively. {These quantities are closely related via the fluctuation-dissipation theorem to the PSD of the system and, at low temperatures, to the MSD of single excitations that we studied in the previous section}. In particular, we have
\begin{equation}\label{eq:4:chipp}
    2T\chi''(\omega) = \omega \mathrm{PSD}(\omega) = \int_0^\infty\mathrm{d}t\sin(\omega t)\dot{M}(t)
    \, ,
\end{equation}
and
\begin{equation}\label{eq:4:chip}
    2T\chi'(\omega) = \int_0^\infty\mathrm{d}t\cos(\omega t)\dot{M}(t)
    \, .
\end{equation}

We start with the {classical version of Eq.~\eqref{eq:4:Drtfic}}, where, {for a single mobile excitation}, $M(t)$ is linear in time. Performing the appropriate integration in Eq.~\eqref{eq:4:chipp} yields $\mathrm{PSD}(\omega) \propto \omega^{-2}$, and therefore an $\omega$-independent $\omega\chi''(\omega)$. Furthermore, Eq.~\eqref{eq:4:chip} yields $\chi'(\omega) = 0$, giving $\phi = \pi/2$. 

We then turn to the RTFIC, where the presence of quantum clusters induces additional structure in $M(t)$ and in the PSD. Using the approximated form of the MSD proposed towards the end of Sec.~\ref{section:3:discussion}, Eq.~\eqref{eq:3:expfit}, the integrals in Eqs.~\eqref{eq:4:chipp} and~\eqref{eq:4:chip} can be explicitly computed: 
\begin{subequations}\label{eq:4:chi2}
\begin{equation}\label{eq:4:chipp2}
    \chi''(\omega) = \beta\frac{D_\infty+D_0\tau^2\omega^2}{\omega(1+\tau^2\omega^2)}
    \, , 
\end{equation}
\begin{equation}\label{eq:4:chip2}
    \chi'(\omega) = \beta\frac{(D_0 - D_\infty)\tau}{1+\tau^2\omega^2}
    \, ,
\end{equation}
\end{subequations}
and thence
\begin{equation}\label{eq:4:phi}
    \tan\phi = \frac{\chi''}{\chi'} = \frac{D_\infty + D_0 \omega^2\tau^2}{(D_0-D_\infty)\omega\tau}
    \, .
\end{equation}
In this case, $\omega\chi''(\omega)$, or equivalently $\omega^2\,\mathrm{PSD}(\omega)$, exhibits a plateau at low- and high-frequencies, with a transition between them around $\omega\sim 10^{0}$, as shown in the inset of Fig.~\ref{fig:PSD}a. Correspondingly, $\phi$ deviates from the classical value of $\pi/2$, with their difference $\pi/2-\phi$ exhibiting a peak when $\omega=\sqrt{D_\infty/D_0}\tau^{-1}$ while vanishing in the limits $\omega\to0$ and $\omega\to\infty$, deep within the two diffusive regimes. 
%
%

\subsection{\label{section:4:results}Results}
We performed Monte-Carlo simulations of the driven model to verify these theoretical predictions made for the RTFIC, namely Eqs.~\eqref{eq:4:chi2} and~\eqref{eq:4:phi}. We start by extracting the magnetization of the system as a function of time, $m(t)$. Details of simulating the time-dependent Hamiltonian Eq.~\eqref{eq:4:Drtfic} are given in App.~\ref{app:F}. An additional step is required to compare with the results in Sec.~\ref{section:4:model}. Due to the introduction of the driving term, 
the eigenstates and their magnetization become dependent on the applied field $B$ and thus implicitly on $t$, giving rise to a time-dependent $m(t)$ even in the absence of any stochastic dynamics. This contribution needs to be accounted for to allow comparison with Eqs.~\eqref{eq:4:chipp2} and~\eqref{eq:4:phi}, which are based on the random walk theory developed for the original Hamiltonian Eq.~\eqref{eq:2:rtfic}. Details for this step are presented in App.~\ref{app:D}. 

We then fit the corrected $m(t)$ to the form $m(t)=m_0\sin(\omega t+\phi)$ to extract $m_0$ and $\phi$, which are used to determine $\chi''(\omega)$. The results obtained for $\omega\chi''(\omega)$ and $\pi/2-\phi$ are presented in Figs.~\ref{fig:4:drtfic}a and~\ref{fig:4:drtfic}b. (Here, $\chi''(\omega)$ per-spin is measured, which differs from its single-excitation value in Eq.~\eqref{eq:4:chipp2} by a temperature-dependent scaling factor.) 

We observe overall agreement with the predictions made in the previous section. 
This is quantitatively checked by fitting the results for $\chi''(\omega)$ using Eq.~\eqref{eq:4:chipp2} (up to a constant scaling factor) and those for $\phi$ using Eq.~\eqref{eq:4:phi}, with the same fitting parameters $D_\infty$, $D_0$ and $\tau$; we then compare the fitted ratio $D_0/D_\infty$ to the one calculated in our random walk analysis in Sec.~\ref{section:3:rw} and App.~\ref{app:C}. The comparison is shown in Fig.~\ref{fig:4:drtfic}c, where we observe good agreement for small values of $r$ and large $T$, with progressive deviations as $r$ is increased and $T$ decreased. This behavior can be attributed to various sources of error, in particular the sensitivity of the quantum clusters to the external field, making investigations with finite $B_0$ inaccurate in nature. The value $B_0=0.02$ used in plotting Fig.~\ref{fig:4:drtfic} is a compromise between this and simulation accuracy. Nonetheless, the overall consistency demonstrates the validity of our theoretical analysis of the effect of the random transverse fields on the magnetic susceptibility $\chi$ of the RTFIC system. 
%
%

\begin{figure*}[htb!]
    \includegraphics[width=0.25\textwidth]{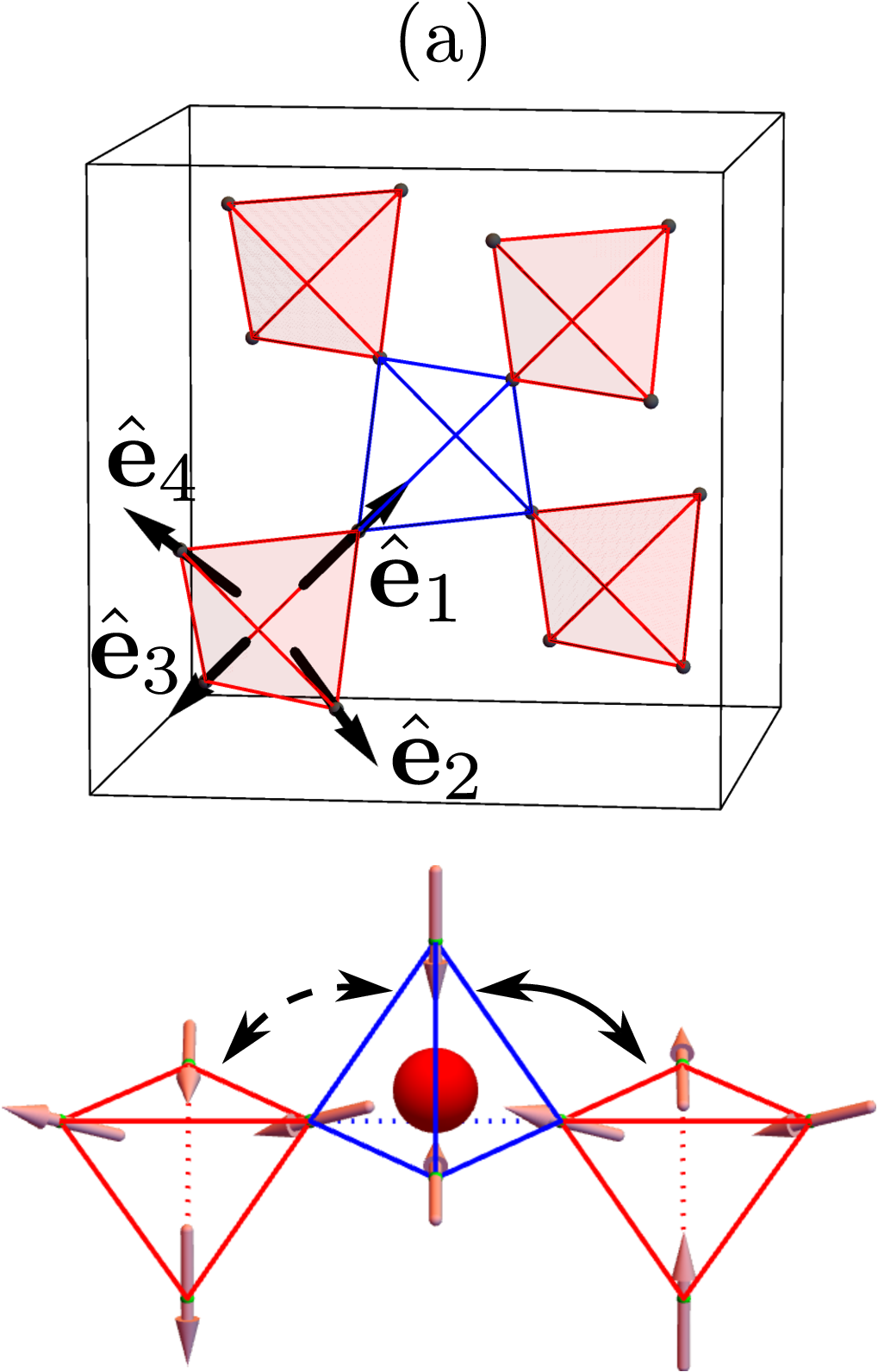}
    \includegraphics[width=0.34\textwidth]{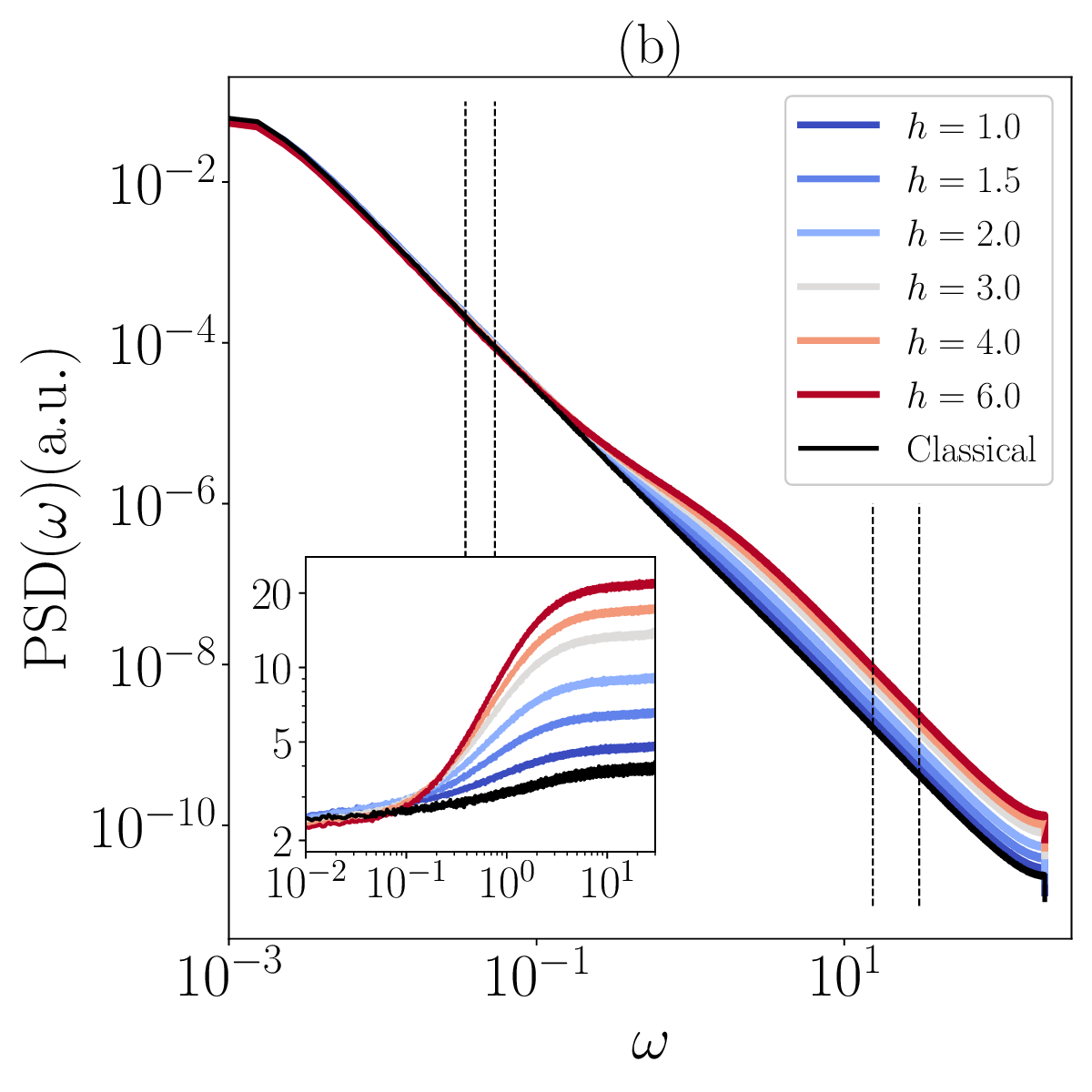}
    \includegraphics[width=0.34\textwidth]{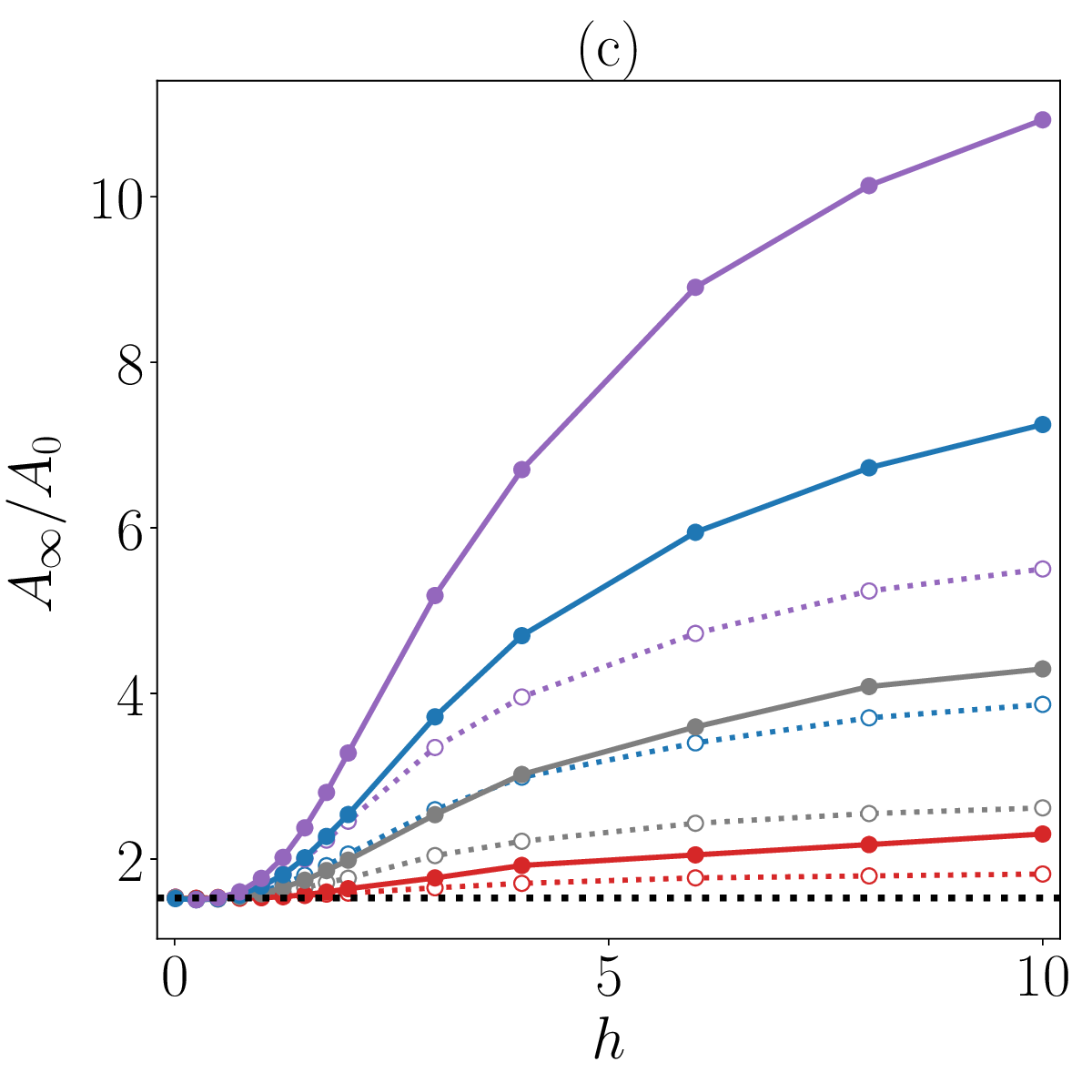}
    \caption{\label{fig:5:rtfsi} Simulation on the three-dimensional CSI and RTFSI models: (a) Top: Unit cell of the pyrochlore lattice, consisting of 16 spin sites and 4 non-touching tetrahedra (red). The directions of the four easy axes are also shown. Bottom: Schematic plot of a hopping monopole in CSI. The middle tetrahedron hosts a monopole (1-in-3-out configuration); while hopping to the right is permitted without increasing the energy, hopping to the left creates a double monopole (all-in configuration) and is hence energetically suppressed at low temperature. (b) $\mathrm{PSD}(\omega)$ with varying values of $h$ for RTFSI, $r=0.08$ and $T=0.3$. The PSD for the {standard model} dynamics of the classical model is also shown for reference { black dash-dot line}. Inset: $\omega^2\,\mathrm{PSD}(\omega)$, focusing on the intermediate region and scaled by a factor of $10^7$. (c) Values of $A_\infty/A_0$, obtained by fitting the two regions of the PSD bounded by the dashed vertical lines to the form $\mathrm{PSD}(\omega)=A\omega^{-2}$, as a function of $h$; results for $r=0.02$ (red), $r=0.04$ (gray), $r=0.06$ (blue), $r=0.08$ (purple), $T=0.3$ (solid lines, filled markers) and $T=0.35$ (dotted lines, empty markers) are shown. The dashed horizontal line represents the value $A_\infty/A_0\approx1.53$ measured for CSI, see main text.}
\end{figure*}

\section{\label{section:5:rtfsi}Random transverse field spin ice}
Up until now, we have been focusing on the one-dimensional RTFIC, whose dynamics can be understood analytically. Here, we generalize our approach to the three-dimensional counterpart of the RTFIC, the random transverse field spin ice (RTFSI) model, to allow for comparison with spin ice experiments and models. 
%
%

\subsection{\label{section:5:setup}Setup}
As shown in Fig.~\ref{fig:5:rtfsi}a, the three-dimensional RTFSI model involves effective spin-$1/2$ degrees of freedom sitting on a pyrochlore lattice. The Hamiltonian reads
\begin{equation}
    \mathcal{H}_{\mathrm{RTFSI}} = \sum_{\langle ij\rangle}S_i^zS_j^z{-h}\sum_{\{\alpha\}}S_\alpha^x\,,
\end{equation}
where the $S_i$'s are now Pauli operators in the local basis. As illustrated in Fig.~\ref{fig:5:rtfsi}a, around a tetrahedron the four local easy-axes are given by 
\begin{equation}
\begin{aligned}
    \hat{\mathbf{e}}_1 &=\frac{1}{\sqrt{3}}(1, 1, 1)\,,\quad
    \hat{\mathbf{e}}_2 =\frac{1}{\sqrt{3}}(1, -1, -1)\,,\quad  \\
    \hat{\mathbf{e}}_3 &=\frac{1}{\sqrt{3}}(-1, 1, -1)\,,\quad 
    \hat{\mathbf{e}}_4 =\frac{1}{\sqrt{3}}(-1, -1, 1)
    \, .
\end{aligned}
\end{equation}
Similar to the RTFIC, the set $\{\alpha\}$ represents the random sites of the pyrochlore lattice affected by the transverse fields. In real materials, they may be induced for instance by oxygen depletion \cite{sala2014vacancy} which has been recently argued to affect sites in pairs \cite{goff}, but we consider here the simpler case of uncorrelated isolated sites as an initial investigation.

We probe the dynamics of the RTFSI model using the same modified Monte-Carlo method outlined at the end of Sec.~\ref{section:2:model}. Namely, we sample the Master equation Eq.~\eqref{eq:2:rme}. Without the random transverse fields, such dynamics would be identical to the `standard model' dynamics of the CSI \cite{jaubert2009signature, nilsson2023dynamics}. This is the starting point of our analysis. For simplicity, a strong magnetic anisotropy is assumed also for the spins affected by the transverse fields, and only their easy-axis ($z$) component is taken to contribute to the total magnetization of the system. The RTFSI system is isotropic in the $x$-, $y$- and $z$-directions; for concreteness, we choose to compute the PSD using the average over $m_x$, $m_y$ and $m_z$, or equivalently the magnetization along the [111] direction. See App.~\ref{app:F} for simulation details. 
%
%

\subsection{\label{section:5:results}Results}
We start by discussing the behavior of the {standard model} dynamics. Similarly to the {classical kinetic Ising chain}, at low temperature the magnetic noise of the CSI in equilibrium is dominated by the contribution from the random walking point-like excitations known as magnetic monopoles. However, such a random walk is constrained: for a given isolated monopole, the move corresponding to flipping the minority spin of the corresponding 1-in-3-out (3-in-1-out) tetrahedron is energetically suppressed (see bottom panel of Fig.~\ref{fig:5:rtfsi}a), as it creates a double-monopole corresponding to an all-out (all-in) tetrahedron. This is different from the 1D scenario where the domain walls can freely move in all directions. 

To study the effect of such a constraint, one can propose a toy random walk problem on the diamond lattice where the walker can always backtrack, but when going forward it finds that one of the three possible directions is blocked at random. This problem can be solved exactly and shown to have diffusive behavior at short and long times, but with different diffusion constants --- in particular, $D_0=3$ while $D_\infty = 2.4$ exactly (notation inherited from Sec.~\ref{section:3:rw}, see App.~\ref{app:E} for details). 

The situation in spin ice is of course different, with important correlations between the blocked directions. This no longer affords access to an exact solution, and we rely on numerics. Our results suggest that similar behavior comes to pass, with the short time diffusive process remaining identical to the above toy model ($D_0=3$), whereas the long time diffusive process is significantly modified, with $D_\infty$ lying in between $1.85$ and $1.90$, up to finite size effects (see App.~\ref{app:E}). 

Extending our arguments from Sec.~\ref{section:3:discussion}, such a deviation between the short- and long-time limit of the monopole random walk implies a PSD profile similar to that depicted in Fig.~\ref{fig:PSD}a --- namely, two distinct diffusive regions with $\mathrm{PSD}\propto \omega^{-2}$ at low and high frequencies. In Fig.~\ref{fig:5:rtfsi}b, we plot the PSD (black lines) as well as $\omega^2\,\mathrm{PSD}(\omega)$ (inset) for CSI; from the latter, one can see the presence of two distinct plateaux at low and high frequencies. Performing a similar analysis to the one described in Sec.~\ref{section:3:psd}, we fit the function $\mathrm{PSD}(\omega)=A\omega^{-2}$ to the low- and high-frequency part of the PSD and extract the proportionality factors $A_0 = \lim_{\omega\to0}\omega^2\, \mathrm{PSD}(\omega)$ and $A_\infty = \lim_{\omega\to\infty}\omega^2\, \mathrm{PSD}(\omega)$. We then use $A_\infty/A_0$ as an indicator of the deviation of the PSD from diffusive behaviors, as described at the end of Sec.~\ref{section:3:discussion}, and we obtain a $A_\infty/A_0\approx 1.53$ for CSI. This is in agreement with the random walk investigations performed above (with $D_0/D_\infty\approx1.6$) and Eq.~\eqref{eq:3:psdmsd}. 

We now turn to discuss the results for the RTFSI model. { Here, we choose to simulate values of $r$ much less than those used in 1D. This is due to the fact that the effects of the quantum clusters are more prominent in the three dimensional spin ice model than in 1D, owing to the larger coordination number (which both leads to larger clusters and a larger number of neighboring classical spins -- an isolated quantum spin, for example, has two neighbors in 1D but six in 3D). 
Moreover, the choice also helps contain the computational efforts required for our study. Specifically, we restrict to $0<r\le0.08$ and we study a broader range of $h$ so as to better resolve the effects of the quantum clusters.} The results are presented in Figs.~\ref{fig:5:rtfsi}b and~\ref{fig:5:rtfsi}c; similarly to the 1D case, the introduction of the random transverse fields increases the deviation between the low and high frequency regimes of the PSD, and hence enlarges $A_\infty/A_0$ with respect to the classical behavior. Fig.~\ref{fig:5:rtfsi}c reveals this quantity to increase as a function of $h$ and $r$ and decrease with $T$, also in agreement with our 1D results (see Fig.~\ref{fig:psdmsd}). 

The resemblance between the results in 1D and 3D allows us to extend the insights gained for the RTFIC in the previous chapters to the RTFSI, namely those in  Secs.~\ref{section:3:rw} and~\ref{section:3:discussion}. The presence of the random transverse fields augments the deviation between the short- and long-time limit of the monopole random walk in the single excitation limit in CSI, which is demonstrated in the PSD data via Eq.~\eqref{eq:3:psdmsd}. 
In Fig.~\ref{fig:5:rtfsi}c, as $h$ is increased, $A_\infty/A_0$ appears to tend to a finite saturation value for the range of $r$ and $T$ probed in our work. 
The $h\to\infty$ limit in the RTFSI corresponds to the quantum spins completely polarized in a direction orthogonal to their easy-axis, and thence vanishing magnetization. Such a scenario is equivalent to non-magnetic dilutions in spin-ice materials~\cite{sala2014vacancy,revell2013evidence, sen2015topological}.
%
%

\section{\label{section:6:discussion}Discussion and outlook}
In this paper, we investigated a random transverse field Ising chain (RTFIC), Eq.~\eqref{eq:2:rtfic}, inspired by the behavior of non-Kramers spin ice materials, and we conducted a comprehensive study of its dynamical properties in the dilute limit. This study can also be seen as a stepping stone to interpolate from classical to quantum behavior. We employed a dissipative dynamical framework \`{a} la Lindblad, Eqs.~\eqref{eq:2:lind} to~\eqref{eq:2:jumpA}, which incorporates classical stochastic single-spin flip dynamics as a limiting case. We used it to develop a modified Monte Carlo method that approximates a stochastic sampling of the Lindblad equation. This approach allows for the computation of dynamical properties in sufficiently large RTFIC systems. 

Using the modified Monte Carlo method, we investigated the magnetic noise of the system in equilibrium by analyzing its power spectral density. Our analysis revealed the emergence of two diffusive regimes that are distinct from the classical limit without transverse fields. Specifically, the PSD shows diffusive (inverse square law) behavior at both low and high frequencies, with different proportionality factors between the two regimes, and a transition region at intermediate frequencies, $\omega \sim 10^0$. We attribute this effect to the separation of short- and long-time behaviors in the random walk of the excitations, which correspondingly manifests as two distinct diffusive regimes in the mean squared displacement. 

Moreover, we uncovered a distinctive profile in the AC susceptibility of our RTFIC. At high frequencies, $\chi''(\omega)$ exhibits an enhancement with respect to the classical case of no transverse fields, while the phase shift $\phi$ deviates from the classical value $\pi/2$ in the transition region between the two diffusive limits. We have also developed analytical understandings of the behavior observed in the numerics. These findings highlight how random disorder in spin systems can significantly influence the fluctuation and response properties by altering the dynamics of the underlying point-like quasiparticles. 

To check the applicability of our results in higher dimensions and other lattice structures, we have performed similar simulations in the three-dimensional pyrochlore lattice, and found similar behavior for the magnetic noise there, despite the differences between random walks in 3D and 1D. 
The models and results provide a valuable stepping stone toward understanding the effects of random transverse fields in spin ice systems, and thence on possible effects of local strains and distortions in non-Kramers pyrochlore magnets such as $\mathrm{Ho_2Ti_2O_7}$. They may also shed light on classical spin ice contexts with other sources of disorders or heterogeneity \cite{lu2024111}.

Looking ahead, an important direction for future work is to interface the quantum dynamics within the clusters with the stochastic dynamics formulated in this paper. In our current framework, we assumed that the quantum clusters of spins acted upon by the transverse field are instantaneously always in one of their eigenstates for the given relevant choice of boundary spins (cf. the end of Sec.~\ref{section:2:model}). Extending our modified Monte Carlo, for example through the use of non-Markovian master equations \cite{yu2004non}, could bridge this gap. Such extension would be needed, for instance, to study `classical impurities' (i.e., spins without a transverse field) in the quantum transverse-field Ising chain (or quantum spin ice), corresponding to the complementary limit $1-r \ll 1$ of our model.

Another possible direction for future work in regard to spin-ice physics would be to integrate the dynamical description presented in this paper with recent discoveries about the correlated quantum tunneling nature of spin flips in spin ice materials~\cite{tomasello2019correlated}. These studies have established a relationship between spin-flip dynamics and the local spin environment, culminating in the intriguing discovery of emergent dynamical fractals underpinning anomalous noise and subdiffusive behavior~\cite{hallen2022dynamical}. We anticipate that combining these two ingredients --- the stochastic and correlated tunneling dynamics --- will contribute to a deeper understanding of the microscopic mechanisms underlying spin-ice dynamics. This integration could also help to understand recent experimental advances and shed light on unresolved questions in spin-ice physics~\cite{wang2021monopolar, morineau2025satisfaction}. 
%
%

\section*{Data Availability Statement}
{ All data produced and presented in this work were obtained by well-known methods that have been described in detail in the main text and App.~\ref{app:F}, and can be readily reproduced. They can also be made available upon reasonable request.} 
%
%

\begin{acknowledgements}
We would like to thank John Chalker, Orazio Scarlatella, Jonathan N. Hall\'{e}n and Nilotpal Chakraborty for useful discussions. 
This work was supported in part by the Engineering and Physical Sciences Research Council (EPSRC) grants No.~EP/T028580/1 and No.~EP/V062654/1 (C.C.). This work was in part supported by the Deutsche
Forschungsgemeinschaft under Grants No. SFB 1143
(Project No. 247310070) and the cluster of excellence
ct.qmat (EXC 2147, Project No. 390858490). 
For the purpose of open access, the authors have applied a Creative Commons Attribution (CC-BY) licence to any Author Accepted Manuscript version arising from this submission.
\end{acknowledgements} 
%
%
\appendix

\section{\label{app:A}
Thermodynamic properties of RTFIC}
In this appendix, we present an analysis of the thermodynamics of the RTFIC, Eq.~\eqref{eq:2:rtfic}, for completeness and to help our quantitative understanding of its dynamics.
%
%

\subsection{Exact diagonalization of the quantum clusters}\label{section:a:ed}
We perform exact diagonalisation of the $n$-cluster Hamiltonian Eq.~\eqref{eq:2:Hn} in Sec.~\ref{section:2:rtfic}, for a range of values of $n$ appropriate for the low density limit of interest in our work. We fix $s_0=1$ without loss of generality due to the overall $\mathbb{Z}_2$ symmetry of the model, while $s_{n+1}=\pm1$ is a free variable. 

In Fig.~\ref{fig:A:thermo}a, we plot the energy levels of the $n=2$-cluster for reference, while keeping our discussion general. We let $|\psi_l^{(n)}\rangle$ to denote the eigenstates of the Hamiltonian $H^{(n)}$ with $s_{n+1}=1$ (where $l=1,2,\dots,2^n$, sorted from low to high energy), and $|\psi_l'^{(n)}\rangle$ to denote the eigenstates with $s_{n+1}=-1$ --- i.e., with anti-aligned boundary spins and hence hosting an excitation. 

For $h=0$, the ground state for $s_{n+1}=1$, $|\psi_1^{(n)}\rangle$, is trivially the $z$-polarized state with $| \langle S_{i}^z\rangle | = 1$ for $1\le i \le n$ with energy $E_1^{n}=-(n+1)$. As $h$ is turned on, $|\psi_1^{(n)}\rangle$ polarizes gradually along the $x$ direction. Let $m_l^{(n)}\equiv\langle\psi_{l}^{(n)}|\sum_{i=1}^{n}S_i^z |\psi_{l}^{(n)}\rangle$ denote the magnetization of the state $|\psi_{l}^{(n)}\rangle$; as shown in Fig.~\ref{fig:A:thermo}a, $m^{n}_1=n$ for $h=0$ and monotonically decreases with $h$. Energetically, $E_1^{(n)}$ receives a second-order perturbative contribution at small $h$ but is linearly dependent on $h$ at large values, when $|\psi_1^{(n)}\rangle$ is almost completely $x$-polarized.

The ground states for $s_{n+1}=-1$, on the other hand, is $n+1$-fold degenerate for $h=0$, corresponding to the $n+1$ possible domain wall positions. As $h$ is turned on, the degeneracy is lifted, with each state receiving a first-order perturbative contribution to its energy. Note that, for $h\ne 0$, the states $\psi_l'^{(n)}$ always have vanishing magnetization ($m_l'^{(n)}=0$), due to the symmetry $S_i^z\rightarrow-S_i^z$ present in the Hamiltonian. 

The discussion above indicates a reduction of the gap between the two lowest-lying states, $|\psi^{(n)}_1\rangle$ and $|\psi'^{(n)}_1\rangle$ as $h$ is turned on, tending to zero in the $h \to \infty$ limit (see Fig.~\ref{fig:A:thermo}). Thermodynamically, the RTFIC hosts `more excitations' (domain walls) than its classical counterpart (i.e., its vanishing field limit) at the same temperature due to the presence of quantum clusters. This observation turns out to be important in our discussions of the thermodynamics below, as well as on the random walk model in App.~\ref{app:C}. 
%
%

\subsection{Heat capacity}
\begin{figure}[tb!]
    \includegraphics[width=\columnwidth]{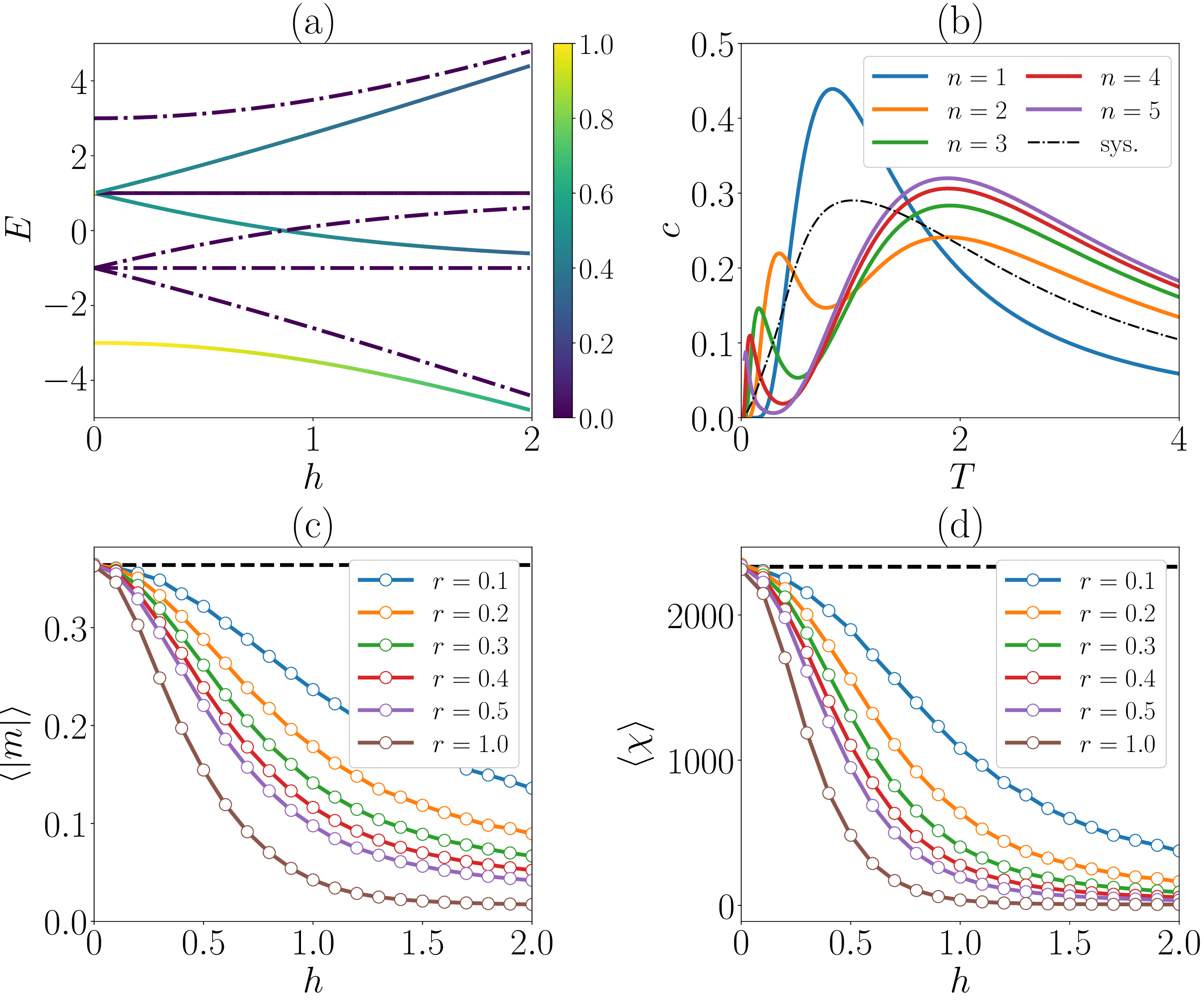}
    \caption{\label{fig:A:thermo}Thermodynamic results for the RTFIC: (a) Exact diagonalization of the $n=2$ cluster, where $4$ energy levels are present for $s_3=+1$ (solid lines) and four for $s_3=-1$ (dash-dotted lines), respectively. The color scale represents the average magnetization per spin of the corresponding state. All states with $s_3=-1$ have vanishing magnetization. (b) Heat capacity per spin, $c_n$, for $\bar{H}^{(n)}$ with $n=1, 2, \dots, 5$; we also plot $c$ for the dual system Hamiltonian $\bar{H}$ with $r=0.3$ calculated from taking the weighted average over the $c_n$'s, using a cutoff of $n_c=8$; in this panel we set $h=2$. SSE results for the correlation function $\langle S_i^zS_{i+x}^z\rangle$ with $r=0.1$ and different values of $h$ are shown in (c), and for the static susceptibility per spin $\chi(h)$ for different values of $r$ in (d); in these two panels we set $T=0.3$. The addition of random transverse fields reduces the correlation length and the magnetic susceptibility of the system.}
\end{figure}
The heat capacity per spin $c$ can be conveniently computed using the Kramers-Wannier duality~\cite{radicevic2018spin}. The duality introduces a set of dual spins $\{\bar{S}_i\}$ living on the dual sites of the original lattice (i.e., the mid points of each bond) and are connected to the original spins by the following relations:
\begin{subequations}
\begin{equation}
    \bar{S}_i^z = S_{i-1}^zS_{i}^z
    \, ,
\end{equation}
\vspace{-2em}
\begin{equation}
    \bar{S}_{i}^x\bar{S}_{i + 1}^x = S_i^x
    \, .
\end{equation}
\end{subequations}

Under this duality, the Hamiltonian Eq.~\eqref{eq:2:rtfic} is transformed into the dual Hamiltonian
\begin{equation}\label{eq:A:kw}
    \bar{H}= -\sum_i\bar{S}^z_i - \sum_{\alpha\in\{\alpha\}}\bar{S}_{\alpha-1}^x\bar{S}_\alpha^{x}
    \, ,
\end{equation}
where the random transverse fields now take the form of random exchange couplings in the $x$-direction. Due to the dilute nature of $\{\alpha\}$ (at $r\ll 1$), $\bar{H}$ decouples into the sum of independent subsystems with Hamiltonians
\begin{equation}\label{eq:A:Hn}
    \bar{H}^{(n)}=-\sum_{i=1}^n\bar{S}_i^z-\sum_{i=1}^{n-1}\bar{S}_i^x\bar{S}_{i+1}^x
    \, .
\end{equation}

For $n\ge 2$, $\bar{H}^{(n)}$ is dual to $H^{(n-1)}$ 
in Eq.~\eqref{eq:2:Hn} and they share the same spectrum. On the other hand, $\bar{H}^{(1)}=-\bar{S}_1^z$ is a simple two-level system, with the excited state signifying an anti-aligned pair of nearest-neighbor spins in the original system, i.e., a domain wall. 
(At low temperatures, the thermal occupancy of the excited states is determined by the corresponding Boltzmann factor $e^{-2\beta}$; in this way, one can then straightforwardly verify the claim about the excitation density and typical separation in the {classical Ising chain} made in Sec.~\ref{section:2:dim}.)

The heat capacity of the dual system can be computed by performing a weighted average over $c_n$, the heat capacity per dual spin for an $n$-spin subsystem with Hamiltonian $\bar{H}^{(n)}$, which can be exactly diagonalized up to some cutoff for $n$. A combinatorial calculation gives the density of $n$-spin subsystems in the dual system as $\bar{p}_n=r^{n-1}(1-r)^2$; thus, $c=\sum_{n=1}^{n_c} n \bar{p}_n c_n$, where $n_c$ is the cutoff. 

In Fig.~\ref{fig:A:thermo}b, we plot the heat capacity for $H^{(n)}$ with fixed $h$ but varying values of $n$ as a function of temperature $T$. For $n\ge 2$, the heat capacity demonstrates bimodal behavior, with the peak at low $T$ representing the transition between the two lowest-lying eigenstates and the one at high $T$ representing all other transitions, with the position of the latter largely independent of $h$. We also plot $c$, calculated with $r = 0.3$. The presence of quantum clusters causes a deviation from the classical result (which coincides with the $n=1$ curve) that takes the form of a broadened peak. 
%
%

\subsection{Quantum Monte Carlo results}
As discussed at the end of App.~\ref{section:a:ed}, there is a reduction in the gap between the two lowest-lying states, $|\psi^{(n)}_1\rangle$ and $|\psi'^{(n)}_1\rangle$, as $h$ is turned on, demonstrating a lowered chemical potential for excitations on the clusters, where they manifest themselves as anti-aligned boundary spins. This signifies a reduced correlation function $C(x)\equiv \langle S_i^zS_{i+x}^z\rangle$ due to the presence of clusters in the RTFIC compared to that of its classical counterpart (no transverse field). Numerically, $C(x)$ can be most conveniently computed using the stochastic series expansion (SSE) quantum Monte-Carlo method~\cite{sandvik2010computational, sandvik2019stochastic}. In Fig.~\ref{fig:A:thermo}c, we present the correlation functions for fixed $r$, $T$ but different values of $h$, where we see a shortened correlation length in the system which verifies the detrimental effect of the clusters to magnetic order.

We also calculate the static susceptibility per-spin $\chi$ using the SSE Monte Carlo via the relation
\begin{equation}
    \chi = \beta \frac{\langle m^2\rangle - \langle m \rangle^2 }{L}\, ,
\end{equation}
where $L$ is the size of the system. For the classical Ising chain one obtains the simple result $\chi=\beta e^{2\beta}$. In Fig.~\ref{fig:A:thermo}d, we plot the statistical average of $\chi$ as a function of $h$ for different values of $r$; we observe a decrease in $\chi$ for increasing $h$ and $r$. Note that, although the results for $r=1$ correspond to those of a quantum transverse-field Ising chain, there is no critical behavior due to the presence of finite $T$. 
%
%

\begin{figure*}[htb!]
\begin{minipage}[]{0.23\linewidth}
\includegraphics[width=\linewidth]{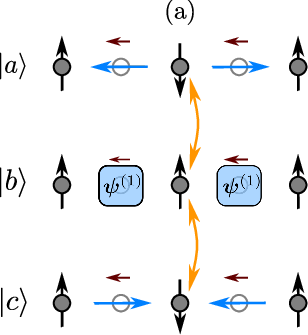}
\end{minipage}
\begin{minipage}[]{0.7\linewidth}
\includegraphics[width=0.46\linewidth]{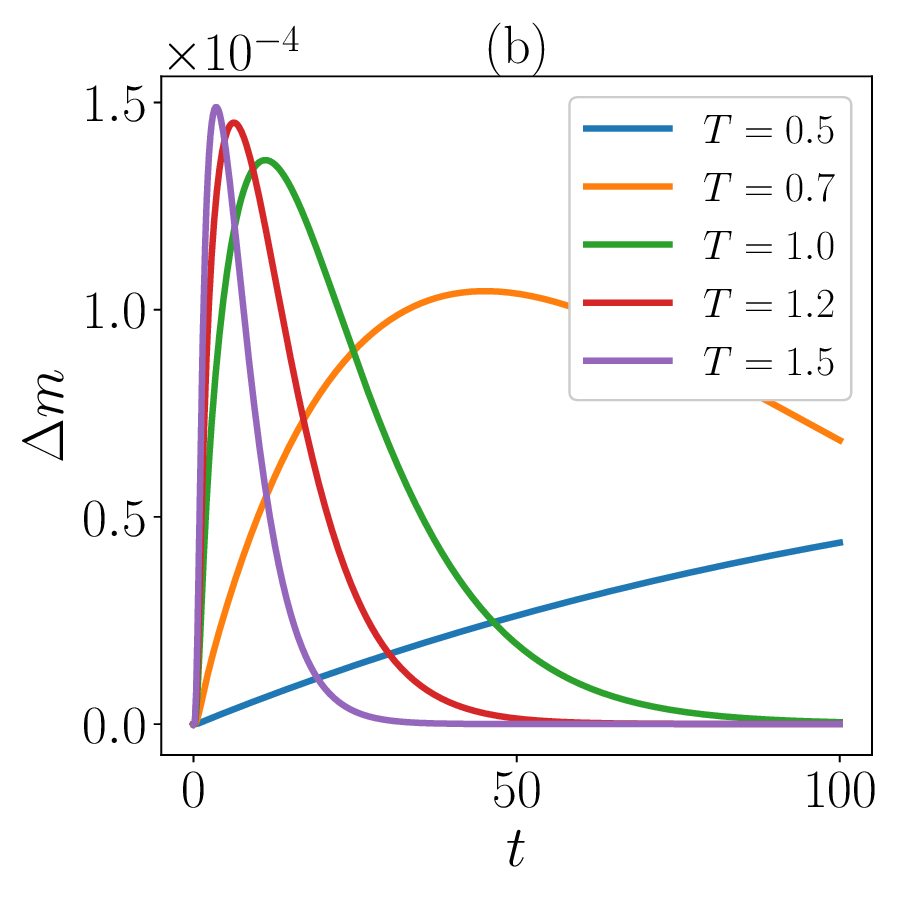}
\includegraphics[width=0.46\linewidth]{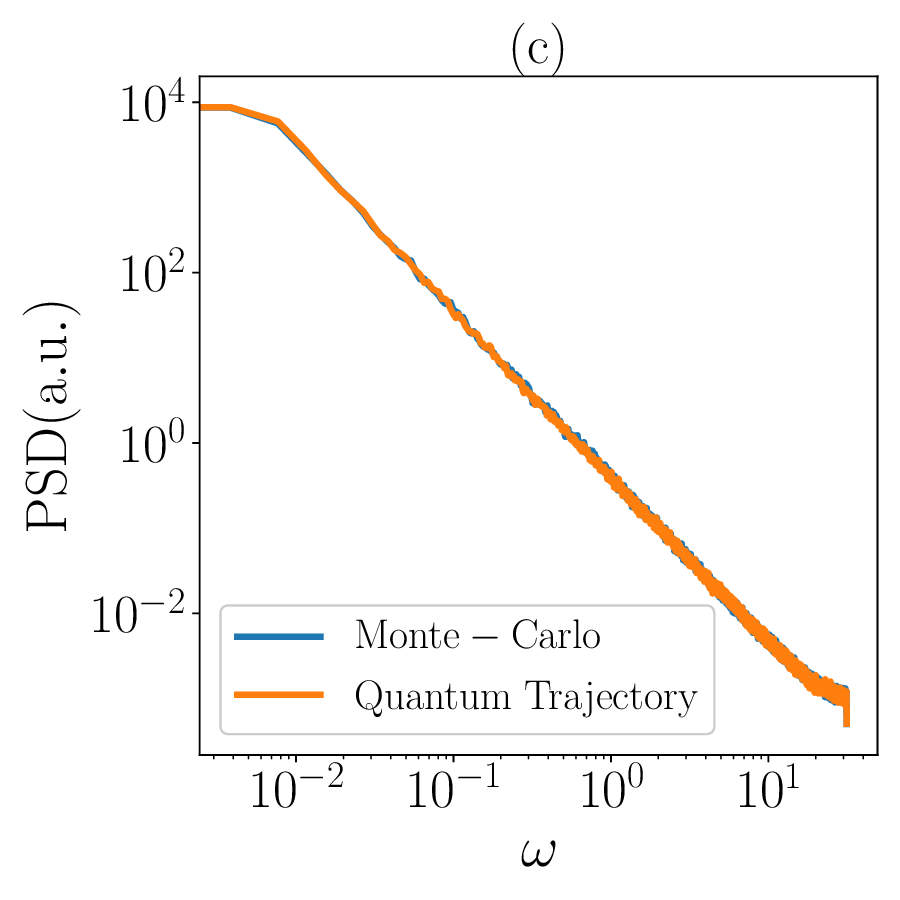}
\end{minipage}
\caption{\label{fig:B:lind} (a) An example scenario discussed in App.~\ref{app:B}, where the top and bottom configurations have identical energies and non-vanishing matrix elements with the middle configuration by flipping the third spin, which is a classical spin situated in between two quantum clusters. The horizontal blue arrows represent spins polarized in the $x$-direction (note the vanishing exchange fields acting on them). (b) Difference $\Delta m$ between the system magnetizations per spin, $m(t)$, calculated from density matrices obtained using the Lindblad equation evolution and the Monte-Carlo master equation evolution, respectively; the system is chosen to be that portrayed in (a), with $h=1.0$. (c) Power spectral density, calculated from both Monte-Carlo simulations and quantum trajectory simulations, for the same system with $h=1.0$ and $T=0.5$.}
\end{figure*}

\subsection{\label{App:A4}Discussion}
The results for $\chi$ allow us to probe the $\omega\to0$ limit of the dynamics which has been largely overlooked in the main text. There, we investigated the RTFIC by probing the random walk of its excitations, neglecting creation and annihilation events which happen on a timescale $\tau_0\sim e^{4/T}$ that diverges as $T$ is decreased. While irrelevant in the regime of $\omega$ discussed in the main text, in the limit $\omega\ll\tau_0^{-1}$ they become important in describing the dynamics of the system, and the random walk picture breaks down.

In the {kinetic classical Ising chain}, the PSD of the system per spin takes the form~\cite{lauritsen1993critical, lauritsen1994spectral, macisaac1992dynamic, nilsson2023dynamics} 
\begin{equation}
    \mathrm{PSD}(\omega)=2T\chi\frac{\tau_0}{1+\omega^2\tau_0^2} 
    \, , 
\end{equation}
where $\chi$ is the static susceptibility of the system. Using the fluctuation-dissipation theorem and the Kramers-Kronig relations~\cite{LANDAU1980333}, one obtains the real and imaginary components of the AC susceptibility: 
\begin{equation}
    \chi''(\omega)=\chi\frac{\omega\tau_0}{1+\omega^2\tau_0^2}
    \, , 
    \quad 
    \chi'(\omega)=\chi\frac{1}{1+\omega^2\tau_0^2} 
    \, . 
\end{equation}
These results are consistent with those derived from standard random walk theory, namely $\mathrm{PSD}(\omega)\propto\omega^{-2}$, $\chi''(\omega)\propto\omega^{-1}$ and $\chi'(\omega)$ vanishing presented in Sec.~\ref{section:4:driven}, in the limit $\omega\tau_0\gg 1$. In the opposite limit $\omega \to 0$, $\chi''(\omega)$ vanishes and $\chi'(\omega)\rightarrow\chi$, and the PSD tends to a constant value proportional to $\chi$. 

The results presented in Fig.~\ref{fig:A:thermo}d, therefore, imply a deviation of the PSD obtained for different values of $h$ and $r$ at $\omega\ll\tau_0^{-1}$, tending to different constant values due to the dependence of $\chi$ on these parameters. Although this is not always seen in our simulations due to limitations on the accessible frequency range, one can already have a glimpse of such effects from the left-most part of the $T=0.5$ curve in Fig.~\ref{fig:PSD}a.
%
%

\section{\label{app:B}
From Lindblad to Monte-Carlo}
In this appendix, we fill in the details in going from Eq.~\eqref{eq:2:pme} to Eq.~\eqref{eq:2:rme} in Sec.~\ref{section:2:rtfic}, where we postulated 
\begin{equation}\label{eq:B:assum}
    \langle a|S_i^x|b\rangle\langle b|S_i^x|c\rangle\delta_{E_a,E_c} \propto \delta_{a, c} \, 
\end{equation}
for the RTFIC.

Violations of this assumption take the form of two distinct eigenstates $|a\rangle$ and $|c\rangle$ with identical energies $E_a = E_c$ that are simultaneously connected to another state $|b\rangle$ by application of an $S_i^x$ operator. To look for the existence of such violations, we consider the following four cases:

\begin{enumerate}[leftmargin=5.5mm]
    \item The $i$th spin is classical, and so are its two neighboring spins. In this case, $S_i^x|a\rangle$ maps to another eigenstate and hence $\langle a|S_i^x|b\rangle\langle b|S_i^x|c\rangle \propto \delta_{a, c}$. Eq.~\eqref{eq:B:assum} holds trivially. 
    \item The $i$th spin is part of a quantum cluster. 
    Application of $S_i^x$ maps the state of this cluster onto some superposition of eigenstates of the cluster, but does not affect other parts of the system.
    As one can see from Fig.~\ref{fig:A:thermo}, the local Hamiltonian of the cluster is non-degenerate once one fixes the orientation of the boundary spins, and therefore $E_a=E_c$ implies $|a\rangle = |c\rangle$.
    \item The $i$th spin is classical, but adjacent to one quantum cluster. Application of $S_i^x$ flips the $i$th spin, and alters the state of the adjacent cluster. For the matrix element factor $\langle a|S_i^x|b\rangle\langle b|S_i^x|c\rangle$ to be nonzero, the orientations of spin $i$ must be identical in $|a\rangle$ and $|c\rangle$. Therefore, using similar arguments, we have once again that $E_a=E_c \Rightarrow |a\rangle = |c\rangle$. 
    \item The $i$th spin is classical, but adjacent to two quantum clusters.
    Once again, we must have identical orientations for spin $i$ in $|a\rangle$ and $|c\rangle$. There now exist cases where two distinct configurations of the clusters have the same total energy --- e.g., when the clusters have the same size, and their two eigenstates are swapped in $|a\rangle$ and $|c\rangle$, an illustration of which is shown in the left panel of Fig.~\ref{fig:B:lind}. In these cases, Eq.~\eqref{eq:B:assum} is violated and the RHS of Eq.~\eqref{eq:2:pme} may involve off-diagonal terms of the density matrix. 
    Despite the locality of the effect of these violations, 
    they necessarily involve one of the two clusters in $|a\rangle$ or $|c\rangle$ not in their respective ground states, $|\psi_1^{(n)}\rangle$ (notation inherited from App.~\ref{app:A}), implying higher energies and thus a lower statistical weight. 
    Hence, we expect Eq.~\eqref{eq:B:assum} to remain a good approximation at low temperatures. 
    Such occasions have been explicitly excluded under the assumptions made towards the single-excitation limit -- see the modeling in App.~\ref{app:C1}. 
\end{enumerate}

To confirm these arguments numerically, we conduct simulations to compare the time evolutions of the same physical quantities under the Lindblad equation Eq.~\eqref{eq:2:lind} with the dissipator Eq.~\eqref{eq:2:dissp} and under the Monte-Carlo dynamics governed by Eq.~\eqref{eq:2:rme}. The system is selected to be the one portrayed in Fig.~\ref{fig:B:lind}a, i.e., a $5$-spin system with a transverse field added on the second and the fourth spin, where we expect violations of Eq.~\eqref{eq:B:assum} to be present. In Fig.~\ref{fig:B:lind}b, we calculate the difference in the system magnetization $m(t)$ computed from solving $\rho(t)$ from the two master equations, upon initializing $\rho(0)=|a\rangle\langle a|$, where $|a\rangle$ is the state with the lowest energy and $m>0$. For $h=1.0$, a vanishing $\Delta m$ is observed, whose peak value seems to decrease with temperature $T$. In Fig.~\ref{fig:B:lind}c, we show the PSD of the 5-spin system with $h=1.0$ and $T=0.5$ computed using both Monte-Carlo and quantum trajectory, a stochastic sampling algorithm with guaranteed fidelity to the Lindblad equation \cite{clerk2010introduction, daley2014quantum} No noticeable differences between the two PSDs can be identified. We therefore expect Eq.~\eqref{eq:B:assum} to be generally a good approximation in our work, especially at low temperatures.
%
%

\section{\label{app:C}
Single-excitation random walk in RTFIC}
In this appendix, we fill in the details omitted in Sec.~\ref{section:3:rw} and provide a study of the random walk model proposed there, governed by Eq.~\eqref{eq:3:rwme}. 
%
%

\subsection{\label{app:C1}Model}
In Sec.~\ref{section:3:rw}, we claimed a one-to-one correspondence between the position of the excitation and the magnetization of the system, despite the presence of quantum clusters which span a non-trivial range of values of their magnetization as well as their spatial size on the lattice. Using the energy level analysis of the clusters in App.~\ref{app:A}, at sufficiently low $T$, we assume that, at any given time, all clusters with aligned boundary spins are in their ground state $|\psi_1^{(n)}\rangle$ (with notation inherited from App.~\ref{app:A}). When an excitation is on the cluster, now with anti-aligned boundary spins, we assume that the cluster is in one of the $n+1$ lowest states, $|\psi'^{(n)}_{1,\dots,n+1}\rangle$, all with vanishing magnetization as mentioned in App.~\ref{app:A}. Therefore, the change of magnetization in the processes of excitations hopping onto/out of a give cluster is determined: it is simply the ground state magnetization of the cluster, $m^{(n)}_0$. This is key to allow us to identify an $n$-cluster as a single site on the random walk lattice, as we did in the main text.

We now turn to the analytical derivation of $W_{ij}$ appearing in the random walk Eq.~\eqref{eq:3:rwme}. This quantity corresponds to the acceptance probability per unit Monte Carlo time, $p$, of flipping the (classical) spin connecting adjacent sites $i$ and $j$, with the initial configuration having an excitation at site $i$; when site $i$ corresponds to a quantum cluster, statistical averaging is performed over the multiple initial configurations corresponding to an excitation at site $i$ to generate a single $W_{ij}$ between each pair of sites. 

To facilitate discussion, we define $s(i)=n$ if site $i$ on the {random walk} lattice represents an $n$-cluster, and $s(i)=0$ if it represents a classical site, and define $\tilde{W}_{s(i), s(j)}\equiv W_{ij}\delta_{i, j\pm 1}$. 
Using Eq.~\eqref{eq:2:qcmcp} in the main text, we have
\begin{equation}\label{eq:C:tr}
    \tilde{W}_{n,n'} = \frac{\displaystyle\sum_{l=1}^{n+1}\sum_{l'=1}^{n'+1} e^{-\beta E_l^{(n)}}
    \min\left(1,\,e^{-\beta \Delta E}\right)\mathcal{M}_{l}^{(n)}\mathcal{M}_{l'}^{(n')}}
    {\displaystyle\sum_{l=1}^{n+1}e^{-\beta E_{l}^{(n)}}}
    \, ,
\end{equation}
where $\Delta E = ( E_{l'}'^{(n')} - E_{1}^{(n')} ) - ( E_{l'}^{(n)} - E_{1}^{(n)} )$ is the energy difference before and after the transition of the excitation from site $i$ to $j$, with $E_0^{(0)} = 0$ and $E_{1}^{(0)} =  2$ for classical sites, and $\mathcal{M}_l^{(n)}$ is the matrix element $|\langle \psi_{l}'^{(n)}| \psi_{1}^{(n)}\rangle|^2$ for $n>1$ and $1$ for $n=0$. Note that in general $\tilde{W}_{n,n'}\ne\tilde{W}_{n',n}$ if $n\ne n'$. 

As mentioned in the main text, we also need to adjust the step lengths between adjacent sites in the {random walk} lattice to account for the (spatial and magnetic) size of the quantum clusters. We define $a_{ij}$ to be the step length between sites $i$ and $j$, representing the difference in magnetization induced via transitions between these sites ($a_{ij}=a_{ji}$). We define $\tilde{a}_{s(i), s(j)} = a_{ij}\delta_{i, j\pm 1}$ as: 
\begin{equation}
    \tilde{a}_{n,n'} = 2+\left(m^{(n)}_0+m^{(n')}_0\right)
    \, ,
\end{equation}
where, as usual, $m_0^{(n)}$ is the ground-state magnetization for $n>1$, and $m^{(0)}_0 = 0$ for classical sites ($n=0$). The additive term on the right hand side comes from flipping the classical spin. 
%
%

\subsection{Diffusion constants}
\begin{figure}[t!]
    \includegraphics[width=0.9\columnwidth]{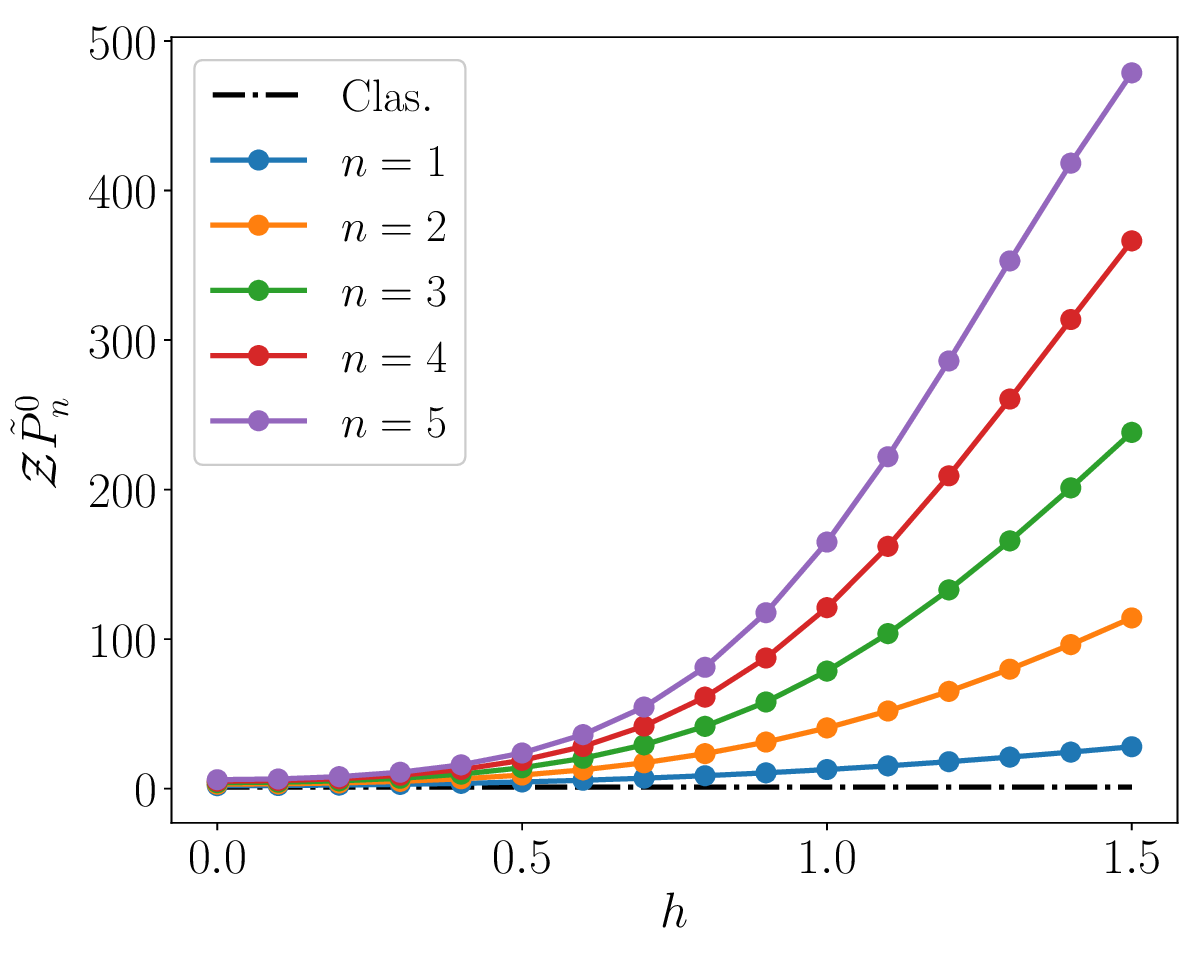}
    \caption{\label{fig:C:time} The value of $\mathcal{Z}P^0_n$, discussed in App.~\ref{app:C}, plotted for several values of $n$ and $T=0.3$. The dash-dotted line represents 
    the classical result 
    discussed in the text.}
\end{figure}
This appendix is dedicated to the solution of the random walk model and derivation of $D_0$ and $D_\infty$. We start by stating the stationary solution of the Master equation Eq.~\eqref{eq:3:rwme}, $\{P_i^0\}$, obtained from setting the LHS to $0$. 
It is proportional to the thermodynamic weight of the configuration corresponding to each site: 
\begin{equation}
    \tilde{P}^0_n = \frac{1}{\mathcal{Z}}
    \displaystyle\sum_{l=1}^{n+1}e^{-\beta (E_{l}'^{(n)}-E_1^{(n)}-2)},\quad \mathcal{Z} = \sum_{i}P^0_i
    \, ,
\end{equation}
where we have defined $\tilde{P}^0_{s(i)} = P^0_i$. (The energy shift $-2$ in the exponent is introduced so that $\mathcal{Z}\tilde{P}_0^0=1$ for simplicity.) The stationary solutions $\tilde{P}^0_n$ is proportional to the mean waiting time for the excitation to move to an adjacent site. In Fig.~\ref{fig:C:time}, we plot the quantity $\mathcal{Z}\tilde{P}^0_n$ as a function of $n$ and $h$. One clearly sees the trapping effect of the quantum clusters demonstrated by an increasing $\mathcal{Z}\tilde{P}^0_n$ as $n$ and $h$ increase. 

We now turn to the two diffusion constants. From the definition of the MSD, Eq.~\eqref{eq:3:msd}, we have \cite{van1982transport}
\begin{equation}\label{eq:C:msdapp}
    M(t) = \sum_{i\ne j} P_i^0 P_{j,i}(t) 
    \left( 
      \sum_{k=\min(i, j)}^{\max(i, j) - 1} a_{k, k+1}
    \right)^2
    \, ,
\end{equation}
where $P_{j,i}(t)$ denotes the conditional probability of finding the random walker at site $j$ at time $t$ given that it is initialized at site $i$ at time $0$. In the limit $t\to 0$, we have $P_{i\pm 1,i} \simeq W_{i, i\pm1} t$ and all other terms vanish, leading to the following expression for $D_0$:
\begin{equation}\label{eq:c:d0}
    D_0 = \left.\frac{\mathrm{d}M(t)}{\mathrm{d}t}\right|_{t\to 0 }=\sum_{i}P^0_i \left(a_{i, i+1}^2W_{i, i+1} + a_{i, i-1}^2W_{i, i-1}\right)
    \, .
\end{equation}
On the other hand, in the limit $t\to\infty$, $P_{j,i}(t)$ gives a finite contribution for a broad range of $j$. Noting that $P_i^0$ and $P_{j,i}(t)$ are independent of $a_{i,j}$, we perform the average over the spatial fluctuations within the parentheses in Eq.~\eqref{eq:C:msdapp}: 
\begin{equation}\label{eq:C:msdavg}
\begin{aligned}
    \lim_{t\to\infty}M(t) 
    &= \sum_{i\ne j} P_i^0P_{j,i}(t)\left(\sum_{k=\min(i, j)}^{\max(i, j) - 1}\bar{a}\right)^2 \\
    &= \bar{a} ^ 2\sum_{i\ne j} P_i^0P_{j,i}(t)|i-j|^2
    \, ,
\end{aligned}
\end{equation}
with
\begin{equation}\label{eq:c:a}
    \bar{a} = \frac{1}{N}\sum_{i} a_{i, i+1}
\end{equation}
being the average step length. Eq.~\eqref{eq:C:msdavg} suggests that the value of the diffusion constant $D_\infty$ is given by the product of $\bar{a}^2$ times the same quantity defined for a random walk with the same Master equation Eq.~\eqref{eq:3:rwme} but happening on a standard lattice (i.e., with uniform unit bond length). The latter quantity can then be calculated using Eqs.~(46) to (50) in Ref.~\onlinecite{derrida1983velocity} and the relationship $\tilde{W}_{n,n'}\tilde{P}_n^0 = \tilde{W}_{n',n}\tilde{P}^0_{n'}$. Altogether, we obtain 
\begin{equation}\label{eq:c:Dinf}
    D_\infty = \left.\frac{\mathrm{d}M(t)}{\mathrm{d}t}\right|_{t\to \infty }=4\bar{a}^2 \left(\sum_{i}\frac{1}{P_i^0W_{i, i+1}}+\frac{1}{P_i^0W_{i, i-1}}\right)^{-1}
    \, .
\end{equation}

These analytical results are used in plotting the inset of Fig.~\ref{fig:rw}b as well as the results for $D_0/D_\infty$ in Fig.~\ref{fig:psdmsd}. There, evaluations of the sums over sites in Eqs.~\eqref{eq:c:d0},~\eqref{eq:c:a} and~\eqref{eq:c:Dinf} are performed on a {random walk} lattice corresponding to a $10^6$-spin system. 
%
%

\section{\label{app:D}
Data processing in simulations of the driven RTFIC}
In this appendix, we explain the procedure used when processing the data for $m(t)$ obtained from Monte-Carlo simulations of the driven RTFIC, discussed in Sec.~\ref{section:4:results}. Consider a single $n$-cluster in a time-independent longitudinal field $B$, with Hamiltonian 
\begin{equation}
    H'^{(n)}=-\left(s_0S_1^z+S_{n}^zs_{n+1}+h\sum_{i=1}^{n}S_i^x+B\sum_{i=1}^{n}S_i^z+\sum_{i=1}^{n-1}S_i^zS_{i+1}^z\right)
    \, .
\end{equation}
The eigenstates $|\psi_m^{(n)}\rangle$ are $B$-dependent and so is their magnetization, $\langle\psi^{(n)}_m|\sum_{i=1}^nS_i^z|\psi_m^{(n)}\rangle$. Therefore, in the driven RTFIC studied throughout Sec.~\ref{section:4:driven}, the magnetization $m(t)$ receives an additional contribution $\Delta m$ coming from the $B$-dependence of the eigenstates (and hence an implicit $t$-dependence in an oscillating driving field) on top of the contribution coming from the dynamics of the excitations. This additional contribution needs to be deducted from $m(t)$ for us to focus on the physics of interest. 

To quantify such contribution, we first compute the statistical average of the magnetization for an $n$-cluster (which is still solvable by ED up to some cutoff $n$). This quantity vanishes trivially for $B=0$ due to the $\mathbb{Z}_2$-symmetry $S_i^z\rightarrow -S_i^z$, but takes a finite value when $B\ne 0$. As the system relaxation time is much greater than the period of the driven field concerned (see the discussion in App.~\ref{App:A4}), the thermal population of each of the eigenstates can be approximately calculated by averaging over the longitudinal field in time, i.e., taking effectively $B=0$. We then have
\begin{equation}\label{eq:D:m}
    m^{(n)}b = \left\langle\sum_{i=1}^nS_i^z\right\rangle = \frac{\displaystyle\sum_{l}e^{-\beta E^{(n)}_l(0)}\langle\psi_{l}^{(n)}|\sum_{i=1}^{n}S_i^z |\psi_{l}^{(n)}\rangle}{\displaystyle\sum_{l}e^{-\beta E^{(n)}_l(0)}}
    \, , 
\end{equation}
where $l$ now labels all possible $4\times2^n$ states, as the cases $s_0=1$ and $s_0=-1$ are no longer equivalent.

We proceed by taking the weighted average over the $m^{(n)}b$'s with different $n$'s scaled by their density, $p_n=r^n(1-r)^2$, up to a cutoff for $n$, to obtain estimates for $\Delta m$ as a function of $B$ and hence implicitly of $t$. Note that, in this process, we do not take into account the effects of the driving field on the classical spins, as these are part of the quasiparticle dynamics that we are interested in.

\begin{figure}[t!]
    \includegraphics[width=\columnwidth]{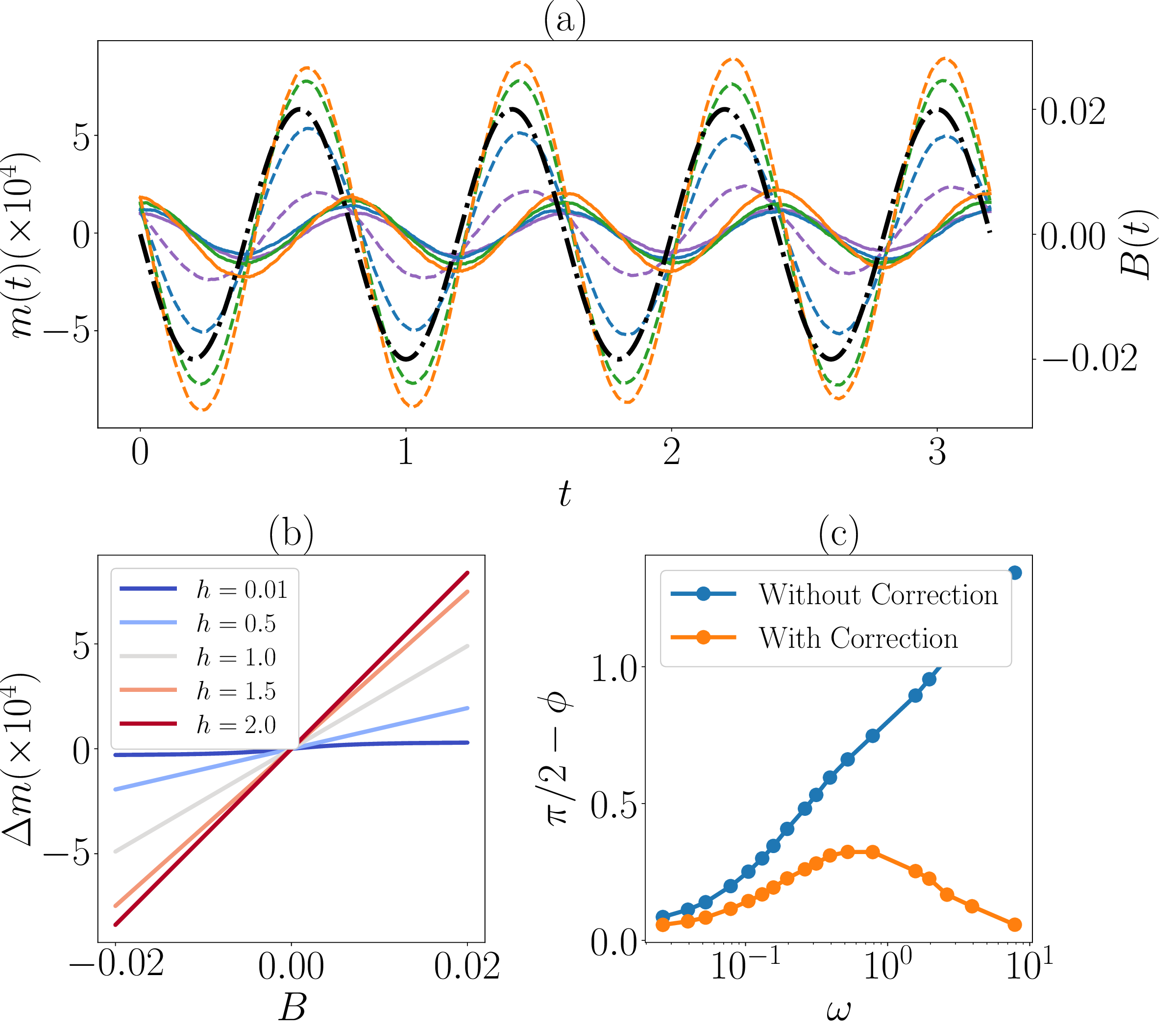}
    \caption{\label{fig:d:drtfic} Data processing of Monte-Carlo simulations of the driven RTFIC, Eq.~\eqref{eq:4:Drtfic}: (a) Results for $m(t)$ simulated with $\omega=2.5\pi$, the largest value considered in this work, and $h=0.5$ (purple), $h=1.0$ (blue), $h=1.5$ (green), and $h=2.0$ (orange). The corrected data (solid lines) and uncorrected data (dashed lines) deviate drastically. The black line represents the driving field, $B(t)$. (b) Calculated values of $\Delta m$ as a function of applied field $B$, with cutoff cluster size $n_c=6$. (c) Values of $\phi$ obtained for $h=2.0$, with and without the correction step. We used $T=0.35$, $r=0.15$ and $B_0=0.02$ throughout.}
\end{figure}

In Fig.~\ref{fig:d:drtfic}b, we plot our results as a function of $B$ with varying $h$ and all other parameters fixed. A linear response to the external field is observed --- apart from the limit $h\to 0$, where our theoretical analysis breaks down --- verifying the validity of using linear response theory and the fluctuation-dissipation theorem in treating the driven RTFIC. One also observes a significant dependence between the proportionality factor and the magnitude of the transverse field, $h$. 

We now discuss the potential impact of $\Delta m$ on our results, had it not been deducted from $m(t)$. $\chi''(\omega)$ is not affected, as it is computed from the out-of-phase component of $m(t)$ while $\Delta m$ is in phase. $\chi'(\omega)$, however, receives an $\omega$-independent contribution on top of its theoretical value, Eq.~\eqref{eq:4:chip2}, which vanishes as $\omega^{-2}$ in the limit $\omega\to\infty$. Therefore, using $m(t)$ without correction yields $\lim_{\omega\to\infty}\phi = 0$, contrary to the predicted value of $\pi/2$ outlined in the main text. To illustrate this point, we plot in Fig.~\ref{fig:d:drtfic}a the results for $m(t)$ obtained with and without removing $\Delta m$ for the largest $\omega$ probed in our investigation. One sees that the uncorrected result is largely in-phase with the driven field $B(t)$ while the corrected results is out-of-phase by approximately a quarter of the period, signifying a phase shift $\phi \simeq \pi/2$. As a further illustration, in Fig.~\ref{fig:d:drtfic}c, we plot the results for $\phi$ obtained with and without such a correction for a specific combination of parameters as an example. As expected, the results for $\phi$ without correction tend to $0$ as $\omega$ increases, indicating that $m(t)$ in this limit is dominated by $\Delta m$ rather than by the quasiparticle dynamics. 
%
%

\section{\label{app:E}
Magnetization dynamics in CSI}
\begin{figure}[tbh!]
    \includegraphics[width=\columnwidth]{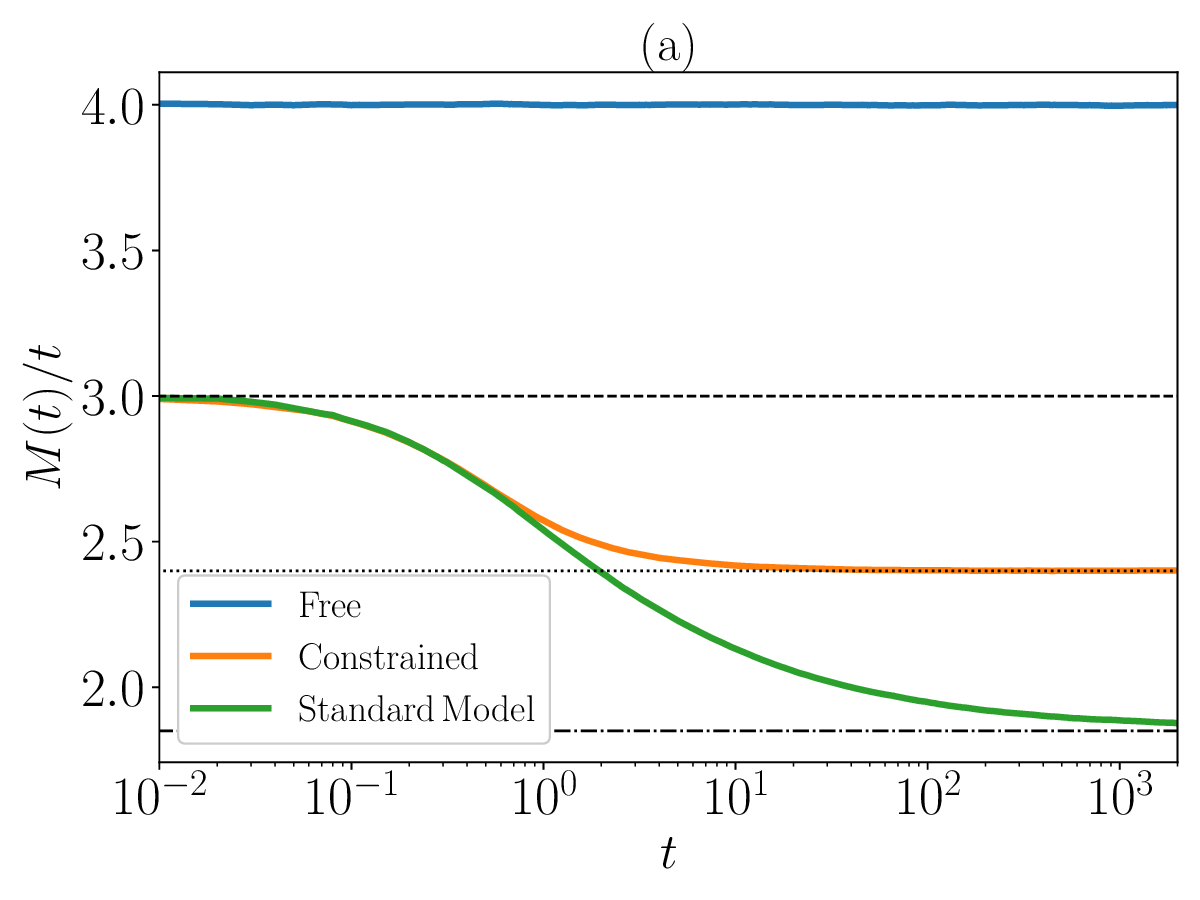}
    \includegraphics[width=\columnwidth]{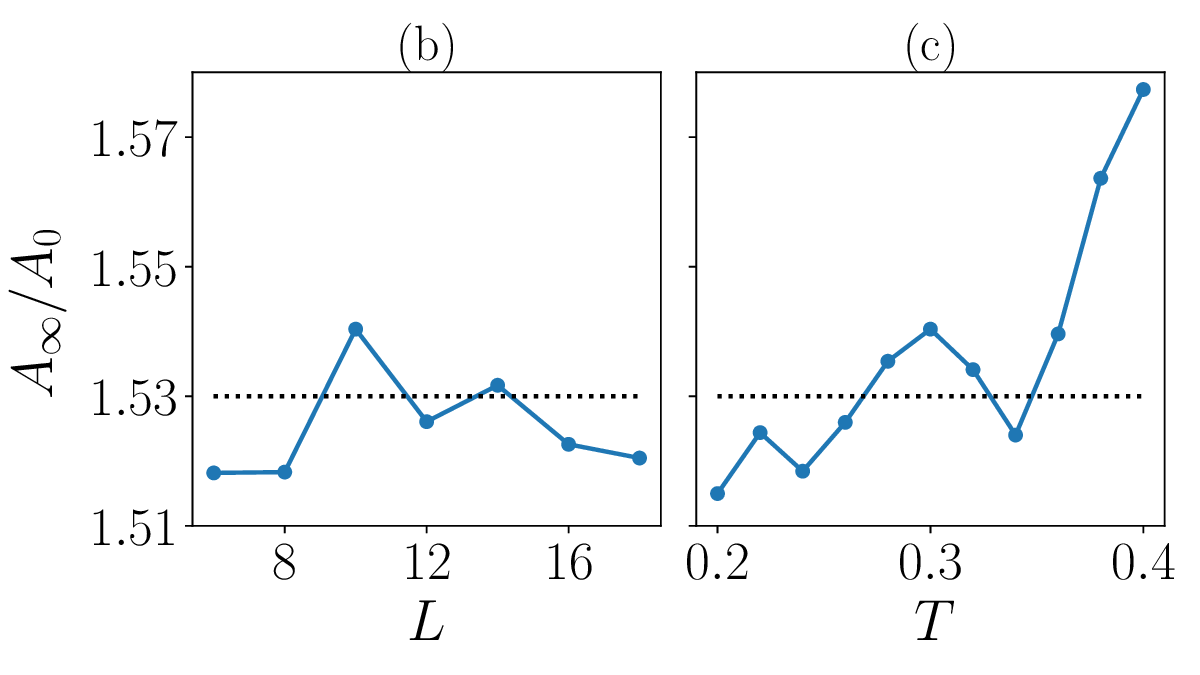}
    \caption{\label{fig:e:rw} Magnetization dynamics in CSI: (a) Results for the free, constrained and {standard model} random walks (see main text) on the diamond lattice. The free walk behaves diffusively at all times, while the constrained and {standard model} random walks are diffusive at short and long times, with a crossover between the two distinct diffusion constants at intermediate times. The three horizontal black lines represent the different diffusive behaviors observed in these models, with $M(t)/t=3$ (dashed), $2.4$ (dotted) and $1.85$ (dash-dotted). (b) and (c) The value of $A_\infty/A_0$ obtained by fitting the low- and high-frequency regions of the PSD (see Sec.~\ref{section:5:rtfsi} and Fig.~\ref{fig:5:rtfsi}b), simulated with system sizes $6\le L\le 18$ and temperatures $0.2\le T\le 0.4$.}
\end{figure}
In this appendix, we study in detail the classical spin ice model discussed in Sec.~\ref{section:5:results} in the main text. At low temperatures, the magnetization dynamics of the system is largely dominated by sparse monopoles performing random walks on the diamond lattice. Ignoring monopole creation and annihilation events, the only difference that remains between the monopolar random walk and a free random walk on the diamond lattice is the presence of blocked spin, as shown in Fig.~\ref{fig:5:rtfsi}a in the main text. 

To understand the effect of such constraints, we simulated and compared the following random walks on the diamond lattice: (i) a free, standard random walk on the diamond lattice with unit bond length, with the transition rates across each bonds set to unity, hereafter referred to as the `free walk'; (ii) a constrained random walk, where the walker is always allowed to backtrack but finds one of the three other directions to be blocked at random (and chosen anew each time, at the time of hopping of the walker), hereafter referred to as the `constrained walk'; (iii) the monopolar random walk happening in CSI in the single-excitation limit, hereafter referred to as the `standard model' random walk. The last walk is simulated following the procedure described in Ref.~\onlinecite{lu2024111}. We randomly initialize a spin system with $40\times40\times40$ unit cells ($10^6$ spins; see App.~\ref{app:F}); after that, perform single-spin flip Monte-Carlo updates at $T=0$ until no monopoles are present in the system, giving us a ground state configuration. Flipping a spin selected at random creates a monopole-anti-monopole pair; by fixing one of them and allowing the other to move in the system, we are able to calculate the MSD $M(t)$, defined at the beginning of Sec.~\ref{section:3:rw} in the main text, from the magnetization change of the system incurred by such a walk. 

Fig.~\ref{fig:e:rw}a summarizes the data obtained for the ratio $M(t)/t$. For the free walk, one has trivially $M(t)/t=4$ from the coordination number of the diamond lattice. For both the constrained walk and the {standard model}, a difference between the short- and long-time diffusive behavior appears. Both exhibit $D_0\equiv\lim_{t\to0}M(t)/t=3$, attributed to the 3 out of 4 available hopping directions the walker can take in its first step. As for $D_\infty\equiv\lim_{t\to\infty}M(t)/t$, the results obtained from our simulations point to $D_\infty =2.4$ for the constrained walk~\cite{haus1987diffusion}, and $D_\infty\approx1.85$ for the {standard model}. In the latter case, however, the presence of finite size effects at $M(t)\gtrsim L^2$ makes the accurate extraction of $D_\infty$ difficult; hence, we estimate a $1.85\le D_\infty\le 1.90$ for {standard model}.

The constrained walk can be exactly solved using the methods developed in Chapter 4 of Ref.~\onlinecite{haus1987diffusion}. In particular, it is a random walk with correlations over two successive steps (as the probability of a transition depends on the previous one --- recall that backtracking is always possible). Eq.~4.25 in the reference gives the exact solution of the MSD of such random walks and resembles Eq.~\eqref{eq:3:expfit} in the main text: 
\begin{equation}
    M(t)=2.4t+0.18(1-\mathrm{e}^{-t/\tau})
    \, ,
\end{equation}
with $\tau=0.3$, from which we trivially have $D_0=3$ and $D_\infty=2.4$. 

The {standard model} random walk differs from the constrained walk by the important aspect of spin correlations \cite{castelnovo2012spin}. In the ground state, they imply correlated blocked directions for the monopolar walker. Moreover, the Dirac string --- a chain of flipped spins along the path --- left by the mobile monopole also implies a memory effect of the random walk dynamics, which grows with time as the monopole leaves behind an increasingly longer trail. As can be seen from Fig.~\ref{fig:e:rw}a, comparing with the constrained walk, such effects seem to come in at around $t\sim10^0$, preventing the walker from exploring farther sites (hence reducing $M(t)/t$) and leaving a long relaxation tail in the MSD of the {standard model} before it converges to the long-time asymptotic diffusive behavior.

The above random walk analysis does not take into account the presence of finite temperature. Moreover, one may worry about the relatively small system sizes accessible in the spin simulation results discussed in the main text ($L<20$, comparing with $L=40$ used in the above {standard model} random walk simulations and $L\sim10^3$ in 1D; see App.~\ref{app:F}). For this reason, in Fig.~\ref{fig:e:rw}b and~\ref{fig:e:rw}c, we investigate the dependence of the shape of the PSD obtained from the spin simulations in the main text with temperature $T$ and system size $L$, by comparing the values $A_\infty/A_0$ using the same method and fitting range as in the main text (see Sec.~\ref{section:5:results} and Fig.~\ref{fig:5:rtfsi}b). As can be seen from the figures, $A_\infty/A_0$ has little variation within the system sizes considered and the temperature range, so long as $T\le0.35$. The upturn at higher temperatures can be attributed to a decreasing magnetic relaxation time which causes the low-frequency region to deviate from the $\mathrm{PSD}\propto\omega^{-2}$ behavior --- see the discussion in Sec.~\ref{App:A4}. 
%
%

\section{\label{app:F}
Simulation details and parameters}
For convenience, we collate and present here all the parameters and details relevant to our numerical simulations. 
%
%

\subsection{Modified Monte-Carlo simulations}
\subsubsection{One-dimensional RTFIC}
To obtain the results presented in Fig.~\ref{fig:PSD}, we performed Monte-Carlo simulations as described at the end of Sec.~\ref{section:2:model}. In the simulations, we consider $N$ spin-flip attempts as 1 MC-iteration, which is then associated with the passage of 1 unit of MC time. We simulate a system with $2000$ spins and periodic boundary conditions, initially in a random state (with classical spins randomly aligned and the quantum clusters in a random eigenstate of their respective Hamiltonians). A cutoff cluster size $n=8$ is used, which is sensible in our range for $r$ (the probability of finding clusters of size $n > 8$ in a $2000$-spin system with $r=0.2$, for example, is $\sim0.01$). After $2\times10^4$ equilibrium iterations, we run the system for an extra $2^{20}\approx10^6$ iterations, measuring the magnetization $m(t)$ every $1/50$ steps (i.e., every $40$ spin-flip attempts). We then employ the \verb|scipy.signal.welch| function~\cite{welch} to compute the PSDs presented in Fig.~\ref{fig:PSD}a using $2^{19}=524288$ periodograms. Each PSD curve is averaged over $12$ MC simulations with different realizations of the random transverse field. {The results in the limit $h\to0$, used in the main text to compare with the classical scenario, are obtained by setting $h=0.01$.}
%
%

\subsubsection{Three-dimensional RTFSI}
To obtain the results presented in Fig.~\ref{fig:5:rtfsi}b and Fig.~\ref{fig:5:rtfsi}c, we performed similar Monte-Carlo simulations as described at Sec.~\ref{section:5:setup}. The setup is similar to the 1D case. The system is selected to be consisting of $10\times10\times10$ unit cells, hence $16000$ spins, with periodic boundary conditions; the cutoff size is selected to be $n=7$, irrespective of the shape of the cluster. After $2\times10^4$ equilibrium iterations, we run the system for an extra $2^{21}\approx2\times10^6$ iterations, measuring the magnetization $m(t)$ every $1/64$ steps (i.e., every 250 spin-flip attempts). The PSD is computed using $2^{18}=262144$ periodograms, and each PSD curve is averaged over $10$ realizations of the random transverse field.
%
%

\subsection{Random-walk simulations}
To obtain the results presented in Fig.~\ref{fig:rw},  we simulate the random walk developed in Sec.~\ref{section:3:rw} and App.~\ref{app:C}. We initialize a system with $6000$ spins and periodic boundary condition and construct the corresponding {random walk} lattice on which our simulation run; a cutoff cluster size $n=10$ is used. A random walker is initialized randomly across the {random walk} lattice sites; in each step, the random walker moves with probability $w\delta t$ to one of its adjacent sites, where $w$ is the corresponding transition rate. We select $\delta t=1/50$. The data presented in Fig.~\ref{fig:rw}b are averaged over $200$ realizations of disorders.
%
%

\subsection{Driven RTFIC simulations}
To obtain the results presented in Fig.~\ref{fig:4:drtfic}, we simulate the Hamiltonian Eq.~\eqref{eq:4:Drtfic} using the Monte-Carlo method. We simulate a system with $1800$ spins and periodic boundary conditions, and select a cutoff $n=8$; on top of the regular Monte-Carlo steps, we also incorporate the changing value of the longitudinal field $B(t)$ in Eq.~\eqref{eq:4:Drtfic}. This is done by approximating $B(t)$ to be constant in a small interval of time $(t,t+\delta t)$ before updating its value. We select $\delta t=1/240t_0$ as a compromise between computational efficiency and data accuracy, where $t_0=2\pi/\omega$ is the period of $B(t)$. In each simulation, we equilibrate the system for $20000$ steps without applying the external longitudinal field, i.e., $B=0$; we then introduce the external field and equilibrate for an extra $30t_0$ to eliminate any transient effects and ensure a steady state of the system. $m(t)$ is then sampled over four periods, which are used to fit the values of $\chi$ and $\phi$ (as discussed at the beginning of Sec.~\ref{section:4:results}). Each data point presented in Figs.~\ref{fig:4:drtfic}a and Figs.~\ref{fig:4:drtfic}b comes from averaging over $3000$ Monte-Carlo simulations with different disorder realizations. 
%
%

\bibliography{ref}

\begin{thebibliography}{75}%
\makeatletter
\providecommand \@ifxundefined [1]{%
 \@ifx{#1\undefined}
}%
\providecommand \@ifnum [1]{%
 \ifnum #1\expandafter \@firstoftwo
 \else \expandafter \@secondoftwo
 \fi
}%
\providecommand \@ifx [1]{%
 \ifx #1\expandafter \@firstoftwo
 \else \expandafter \@secondoftwo
 \fi
}%
\providecommand \natexlab [1]{#1}%
\providecommand \enquote  [1]{``#1''}%
\providecommand \bibnamefont  [1]{#1}%
\providecommand \bibfnamefont [1]{#1}%
\providecommand \citenamefont [1]{#1}%
\providecommand \href@noop [0]{\@secondoftwo}%
\providecommand \href [0]{\begingroup \@sanitize@url \@href}%
\providecommand \@href[1]{\@@startlink{#1}\@@href}%
\providecommand \@@href[1]{\endgroup#1\@@endlink}%
\providecommand \@sanitize@url [0]{\catcode `\\12\catcode `\$12\catcode `\&12\catcode `\#12\catcode `\^12\catcode `\_12\catcode `\%12\relax}%
\providecommand \@@startlink[1]{}%
\providecommand \@@endlink[0]{}%
\providecommand \url  [0]{\begingroup\@sanitize@url \@url }%
\providecommand \@url [1]{\endgroup\@href {#1}{\urlprefix }}%
\providecommand \urlprefix  [0]{URL }%
\providecommand \Eprint [0]{\href }%
\providecommand \doibase [0]{https://doi.org/}%
\providecommand \selectlanguage [0]{\@gobble}%
\providecommand \bibinfo  [0]{\@secondoftwo}%
\providecommand \bibfield  [0]{\@secondoftwo}%
\providecommand \translation [1]{[#1]}%
\providecommand \BibitemOpen [0]{}%
\providecommand \bibitemStop [0]{}%
\providecommand \bibitemNoStop [0]{.\EOS\space}%
\providecommand \EOS [0]{\spacefactor3000\relax}%
\providecommand \BibitemShut  [1]{\csname bibitem#1\endcsname}%
\let\auto@bib@innerbib\@empty
\bibitem [{\citenamefont {Savary}\ and\ \citenamefont {Balents}(2016{\natexlab{a}})}]{savary2016quantum}%
  \BibitemOpen
  \bibfield  {author} {\bibinfo {author} {\bibfnamefont {L.}~\bibnamefont {Savary}}\ and\ \bibinfo {author} {\bibfnamefont {L.}~\bibnamefont {Balents}},\ }\bibfield  {title} {\bibinfo {title} {Quantum spin liquids: a review},\ }\href@noop {} {\bibfield  {journal} {\bibinfo  {journal} {Reports on Progress in Physics}\ }\textbf {\bibinfo {volume} {80}},\ \bibinfo {pages} {016502} (\bibinfo {year} {2016}{\natexlab{a}})}\BibitemShut {NoStop}%
\bibitem [{\citenamefont {Knolle}\ and\ \citenamefont {Moessner}(2019)}]{Knolle_2019}%
  \BibitemOpen
  \bibfield  {author} {\bibinfo {author} {\bibfnamefont {J.}~\bibnamefont {Knolle}}\ and\ \bibinfo {author} {\bibfnamefont {R.}~\bibnamefont {Moessner}},\ }\bibfield  {title} {\bibinfo {title} {A field guide to spin liquids},\ }\href {https://doi.org/10.1146/annurev-conmatphys-031218-013401} {\bibfield  {journal} {\bibinfo  {journal} {Annual Review of Condensed Matter Physics}\ }\textbf {\bibinfo {volume} {10}},\ \bibinfo {pages} {451–472} (\bibinfo {year} {2019})}\BibitemShut {NoStop}%
\bibitem [{\citenamefont {Broholm}\ \emph {et~al.}(2020)\citenamefont {Broholm}, \citenamefont {Cava}, \citenamefont {Kivelson}, \citenamefont {Nocera}, \citenamefont {Norman},\ and\ \citenamefont {Senthil}}]{broholm2020quantum}%
  \BibitemOpen
  \bibfield  {author} {\bibinfo {author} {\bibfnamefont {C.}~\bibnamefont {Broholm}}, \bibinfo {author} {\bibfnamefont {R.}~\bibnamefont {Cava}}, \bibinfo {author} {\bibfnamefont {S.}~\bibnamefont {Kivelson}}, \bibinfo {author} {\bibfnamefont {D.}~\bibnamefont {Nocera}}, \bibinfo {author} {\bibfnamefont {M.}~\bibnamefont {Norman}},\ and\ \bibinfo {author} {\bibfnamefont {T.}~\bibnamefont {Senthil}},\ }\bibfield  {title} {\bibinfo {title} {Quantum spin liquids},\ }\href@noop {} {\bibfield  {journal} {\bibinfo  {journal} {Science}\ }\textbf {\bibinfo {volume} {367}},\ \bibinfo {pages} {eaay0668} (\bibinfo {year} {2020})}\BibitemShut {NoStop}%
\bibitem [{\citenamefont {Knolle}\ \emph {et~al.}(2014)\citenamefont {Knolle}, \citenamefont {Kovrizhin}, \citenamefont {Chalker},\ and\ \citenamefont {Moessner}}]{knolle2014dynamics}%
  \BibitemOpen
  \bibfield  {author} {\bibinfo {author} {\bibfnamefont {J.}~\bibnamefont {Knolle}}, \bibinfo {author} {\bibfnamefont {D.}~\bibnamefont {Kovrizhin}}, \bibinfo {author} {\bibfnamefont {J.}~\bibnamefont {Chalker}},\ and\ \bibinfo {author} {\bibfnamefont {R.}~\bibnamefont {Moessner}},\ }\bibfield  {title} {\bibinfo {title} {Dynamics of a two-dimensional quantum spin liquid: signatures of emergent majorana fermions and fluxes},\ }\href@noop {} {\bibfield  {journal} {\bibinfo  {journal} {Physical Review Letters}\ }\textbf {\bibinfo {volume} {112}},\ \bibinfo {pages} {207203} (\bibinfo {year} {2014})}\BibitemShut {NoStop}%
\bibitem [{\citenamefont {Knolle}(2016)}]{knolle2016dynamics}%
  \BibitemOpen
  \bibfield  {author} {\bibinfo {author} {\bibfnamefont {J.}~\bibnamefont {Knolle}},\ }\href@noop {} {\emph {\bibinfo {title} {Dynamics of a Quantum Spin Liquid}}}\ (\bibinfo  {publisher} {Springer},\ \bibinfo {year} {2016})\BibitemShut {NoStop}%
\bibitem [{\citenamefont {Udagawa}\ \emph {et~al.}(2016)\citenamefont {Udagawa}, \citenamefont {Jaubert}, \citenamefont {Castelnovo},\ and\ \citenamefont {Moessner}}]{udagawa2016out}%
  \BibitemOpen
  \bibfield  {author} {\bibinfo {author} {\bibfnamefont {M.}~\bibnamefont {Udagawa}}, \bibinfo {author} {\bibfnamefont {L.}~\bibnamefont {Jaubert}}, \bibinfo {author} {\bibfnamefont {C.}~\bibnamefont {Castelnovo}},\ and\ \bibinfo {author} {\bibfnamefont {R.}~\bibnamefont {Moessner}},\ }\bibfield  {title} {\bibinfo {title} {Out-of-equilibrium dynamics and extended textures of topological defects in spin ice},\ }\href@noop {} {\bibfield  {journal} {\bibinfo  {journal} {Physical Review B}\ }\textbf {\bibinfo {volume} {94}},\ \bibinfo {pages} {104416} (\bibinfo {year} {2016})}\BibitemShut {NoStop}%
\bibitem [{\citenamefont {Gohlke}\ \emph {et~al.}(2018)\citenamefont {Gohlke}, \citenamefont {Moessner},\ and\ \citenamefont {Pollmann}}]{gohlke2018dynamical}%
  \BibitemOpen
  \bibfield  {author} {\bibinfo {author} {\bibfnamefont {M.}~\bibnamefont {Gohlke}}, \bibinfo {author} {\bibfnamefont {R.}~\bibnamefont {Moessner}},\ and\ \bibinfo {author} {\bibfnamefont {F.}~\bibnamefont {Pollmann}},\ }\bibfield  {title} {\bibinfo {title} {Dynamical and topological properties of the kitaev model in a [111] magnetic field},\ }\href@noop {} {\bibfield  {journal} {\bibinfo  {journal} {Physical Review B}\ }\textbf {\bibinfo {volume} {98}},\ \bibinfo {pages} {014418} (\bibinfo {year} {2018})}\BibitemShut {NoStop}%
\bibitem [{\citenamefont {Revell}\ \emph {et~al.}(2013)\citenamefont {Revell}, \citenamefont {Yaraskavitch}, \citenamefont {Mason}, \citenamefont {Ross}, \citenamefont {Noad}, \citenamefont {Dabkowska}, \citenamefont {Gaulin}, \citenamefont {Henelius},\ and\ \citenamefont {Kycia}}]{revell2013evidence}%
  \BibitemOpen
  \bibfield  {author} {\bibinfo {author} {\bibfnamefont {H.}~\bibnamefont {Revell}}, \bibinfo {author} {\bibfnamefont {L.}~\bibnamefont {Yaraskavitch}}, \bibinfo {author} {\bibfnamefont {J.}~\bibnamefont {Mason}}, \bibinfo {author} {\bibfnamefont {K.}~\bibnamefont {Ross}}, \bibinfo {author} {\bibfnamefont {H.}~\bibnamefont {Noad}}, \bibinfo {author} {\bibfnamefont {H.}~\bibnamefont {Dabkowska}}, \bibinfo {author} {\bibfnamefont {B.}~\bibnamefont {Gaulin}}, \bibinfo {author} {\bibfnamefont {P.}~\bibnamefont {Henelius}},\ and\ \bibinfo {author} {\bibfnamefont {J.}~\bibnamefont {Kycia}},\ }\bibfield  {title} {\bibinfo {title} {Evidence of impurity and boundary effects on magnetic monopole dynamics in spin ice},\ }\href@noop {} {\bibfield  {journal} {\bibinfo  {journal} {Nature Physics}\ }\textbf {\bibinfo {volume} {9}},\ \bibinfo {pages} {34} (\bibinfo {year} {2013})}\BibitemShut {NoStop}%
\bibitem [{\citenamefont {Samarakoon}\ \emph {et~al.}(2022)\citenamefont {Samarakoon}, \citenamefont {Grigera}, \citenamefont {Tennant}, \citenamefont {Kirste}, \citenamefont {Klemke}, \citenamefont {Strehlow}, \citenamefont {Meissner}, \citenamefont {Hall{\'e}n}, \citenamefont {Jaubert}, \citenamefont {Castelnovo} \emph {et~al.}}]{samarakoon2022anomalous}%
  \BibitemOpen
  \bibfield  {author} {\bibinfo {author} {\bibfnamefont {A.~M.}\ \bibnamefont {Samarakoon}}, \bibinfo {author} {\bibfnamefont {S.}~\bibnamefont {Grigera}}, \bibinfo {author} {\bibfnamefont {D.~A.}\ \bibnamefont {Tennant}}, \bibinfo {author} {\bibfnamefont {A.}~\bibnamefont {Kirste}}, \bibinfo {author} {\bibfnamefont {B.}~\bibnamefont {Klemke}}, \bibinfo {author} {\bibfnamefont {P.}~\bibnamefont {Strehlow}}, \bibinfo {author} {\bibfnamefont {M.}~\bibnamefont {Meissner}}, \bibinfo {author} {\bibfnamefont {J.~N.}\ \bibnamefont {Hall{\'e}n}}, \bibinfo {author} {\bibfnamefont {L.}~\bibnamefont {Jaubert}}, \bibinfo {author} {\bibfnamefont {C.}~\bibnamefont {Castelnovo}}, \emph {et~al.},\ }\bibfield  {title} {\bibinfo {title} {Anomalous magnetic noise in an imperfectly flat landscape in the topological magnet dy2ti2o7},\ }\href@noop {} {\bibfield  {journal} {\bibinfo  {journal} {Proceedings of the National Academy of Sciences}\ }\textbf {\bibinfo {volume} {119}},\ \bibinfo {pages} {e2117453119} (\bibinfo {year} {2022})}\BibitemShut {NoStop}%
\bibitem [{\citenamefont {Hall{\'e}n}\ \emph {et~al.}(2022)\citenamefont {Hall{\'e}n}, \citenamefont {Grigera}, \citenamefont {Tennant}, \citenamefont {Castelnovo},\ and\ \citenamefont {Moessner}}]{hallen2022dynamical}%
  \BibitemOpen
  \bibfield  {author} {\bibinfo {author} {\bibfnamefont {J.~N.}\ \bibnamefont {Hall{\'e}n}}, \bibinfo {author} {\bibfnamefont {S.~A.}\ \bibnamefont {Grigera}}, \bibinfo {author} {\bibfnamefont {D.~A.}\ \bibnamefont {Tennant}}, \bibinfo {author} {\bibfnamefont {C.}~\bibnamefont {Castelnovo}},\ and\ \bibinfo {author} {\bibfnamefont {R.}~\bibnamefont {Moessner}},\ }\bibfield  {title} {\bibinfo {title} {Dynamical fractal and anomalous noise in a clean magnetic crystal},\ }\href@noop {} {\bibfield  {journal} {\bibinfo  {journal} {Science}\ }\textbf {\bibinfo {volume} {378}},\ \bibinfo {pages} {1218} (\bibinfo {year} {2022})}\BibitemShut {NoStop}%
\bibitem [{\citenamefont {Hsu}\ \emph {et~al.}(2024)\citenamefont {Hsu}, \citenamefont {Takahashi}, \citenamefont {Jerzembeck}, \citenamefont {Dasini}, \citenamefont {Carroll}, \citenamefont {Dusad}, \citenamefont {Ward}, \citenamefont {Dawson}, \citenamefont {Sharma}, \citenamefont {Luke} \emph {et~al.}}]{hsu2024dichotomous}%
  \BibitemOpen
  \bibfield  {author} {\bibinfo {author} {\bibfnamefont {C.-C.}\ \bibnamefont {Hsu}}, \bibinfo {author} {\bibfnamefont {H.}~\bibnamefont {Takahashi}}, \bibinfo {author} {\bibfnamefont {F.}~\bibnamefont {Jerzembeck}}, \bibinfo {author} {\bibfnamefont {J.}~\bibnamefont {Dasini}}, \bibinfo {author} {\bibfnamefont {C.}~\bibnamefont {Carroll}}, \bibinfo {author} {\bibfnamefont {R.}~\bibnamefont {Dusad}}, \bibinfo {author} {\bibfnamefont {J.}~\bibnamefont {Ward}}, \bibinfo {author} {\bibfnamefont {C.}~\bibnamefont {Dawson}}, \bibinfo {author} {\bibfnamefont {S.}~\bibnamefont {Sharma}}, \bibinfo {author} {\bibfnamefont {G.~M.}\ \bibnamefont {Luke}}, \emph {et~al.},\ }\bibfield  {title} {\bibinfo {title} {Dichotomous dynamics of magnetic monopole fluids},\ }\href@noop {} {\bibfield  {journal} {\bibinfo  {journal} {Proceedings of the National Academy of Sciences}\ }\textbf {\bibinfo {volume} {121}},\ \bibinfo {pages} {e2320384121} (\bibinfo {year} {2024})}\BibitemShut {NoStop}%
\bibitem [{\citenamefont {Misguich}\ and\ \citenamefont {Mila}(2008)}]{misguich2008quantum}%
  \BibitemOpen
  \bibfield  {author} {\bibinfo {author} {\bibfnamefont {G.}~\bibnamefont {Misguich}}\ and\ \bibinfo {author} {\bibfnamefont {F.}~\bibnamefont {Mila}},\ }\bibfield  {title} {\bibinfo {title} {Quantum dimer model on the triangular lattice: Semiclassical and variational approaches to vison dispersion and condensation},\ }\href@noop {} {\bibfield  {journal} {\bibinfo  {journal} {Physical Review B—Condensed Matter and Materials Physics}\ }\textbf {\bibinfo {volume} {77}},\ \bibinfo {pages} {134421} (\bibinfo {year} {2008})}\BibitemShut {NoStop}%
\bibitem [{\citenamefont {Yan}\ \emph {et~al.}(2021)\citenamefont {Yan}, \citenamefont {Wang}, \citenamefont {Ma}, \citenamefont {Qi},\ and\ \citenamefont {Meng}}]{yan2021topological}%
  \BibitemOpen
  \bibfield  {author} {\bibinfo {author} {\bibfnamefont {Z.}~\bibnamefont {Yan}}, \bibinfo {author} {\bibfnamefont {Y.-C.}\ \bibnamefont {Wang}}, \bibinfo {author} {\bibfnamefont {N.}~\bibnamefont {Ma}}, \bibinfo {author} {\bibfnamefont {Y.}~\bibnamefont {Qi}},\ and\ \bibinfo {author} {\bibfnamefont {Z.~Y.}\ \bibnamefont {Meng}},\ }\bibfield  {title} {\bibinfo {title} {Topological phase transition and single/multi anyon dynamics of z 2 spin liquid},\ }\href@noop {} {\bibfield  {journal} {\bibinfo  {journal} {npj Quantum Materials}\ }\textbf {\bibinfo {volume} {6}},\ \bibinfo {pages} {39} (\bibinfo {year} {2021})}\BibitemShut {NoStop}%
\bibitem [{\citenamefont {Ritort}\ and\ \citenamefont {Sollich}(2003)}]{ritort2003glassy}%
  \BibitemOpen
  \bibfield  {author} {\bibinfo {author} {\bibfnamefont {F.}~\bibnamefont {Ritort}}\ and\ \bibinfo {author} {\bibfnamefont {P.}~\bibnamefont {Sollich}},\ }\bibfield  {title} {\bibinfo {title} {Glassy dynamics of kinetically constrained models},\ }\href@noop {} {\bibfield  {journal} {\bibinfo  {journal} {Advances in physics}\ }\textbf {\bibinfo {volume} {52}},\ \bibinfo {pages} {219} (\bibinfo {year} {2003})}\BibitemShut {NoStop}%
\bibitem [{\citenamefont {Willans}\ \emph {et~al.}(2011)\citenamefont {Willans}, \citenamefont {Chalker},\ and\ \citenamefont {Moessner}}]{willans2011site}%
  \BibitemOpen
  \bibfield  {author} {\bibinfo {author} {\bibfnamefont {A.}~\bibnamefont {Willans}}, \bibinfo {author} {\bibfnamefont {J.}~\bibnamefont {Chalker}},\ and\ \bibinfo {author} {\bibfnamefont {R.}~\bibnamefont {Moessner}},\ }\bibfield  {title} {\bibinfo {title} {Site dilution in the kitaev honeycomb model},\ }\href@noop {} {\bibfield  {journal} {\bibinfo  {journal} {Physical Review B—Condensed Matter and Materials Physics}\ }\textbf {\bibinfo {volume} {84}},\ \bibinfo {pages} {115146} (\bibinfo {year} {2011})}\BibitemShut {NoStop}%
\bibitem [{\citenamefont {Petrova}\ \emph {et~al.}(2015)\citenamefont {Petrova}, \citenamefont {Moessner},\ and\ \citenamefont {Sondhi}}]{petrova2015hydrogenic}%
  \BibitemOpen
  \bibfield  {author} {\bibinfo {author} {\bibfnamefont {O.}~\bibnamefont {Petrova}}, \bibinfo {author} {\bibfnamefont {R.}~\bibnamefont {Moessner}},\ and\ \bibinfo {author} {\bibfnamefont {S.~L.}\ \bibnamefont {Sondhi}},\ }\bibfield  {title} {\bibinfo {title} {Hydrogenic states of monopoles in diluted quantum spin ice},\ }\href@noop {} {\bibfield  {journal} {\bibinfo  {journal} {Physical Review B}\ }\textbf {\bibinfo {volume} {92}},\ \bibinfo {pages} {100401} (\bibinfo {year} {2015})}\BibitemShut {NoStop}%
\bibitem [{\citenamefont {Andreanov}\ and\ \citenamefont {McClarty}(2015)}]{andreanov2015order}%
  \BibitemOpen
  \bibfield  {author} {\bibinfo {author} {\bibfnamefont {A.}~\bibnamefont {Andreanov}}\ and\ \bibinfo {author} {\bibfnamefont {P.}~\bibnamefont {McClarty}},\ }\bibfield  {title} {\bibinfo {title} {Order induced by dilution in pyrochlore xy antiferromagnets},\ }\href@noop {} {\bibfield  {journal} {\bibinfo  {journal} {Physical Review B}\ }\textbf {\bibinfo {volume} {91}},\ \bibinfo {pages} {064401} (\bibinfo {year} {2015})}\BibitemShut {NoStop}%
\bibitem [{\citenamefont {Furukawa}\ \emph {et~al.}(2015)\citenamefont {Furukawa}, \citenamefont {Miyagawa}, \citenamefont {Itou}, \citenamefont {Ito}, \citenamefont {Taniguchi}, \citenamefont {Saito}, \citenamefont {Iguchi}, \citenamefont {Sasaki},\ and\ \citenamefont {Kanoda}}]{furukawa2015quantum}%
  \BibitemOpen
  \bibfield  {author} {\bibinfo {author} {\bibfnamefont {T.}~\bibnamefont {Furukawa}}, \bibinfo {author} {\bibfnamefont {K.}~\bibnamefont {Miyagawa}}, \bibinfo {author} {\bibfnamefont {T.}~\bibnamefont {Itou}}, \bibinfo {author} {\bibfnamefont {M.}~\bibnamefont {Ito}}, \bibinfo {author} {\bibfnamefont {H.}~\bibnamefont {Taniguchi}}, \bibinfo {author} {\bibfnamefont {M.}~\bibnamefont {Saito}}, \bibinfo {author} {\bibfnamefont {S.}~\bibnamefont {Iguchi}}, \bibinfo {author} {\bibfnamefont {T.}~\bibnamefont {Sasaki}},\ and\ \bibinfo {author} {\bibfnamefont {K.}~\bibnamefont {Kanoda}},\ }\bibfield  {title} {\bibinfo {title} {Quantum spin liquid emerging from antiferromagnetic order by introducing disorder},\ }\href@noop {} {\bibfield  {journal} {\bibinfo  {journal} {Physical review letters}\ }\textbf {\bibinfo {volume} {115}},\ \bibinfo {pages} {077001} (\bibinfo {year} {2015})}\BibitemShut {NoStop}%
\bibitem [{\citenamefont {Zhu}\ \emph {et~al.}(2017)\citenamefont {Zhu}, \citenamefont {Maksimov}, \citenamefont {White},\ and\ \citenamefont {Chernyshev}}]{zhu2017disorder}%
  \BibitemOpen
  \bibfield  {author} {\bibinfo {author} {\bibfnamefont {Z.}~\bibnamefont {Zhu}}, \bibinfo {author} {\bibfnamefont {P.}~\bibnamefont {Maksimov}}, \bibinfo {author} {\bibfnamefont {S.~R.}\ \bibnamefont {White}},\ and\ \bibinfo {author} {\bibfnamefont {A.}~\bibnamefont {Chernyshev}},\ }\bibfield  {title} {\bibinfo {title} {Disorder-induced mimicry of a spin liquid in ybmggao 4},\ }\href@noop {} {\bibfield  {journal} {\bibinfo  {journal} {Physical review letters}\ }\textbf {\bibinfo {volume} {119}},\ \bibinfo {pages} {157201} (\bibinfo {year} {2017})}\BibitemShut {NoStop}%
\bibitem [{\citenamefont {Udagawa}\ \emph {et~al.}(2021)\citenamefont {Udagawa}, \citenamefont {Jaubert} \emph {et~al.}}]{udagawa2021spin}%
  \BibitemOpen
  \bibfield  {author} {\bibinfo {author} {\bibfnamefont {M.}~\bibnamefont {Udagawa}}, \bibinfo {author} {\bibfnamefont {L.}~\bibnamefont {Jaubert}}, \emph {et~al.},\ }\href@noop {} {\emph {\bibinfo {title} {Spin Ice}}}\ (\bibinfo  {publisher} {Springer},\ \bibinfo {year} {2021})\BibitemShut {NoStop}%
\bibitem [{\citenamefont {Castelnovo}\ \emph {et~al.}(2008)\citenamefont {Castelnovo}, \citenamefont {Moessner},\ and\ \citenamefont {Sondhi}}]{castelnovo2008magnetic}%
  \BibitemOpen
  \bibfield  {author} {\bibinfo {author} {\bibfnamefont {C.}~\bibnamefont {Castelnovo}}, \bibinfo {author} {\bibfnamefont {R.}~\bibnamefont {Moessner}},\ and\ \bibinfo {author} {\bibfnamefont {S.~L.}\ \bibnamefont {Sondhi}},\ }\bibfield  {title} {\bibinfo {title} {Magnetic monopoles in spin ice},\ }\href@noop {} {\bibfield  {journal} {\bibinfo  {journal} {Nature}\ }\textbf {\bibinfo {volume} {451}},\ \bibinfo {pages} {42} (\bibinfo {year} {2008})}\BibitemShut {NoStop}%
\bibitem [{\citenamefont {Castelnovo}\ \emph {et~al.}(2012)\citenamefont {Castelnovo}, \citenamefont {Moessner},\ and\ \citenamefont {Sondhi}}]{castelnovo2012spin}%
  \BibitemOpen
  \bibfield  {author} {\bibinfo {author} {\bibfnamefont {C.}~\bibnamefont {Castelnovo}}, \bibinfo {author} {\bibfnamefont {R.}~\bibnamefont {Moessner}},\ and\ \bibinfo {author} {\bibfnamefont {S.~L.}\ \bibnamefont {Sondhi}},\ }\bibfield  {title} {\bibinfo {title} {Spin ice, fractionalization, and topological order},\ }\href@noop {} {\bibfield  {journal} {\bibinfo  {journal} {Annu. Rev. Condens. Matter Phys.}\ }\textbf {\bibinfo {volume} {3}},\ \bibinfo {pages} {35} (\bibinfo {year} {2012})}\BibitemShut {NoStop}%
\bibitem [{\citenamefont {Kolland}\ \emph {et~al.}(2012)\citenamefont {Kolland}, \citenamefont {Breunig}, \citenamefont {Valldor}, \citenamefont {Hiertz}, \citenamefont {Frielingsdorf},\ and\ \citenamefont {Lorenz}}]{kolland2012thermal}%
  \BibitemOpen
  \bibfield  {author} {\bibinfo {author} {\bibfnamefont {G.}~\bibnamefont {Kolland}}, \bibinfo {author} {\bibfnamefont {O.}~\bibnamefont {Breunig}}, \bibinfo {author} {\bibfnamefont {M.}~\bibnamefont {Valldor}}, \bibinfo {author} {\bibfnamefont {M.}~\bibnamefont {Hiertz}}, \bibinfo {author} {\bibfnamefont {J.}~\bibnamefont {Frielingsdorf}},\ and\ \bibinfo {author} {\bibfnamefont {T.}~\bibnamefont {Lorenz}},\ }\bibfield  {title} {\bibinfo {title} {Thermal conductivity and specific heat of the spin-ice compound dy 2 ti 2 o 7: Experimental evidence for monopole heat transport},\ }\href@noop {} {\bibfield  {journal} {\bibinfo  {journal} {Physical Review B—Condensed Matter and Materials Physics}\ }\textbf {\bibinfo {volume} {86}},\ \bibinfo {pages} {060402} (\bibinfo {year} {2012})}\BibitemShut {NoStop}%
\bibitem [{\citenamefont {Bramwell}\ \emph {et~al.}(2009)\citenamefont {Bramwell}, \citenamefont {Giblin}, \citenamefont {Calder}, \citenamefont {Aldus}, \citenamefont {Prabhakaran},\ and\ \citenamefont {Fennell}}]{bramwell2009measurement}%
  \BibitemOpen
  \bibfield  {author} {\bibinfo {author} {\bibfnamefont {S.~T.}\ \bibnamefont {Bramwell}}, \bibinfo {author} {\bibfnamefont {S.}~\bibnamefont {Giblin}}, \bibinfo {author} {\bibfnamefont {S.}~\bibnamefont {Calder}}, \bibinfo {author} {\bibfnamefont {R.}~\bibnamefont {Aldus}}, \bibinfo {author} {\bibfnamefont {D.}~\bibnamefont {Prabhakaran}},\ and\ \bibinfo {author} {\bibfnamefont {T.}~\bibnamefont {Fennell}},\ }\bibfield  {title} {\bibinfo {title} {Measurement of the charge and current of magnetic monopoles in spin ice},\ }\href@noop {} {\bibfield  {journal} {\bibinfo  {journal} {Nature}\ }\textbf {\bibinfo {volume} {461}},\ \bibinfo {pages} {956} (\bibinfo {year} {2009})}\BibitemShut {NoStop}%
\bibitem [{\citenamefont {Castelnovo}\ \emph {et~al.}(2010)\citenamefont {Castelnovo}, \citenamefont {Moessner},\ and\ \citenamefont {Sondhi}}]{castelnovo2010thermal}%
  \BibitemOpen
  \bibfield  {author} {\bibinfo {author} {\bibfnamefont {C.}~\bibnamefont {Castelnovo}}, \bibinfo {author} {\bibfnamefont {R.}~\bibnamefont {Moessner}},\ and\ \bibinfo {author} {\bibfnamefont {S.~L.}\ \bibnamefont {Sondhi}},\ }\bibfield  {title} {\bibinfo {title} {Thermal quenches in spin ice},\ }\href@noop {} {\bibfield  {journal} {\bibinfo  {journal} {Physical review letters}\ }\textbf {\bibinfo {volume} {104}},\ \bibinfo {pages} {107201} (\bibinfo {year} {2010})}\BibitemShut {NoStop}%
\bibitem [{\citenamefont {Mostame}\ \emph {et~al.}(2014)\citenamefont {Mostame}, \citenamefont {Castelnovo}, \citenamefont {Moessner},\ and\ \citenamefont {Sondhi}}]{mostame2014tunable}%
  \BibitemOpen
  \bibfield  {author} {\bibinfo {author} {\bibfnamefont {S.}~\bibnamefont {Mostame}}, \bibinfo {author} {\bibfnamefont {C.}~\bibnamefont {Castelnovo}}, \bibinfo {author} {\bibfnamefont {R.}~\bibnamefont {Moessner}},\ and\ \bibinfo {author} {\bibfnamefont {S.~L.}\ \bibnamefont {Sondhi}},\ }\bibfield  {title} {\bibinfo {title} {Tunable nonequilibrium dynamics of field quenches in spin ice},\ }\href@noop {} {\bibfield  {journal} {\bibinfo  {journal} {Proceedings of the National Academy of Sciences}\ }\textbf {\bibinfo {volume} {111}},\ \bibinfo {pages} {640} (\bibinfo {year} {2014})}\BibitemShut {NoStop}%
\bibitem [{\citenamefont {Savary}\ and\ \citenamefont {Balents}(2016{\natexlab{b}})}]{savary2016disorder}%
  \BibitemOpen
  \bibfield  {author} {\bibinfo {author} {\bibfnamefont {L.}~\bibnamefont {Savary}}\ and\ \bibinfo {author} {\bibfnamefont {L.}~\bibnamefont {Balents}},\ }\bibfield  {title} {\bibinfo {title} {Disorder-induced entanglement in spin ice pyrochlores},\ }\href@noop {} {\bibfield  {journal} {\bibinfo  {journal} {arXiv preprint arXiv:1604.04630}\ } (\bibinfo {year} {2016}{\natexlab{b}})}\BibitemShut {NoStop}%
\bibitem [{\citenamefont {Benton}(2018)}]{benton2018instabilities}%
  \BibitemOpen
  \bibfield  {author} {\bibinfo {author} {\bibfnamefont {O.}~\bibnamefont {Benton}},\ }\bibfield  {title} {\bibinfo {title} {Instabilities of a u (1) quantum spin liquid in disordered non-kramers pyrochlores},\ }\href@noop {} {\bibfield  {journal} {\bibinfo  {journal} {Physical Review Letters}\ }\textbf {\bibinfo {volume} {121}},\ \bibinfo {pages} {037203} (\bibinfo {year} {2018})}\BibitemShut {NoStop}%
\bibitem [{\citenamefont {Lindblad}(1976)}]{lindblad1976generators}%
  \BibitemOpen
  \bibfield  {author} {\bibinfo {author} {\bibfnamefont {G.}~\bibnamefont {Lindblad}},\ }\bibfield  {title} {\bibinfo {title} {On the generators of quantum dynamical semigroups},\ }\href@noop {} {\bibfield  {journal} {\bibinfo  {journal} {Communications in mathematical physics}\ }\textbf {\bibinfo {volume} {48}},\ \bibinfo {pages} {119} (\bibinfo {year} {1976})}\BibitemShut {NoStop}%
\bibitem [{\citenamefont {Gorini}\ \emph {et~al.}(1976)\citenamefont {Gorini}, \citenamefont {Kossakowski},\ and\ \citenamefont {Sudarshan}}]{gorini1976completely}%
  \BibitemOpen
  \bibfield  {author} {\bibinfo {author} {\bibfnamefont {V.}~\bibnamefont {Gorini}}, \bibinfo {author} {\bibfnamefont {A.}~\bibnamefont {Kossakowski}},\ and\ \bibinfo {author} {\bibfnamefont {E.~C.~G.}\ \bibnamefont {Sudarshan}},\ }\bibfield  {title} {\bibinfo {title} {Completely positive dynamical semigroups of n-level systems},\ }\href@noop {} {\bibfield  {journal} {\bibinfo  {journal} {Journal of Mathematical Physics}\ }\textbf {\bibinfo {volume} {17}},\ \bibinfo {pages} {821} (\bibinfo {year} {1976})}\BibitemShut {NoStop}%
\bibitem [{\citenamefont {Dusad}\ \emph {et~al.}(2019)\citenamefont {Dusad}, \citenamefont {Kirschner}, \citenamefont {Hoke}, \citenamefont {Roberts}, \citenamefont {Eyal}, \citenamefont {Flicker}, \citenamefont {Luke}, \citenamefont {Blundell},\ and\ \citenamefont {Davis}}]{dusad2019magnetic}%
  \BibitemOpen
  \bibfield  {author} {\bibinfo {author} {\bibfnamefont {R.}~\bibnamefont {Dusad}}, \bibinfo {author} {\bibfnamefont {F.~K.}\ \bibnamefont {Kirschner}}, \bibinfo {author} {\bibfnamefont {J.~C.}\ \bibnamefont {Hoke}}, \bibinfo {author} {\bibfnamefont {B.~R.}\ \bibnamefont {Roberts}}, \bibinfo {author} {\bibfnamefont {A.}~\bibnamefont {Eyal}}, \bibinfo {author} {\bibfnamefont {F.}~\bibnamefont {Flicker}}, \bibinfo {author} {\bibfnamefont {G.~M.}\ \bibnamefont {Luke}}, \bibinfo {author} {\bibfnamefont {S.~J.}\ \bibnamefont {Blundell}},\ and\ \bibinfo {author} {\bibfnamefont {J.~S.}\ \bibnamefont {Davis}},\ }\bibfield  {title} {\bibinfo {title} {Magnetic monopole noise},\ }\href@noop {} {\bibfield  {journal} {\bibinfo  {journal} {Nature}\ }\textbf {\bibinfo {volume} {571}},\ \bibinfo {pages} {234} (\bibinfo {year} {2019})}\BibitemShut {NoStop}%
\bibitem [{\citenamefont {Morineau}\ \emph {et~al.}(2025)\citenamefont {Morineau}, \citenamefont {Cathelin}, \citenamefont {Holdsworth}, \citenamefont {Giblin}, \citenamefont {Balakhrishnan}, \citenamefont {Matsuhira}, \citenamefont {Paulsen},\ and\ \citenamefont {Lhotel}}]{morineau2025satisfaction}%
  \BibitemOpen
  \bibfield  {author} {\bibinfo {author} {\bibfnamefont {F.}~\bibnamefont {Morineau}}, \bibinfo {author} {\bibfnamefont {V.}~\bibnamefont {Cathelin}}, \bibinfo {author} {\bibfnamefont {P.}~\bibnamefont {Holdsworth}}, \bibinfo {author} {\bibfnamefont {S.}~\bibnamefont {Giblin}}, \bibinfo {author} {\bibfnamefont {G.}~\bibnamefont {Balakhrishnan}}, \bibinfo {author} {\bibfnamefont {K.}~\bibnamefont {Matsuhira}}, \bibinfo {author} {\bibfnamefont {C.}~\bibnamefont {Paulsen}},\ and\ \bibinfo {author} {\bibfnamefont {E.}~\bibnamefont {Lhotel}},\ }\bibfield  {title} {\bibinfo {title} {Satisfaction and violation of the fluctuation-dissipation relation in spin ice materials},\ }\href@noop {} {\bibfield  {journal} {\bibinfo  {journal} {Physical Review Letters}\ }\textbf {\bibinfo {volume} {134}},\ \bibinfo {pages} {096702} (\bibinfo {year} {2025})}\BibitemShut {NoStop}%
\bibitem [{\citenamefont {Hastings}(1970)}]{hastings1970monte}%
  \BibitemOpen
  \bibfield  {author} {\bibinfo {author} {\bibfnamefont {W.~K.}\ \bibnamefont {Hastings}},\ }\bibfield  {title} {\bibinfo {title} {Monte carlo sampling methods using markov chains and their applications},\ }\href@noop {} {\bibfield  {journal} {\bibinfo  {journal} {Biometrika}\ }\textbf {\bibinfo {volume} {57}},\ \bibinfo {pages} {97} (\bibinfo {year} {1970})}\BibitemShut {NoStop}%
\bibitem [{\citenamefont {Augusiak}\ \emph {et~al.}(2010)\citenamefont {Augusiak}, \citenamefont {Cucchietti}, \citenamefont {Haake},\ and\ \citenamefont {Lewenstein}}]{augusiak2010quantum}%
  \BibitemOpen
  \bibfield  {author} {\bibinfo {author} {\bibfnamefont {R.}~\bibnamefont {Augusiak}}, \bibinfo {author} {\bibfnamefont {F.}~\bibnamefont {Cucchietti}}, \bibinfo {author} {\bibfnamefont {F.}~\bibnamefont {Haake}},\ and\ \bibinfo {author} {\bibfnamefont {M.}~\bibnamefont {Lewenstein}},\ }\bibfield  {title} {\bibinfo {title} {Quantum kinetic ising models},\ }\href@noop {} {\bibfield  {journal} {\bibinfo  {journal} {New Journal of Physics}\ }\textbf {\bibinfo {volume} {12}},\ \bibinfo {pages} {025021} (\bibinfo {year} {2010})}\BibitemShut {NoStop}%
\bibitem [{\citenamefont {Krapivsky}\ \emph {et~al.}(2010)\citenamefont {Krapivsky}, \citenamefont {Redner},\ and\ \citenamefont {Ben-Naim}}]{krapivsky2010kinetic}%
  \BibitemOpen
  \bibfield  {author} {\bibinfo {author} {\bibfnamefont {P.~L.}\ \bibnamefont {Krapivsky}}, \bibinfo {author} {\bibfnamefont {S.}~\bibnamefont {Redner}},\ and\ \bibinfo {author} {\bibfnamefont {E.}~\bibnamefont {Ben-Naim}},\ }\href@noop {} {\emph {\bibinfo {title} {A kinetic view of statistical physics}}}\ (\bibinfo  {publisher} {Cambridge University Press},\ \bibinfo {year} {2010})\BibitemShut {NoStop}%
\bibitem [{\citenamefont {Suzuki}\ \emph {et~al.}(2013)\citenamefont {Suzuki}, \citenamefont {Inoue},\ and\ \citenamefont {Chakrabarti}}]{Suzuki2013}%
  \BibitemOpen
  \bibfield  {author} {\bibinfo {author} {\bibfnamefont {S.}~\bibnamefont {Suzuki}}, \bibinfo {author} {\bibfnamefont {J.-i.}\ \bibnamefont {Inoue}},\ and\ \bibinfo {author} {\bibfnamefont {B.~K.}\ \bibnamefont {Chakrabarti}},\ }\bibinfo {title} {Dilute and random transverse ising systems},\ in\ \href {https://doi.org/10.1007/978-3-642-33039-1_5} {\emph {\bibinfo {booktitle} {Quantum Ising Phases and Transitions in Transverse Ising Models}}}\ (\bibinfo  {publisher} {Springer Berlin Heidelberg},\ \bibinfo {address} {Berlin, Heidelberg},\ \bibinfo {year} {2013})\ pp.\ \bibinfo {pages} {105--122}\BibitemShut {NoStop}%
\bibitem [{\citenamefont {Breuer}\ and\ \citenamefont {Petruccione}(2002)}]{breuer2002theory}%
  \BibitemOpen
  \bibfield  {author} {\bibinfo {author} {\bibfnamefont {H.-P.}\ \bibnamefont {Breuer}}\ and\ \bibinfo {author} {\bibfnamefont {F.}~\bibnamefont {Petruccione}},\ }\href@noop {} {\emph {\bibinfo {title} {The theory of open quantum systems}}}\ (\bibinfo  {publisher} {Oxford University Press, USA},\ \bibinfo {year} {2002})\BibitemShut {NoStop}%
\bibitem [{\citenamefont {Landi}\ \emph {et~al.}(2022)\citenamefont {Landi}, \citenamefont {Poletti},\ and\ \citenamefont {Schaller}}]{landi2022nonequilibrium}%
  \BibitemOpen
  \bibfield  {author} {\bibinfo {author} {\bibfnamefont {G.~T.}\ \bibnamefont {Landi}}, \bibinfo {author} {\bibfnamefont {D.}~\bibnamefont {Poletti}},\ and\ \bibinfo {author} {\bibfnamefont {G.}~\bibnamefont {Schaller}},\ }\bibfield  {title} {\bibinfo {title} {Nonequilibrium boundary-driven quantum systems: Models, methods, and properties},\ }\href@noop {} {\bibfield  {journal} {\bibinfo  {journal} {Reviews of Modern Physics}\ }\textbf {\bibinfo {volume} {94}},\ \bibinfo {pages} {045006} (\bibinfo {year} {2022})}\BibitemShut {NoStop}%
\bibitem [{\citenamefont {Weisbrich}\ \emph {et~al.}(2018)\citenamefont {Weisbrich}, \citenamefont {Saussol}, \citenamefont {Belzig},\ and\ \citenamefont {Rastelli}}]{weisbrich2018decoherence}%
  \BibitemOpen
  \bibfield  {author} {\bibinfo {author} {\bibfnamefont {H.}~\bibnamefont {Weisbrich}}, \bibinfo {author} {\bibfnamefont {C.}~\bibnamefont {Saussol}}, \bibinfo {author} {\bibfnamefont {W.}~\bibnamefont {Belzig}},\ and\ \bibinfo {author} {\bibfnamefont {G.}~\bibnamefont {Rastelli}},\ }\bibfield  {title} {\bibinfo {title} {Decoherence in the quantum ising model with transverse dissipative interaction in the strong-coupling regime},\ }\href@noop {} {\bibfield  {journal} {\bibinfo  {journal} {Physical Review A}\ }\textbf {\bibinfo {volume} {98}},\ \bibinfo {pages} {052109} (\bibinfo {year} {2018})}\BibitemShut {NoStop}%
\bibitem [{\citenamefont {Rajagopal}(1998)}]{rajagopal1998principle}%
  \BibitemOpen
  \bibfield  {author} {\bibinfo {author} {\bibfnamefont {A.}~\bibnamefont {Rajagopal}},\ }\bibfield  {title} {\bibinfo {title} {The principle of detailed balance and the lindblad dissipative quantum dynamics},\ }\href@noop {} {\bibfield  {journal} {\bibinfo  {journal} {Physics Letters A}\ }\textbf {\bibinfo {volume} {246}},\ \bibinfo {pages} {237} (\bibinfo {year} {1998})}\BibitemShut {NoStop}%
\bibitem [{Note1()}]{Note1}%
  \BibitemOpen
  \bibinfo {note} {This $\omega $ is not to be confused with the angular frequency used in the power spectral density study later on in the manuscript.}\BibitemShut {Stop}%
\bibitem [{\citenamefont {Wiener}(1930)}]{wiener1930generalized}%
  \BibitemOpen
  \bibfield  {author} {\bibinfo {author} {\bibfnamefont {N.}~\bibnamefont {Wiener}},\ }\bibfield  {title} {\bibinfo {title} {Generalized harmonic analysis},\ }\href@noop {} {\bibfield  {journal} {\bibinfo  {journal} {Acta mathematica}\ }\textbf {\bibinfo {volume} {55}},\ \bibinfo {pages} {117} (\bibinfo {year} {1930})}\BibitemShut {NoStop}%
\bibitem [{\citenamefont {Cardiner}(1985)}]{cardiner1985handbook}%
  \BibitemOpen
  \bibfield  {author} {\bibinfo {author} {\bibfnamefont {C.}~\bibnamefont {Cardiner}},\ }\href@noop {} {\bibinfo {title} {Handbook of stochastic methods}} (\bibinfo {year} {1985})\BibitemShut {NoStop}%
\bibitem [{\citenamefont {Welch}(1967)}]{welch1967use}%
  \BibitemOpen
  \bibfield  {author} {\bibinfo {author} {\bibfnamefont {P.}~\bibnamefont {Welch}},\ }\bibfield  {title} {\bibinfo {title} {The use of fast fourier transform for the estimation of power spectra: a method based on time averaging over short, modified periodograms},\ }\href@noop {} {\bibfield  {journal} {\bibinfo  {journal} {IEEE Transactions on audio and electroacoustics}\ }\textbf {\bibinfo {volume} {15}},\ \bibinfo {pages} {70} (\bibinfo {year} {1967})}\BibitemShut {NoStop}%
\bibitem [{\citenamefont {Chandler}(1987)}]{chandler1987introduction}%
  \BibitemOpen
  \bibfield  {author} {\bibinfo {author} {\bibfnamefont {D.}~\bibnamefont {Chandler}},\ }\bibfield  {title} {\bibinfo {title} {Introduction to modern statistical},\ }\href@noop {} {\bibfield  {journal} {\bibinfo  {journal} {Mechanics. Oxford University Press, Oxford, UK}\ }\textbf {\bibinfo {volume} {5}},\ \bibinfo {pages} {11} (\bibinfo {year} {1987})}\BibitemShut {NoStop}%
\bibitem [{\citenamefont {Nishi}\ \emph {et~al.}(2018)\citenamefont {Nishi}, \citenamefont {Kilfoil}, \citenamefont {Schmidt},\ and\ \citenamefont {MacKintosh}}]{nishi2018symmetrical}%
  \BibitemOpen
  \bibfield  {author} {\bibinfo {author} {\bibfnamefont {K.}~\bibnamefont {Nishi}}, \bibinfo {author} {\bibfnamefont {M.~L.}\ \bibnamefont {Kilfoil}}, \bibinfo {author} {\bibfnamefont {C.~F.}\ \bibnamefont {Schmidt}},\ and\ \bibinfo {author} {\bibfnamefont {F.~C.}\ \bibnamefont {MacKintosh}},\ }\bibfield  {title} {\bibinfo {title} {A symmetrical method to obtain shear moduli from microrheology},\ }\href@noop {} {\bibfield  {journal} {\bibinfo  {journal} {Soft matter}\ }\textbf {\bibinfo {volume} {14}},\ \bibinfo {pages} {3716} (\bibinfo {year} {2018})}\BibitemShut {NoStop}%
\bibitem [{\citenamefont {Haus}\ and\ \citenamefont {Kehr}(1987)}]{haus1987diffusion}%
  \BibitemOpen
  \bibfield  {author} {\bibinfo {author} {\bibfnamefont {J.~W.}\ \bibnamefont {Haus}}\ and\ \bibinfo {author} {\bibfnamefont {K.~W.}\ \bibnamefont {Kehr}},\ }\bibfield  {title} {\bibinfo {title} {Diffusion in regular and disordered lattices},\ }\href@noop {} {\bibfield  {journal} {\bibinfo  {journal} {Physics Reports}\ }\textbf {\bibinfo {volume} {150}},\ \bibinfo {pages} {263} (\bibinfo {year} {1987})}\BibitemShut {NoStop}%
\bibitem [{\citenamefont {Choudhury}\ \emph {et~al.}(2017)\citenamefont {Choudhury}, \citenamefont {Straube}, \citenamefont {Fischer}, \citenamefont {Gibbs},\ and\ \citenamefont {H{\"o}fling}}]{choudhury2017active}%
  \BibitemOpen
  \bibfield  {author} {\bibinfo {author} {\bibfnamefont {U.}~\bibnamefont {Choudhury}}, \bibinfo {author} {\bibfnamefont {A.~V.}\ \bibnamefont {Straube}}, \bibinfo {author} {\bibfnamefont {P.}~\bibnamefont {Fischer}}, \bibinfo {author} {\bibfnamefont {J.~G.}\ \bibnamefont {Gibbs}},\ and\ \bibinfo {author} {\bibfnamefont {F.}~\bibnamefont {H{\"o}fling}},\ }\bibfield  {title} {\bibinfo {title} {Active colloidal propulsion over a crystalline surface},\ }\href@noop {} {\bibfield  {journal} {\bibinfo  {journal} {New Journal of Physics}\ }\textbf {\bibinfo {volume} {19}},\ \bibinfo {pages} {125010} (\bibinfo {year} {2017})}\BibitemShut {NoStop}%
\bibitem [{\citenamefont {Su}\ \emph {et~al.}(2017)\citenamefont {Su}, \citenamefont {Lai}, \citenamefont {Ackerson}, \citenamefont {Cao}, \citenamefont {Han},\ and\ \citenamefont {Tong}}]{su2017colloidal}%
  \BibitemOpen
  \bibfield  {author} {\bibinfo {author} {\bibfnamefont {Y.}~\bibnamefont {Su}}, \bibinfo {author} {\bibfnamefont {P.-Y.}\ \bibnamefont {Lai}}, \bibinfo {author} {\bibfnamefont {B.~J.}\ \bibnamefont {Ackerson}}, \bibinfo {author} {\bibfnamefont {X.}~\bibnamefont {Cao}}, \bibinfo {author} {\bibfnamefont {Y.}~\bibnamefont {Han}},\ and\ \bibinfo {author} {\bibfnamefont {P.}~\bibnamefont {Tong}},\ }\bibfield  {title} {\bibinfo {title} {Colloidal diffusion over a quasicrystalline-patterned surface},\ }\href@noop {} {\bibfield  {journal} {\bibinfo  {journal} {The Journal of Chemical Physics}\ }\textbf {\bibinfo {volume} {146}} (\bibinfo {year} {2017})}\BibitemShut {NoStop}%
\bibitem [{\citenamefont {Hanes}\ and\ \citenamefont {Egelhaaf}(2012)}]{hanes2012dynamics}%
  \BibitemOpen
  \bibfield  {author} {\bibinfo {author} {\bibfnamefont {R.~D.}\ \bibnamefont {Hanes}}\ and\ \bibinfo {author} {\bibfnamefont {S.~U.}\ \bibnamefont {Egelhaaf}},\ }\bibfield  {title} {\bibinfo {title} {Dynamics of individual colloidal particles in one-dimensional random potentials: a simulation study},\ }\href@noop {} {\bibfield  {journal} {\bibinfo  {journal} {Journal of Physics: Condensed Matter}\ }\textbf {\bibinfo {volume} {24}},\ \bibinfo {pages} {464116} (\bibinfo {year} {2012})}\BibitemShut {NoStop}%
\bibitem [{\citenamefont {Evers}\ \emph {et~al.}(2013)\citenamefont {Evers}, \citenamefont {Zunke}, \citenamefont {Hanes}, \citenamefont {Bewerunge}, \citenamefont {Ladadwa}, \citenamefont {Heuer},\ and\ \citenamefont {Egelhaaf}}]{evers2013particle}%
  \BibitemOpen
  \bibfield  {author} {\bibinfo {author} {\bibfnamefont {F.}~\bibnamefont {Evers}}, \bibinfo {author} {\bibfnamefont {C.}~\bibnamefont {Zunke}}, \bibinfo {author} {\bibfnamefont {R.~D.}\ \bibnamefont {Hanes}}, \bibinfo {author} {\bibfnamefont {J.}~\bibnamefont {Bewerunge}}, \bibinfo {author} {\bibfnamefont {I.}~\bibnamefont {Ladadwa}}, \bibinfo {author} {\bibfnamefont {A.}~\bibnamefont {Heuer}},\ and\ \bibinfo {author} {\bibfnamefont {S.~U.}\ \bibnamefont {Egelhaaf}},\ }\bibfield  {title} {\bibinfo {title} {Particle dynamics in two-dimensional random-energy landscapes: Experiments and simulations},\ }\href@noop {} {\bibfield  {journal} {\bibinfo  {journal} {Physical Review E—Statistical, Nonlinear, and Soft Matter Physics}\ }\textbf {\bibinfo {volume} {88}},\ \bibinfo {pages} {022125} (\bibinfo {year} {2013})}\BibitemShut {NoStop}%
\bibitem [{\citenamefont {Fulde}\ \emph {et~al.}(1975)\citenamefont {Fulde}, \citenamefont {Pietronero}, \citenamefont {Schneider},\ and\ \citenamefont {Str{\"a}ssler}}]{fulde1975problem}%
  \BibitemOpen
  \bibfield  {author} {\bibinfo {author} {\bibfnamefont {P.}~\bibnamefont {Fulde}}, \bibinfo {author} {\bibfnamefont {L.}~\bibnamefont {Pietronero}}, \bibinfo {author} {\bibfnamefont {W.}~\bibnamefont {Schneider}},\ and\ \bibinfo {author} {\bibfnamefont {S.}~\bibnamefont {Str{\"a}ssler}},\ }\bibfield  {title} {\bibinfo {title} {Problem of brownian motion in a periodic potential},\ }\href@noop {} {\bibfield  {journal} {\bibinfo  {journal} {Physical Review Letters}\ }\textbf {\bibinfo {volume} {35}},\ \bibinfo {pages} {1776} (\bibinfo {year} {1975})}\BibitemShut {NoStop}%
\bibitem [{\citenamefont {Festa}\ and\ \citenamefont {d'Agliano}(1978)}]{festa1978diffusion}%
  \BibitemOpen
  \bibfield  {author} {\bibinfo {author} {\bibfnamefont {R.}~\bibnamefont {Festa}}\ and\ \bibinfo {author} {\bibfnamefont {E.~G.}\ \bibnamefont {d'Agliano}},\ }\bibfield  {title} {\bibinfo {title} {Diffusion coefficient for a brownian particle in a periodic field of force i. large friction limit},\ }\href@noop {} {\bibfield  {journal} {\bibinfo  {journal} {Physica A Statistical Mechanics and its Applications}\ }\textbf {\bibinfo {volume} {90}},\ \bibinfo {pages} {229} (\bibinfo {year} {1978})}\BibitemShut {NoStop}%
\bibitem [{\citenamefont {Nilsson~Hall{\'e}n}(2023)}]{nilsson2023dynamics}%
  \BibitemOpen
  \bibfield  {author} {\bibinfo {author} {\bibfnamefont {E.~J.}\ \bibnamefont {Nilsson~Hall{\'e}n}},\ }\emph {\bibinfo {title} {Dynamics of frustrated magnetic systems--Emergent fractals and anomalous magnetic noise in spin ice}},\ \href@noop {} {Ph.D. thesis},\ \bibinfo  {school} {University of Cambridge} (\bibinfo {year} {2023})\BibitemShut {NoStop}%
\bibitem [{\citenamefont {Chakrabarti}\ and\ \citenamefont {Acharyya}(1999)}]{chakrabarti1999dynamic}%
  \BibitemOpen
  \bibfield  {author} {\bibinfo {author} {\bibfnamefont {B.~K.}\ \bibnamefont {Chakrabarti}}\ and\ \bibinfo {author} {\bibfnamefont {M.}~\bibnamefont {Acharyya}},\ }\bibfield  {title} {\bibinfo {title} {Dynamic transitions and hysteresis},\ }\href@noop {} {\bibfield  {journal} {\bibinfo  {journal} {Reviews of Modern Physics}\ }\textbf {\bibinfo {volume} {71}},\ \bibinfo {pages} {847} (\bibinfo {year} {1999})}\BibitemShut {NoStop}%
\bibitem [{\citenamefont {Sala}\ \emph {et~al.}(2014)\citenamefont {Sala}, \citenamefont {Gutmann}, \citenamefont {Prabhakaran}, \citenamefont {Pomaranski}, \citenamefont {Mitchelitis}, \citenamefont {Kycia}, \citenamefont {Porter}, \citenamefont {Castelnovo},\ and\ \citenamefont {Goff}}]{sala2014vacancy}%
  \BibitemOpen
  \bibfield  {author} {\bibinfo {author} {\bibfnamefont {G.}~\bibnamefont {Sala}}, \bibinfo {author} {\bibfnamefont {M.}~\bibnamefont {Gutmann}}, \bibinfo {author} {\bibfnamefont {D.}~\bibnamefont {Prabhakaran}}, \bibinfo {author} {\bibfnamefont {D.}~\bibnamefont {Pomaranski}}, \bibinfo {author} {\bibfnamefont {C.}~\bibnamefont {Mitchelitis}}, \bibinfo {author} {\bibfnamefont {J.}~\bibnamefont {Kycia}}, \bibinfo {author} {\bibfnamefont {D.}~\bibnamefont {Porter}}, \bibinfo {author} {\bibfnamefont {C.}~\bibnamefont {Castelnovo}},\ and\ \bibinfo {author} {\bibfnamefont {J.}~\bibnamefont {Goff}},\ }\bibfield  {title} {\bibinfo {title} {Vacancy defects and monopole dynamics in oxygen-deficient pyrochlores},\ }\href@noop {} {\bibfield  {journal} {\bibinfo  {journal} {Nature Materials}\ }\textbf {\bibinfo {volume} {13}},\ \bibinfo {pages} {488} (\bibinfo {year} {2014})}\BibitemShut {NoStop}%
\bibitem [{\citenamefont {Goff}(2023)}]{goff}%
  \BibitemOpen
  \bibfield  {author} {\bibinfo {author} {\bibfnamefont {J.}~\bibnamefont {Goff}},\ }\href@noop {} {}\bibinfo {howpublished} {Private communication.} (\bibinfo {year} {2023})\BibitemShut {NoStop}%
\bibitem [{\citenamefont {Jaubert}\ and\ \citenamefont {Holdsworth}(2009)}]{jaubert2009signature}%
  \BibitemOpen
  \bibfield  {author} {\bibinfo {author} {\bibfnamefont {L.~D.}\ \bibnamefont {Jaubert}}\ and\ \bibinfo {author} {\bibfnamefont {P.~C.}\ \bibnamefont {Holdsworth}},\ }\bibfield  {title} {\bibinfo {title} {Signature of magnetic monopole and dirac string dynamics in spin ice},\ }\href@noop {} {\bibfield  {journal} {\bibinfo  {journal} {Nature Physics}\ }\textbf {\bibinfo {volume} {5}},\ \bibinfo {pages} {258} (\bibinfo {year} {2009})}\BibitemShut {NoStop}%
\bibitem [{\citenamefont {Sen}\ and\ \citenamefont {Moessner}(2015)}]{sen2015topological}%
  \BibitemOpen
  \bibfield  {author} {\bibinfo {author} {\bibfnamefont {A.}~\bibnamefont {Sen}}\ and\ \bibinfo {author} {\bibfnamefont {R.}~\bibnamefont {Moessner}},\ }\bibfield  {title} {\bibinfo {title} {Topological spin glass in diluted spin ice},\ }\href@noop {} {\bibfield  {journal} {\bibinfo  {journal} {Physical review letters}\ }\textbf {\bibinfo {volume} {114}},\ \bibinfo {pages} {247207} (\bibinfo {year} {2015})}\BibitemShut {NoStop}%
\bibitem [{\citenamefont {Lu}\ \emph {et~al.}(2024)\citenamefont {Lu}, \citenamefont {Sch{\"a}fer}, \citenamefont {Hall{\'e}n},\ and\ \citenamefont {Laumann}}]{lu2024111}%
  \BibitemOpen
  \bibfield  {author} {\bibinfo {author} {\bibfnamefont {Z.}~\bibnamefont {Lu}}, \bibinfo {author} {\bibfnamefont {R.}~\bibnamefont {Sch{\"a}fer}}, \bibinfo {author} {\bibfnamefont {J.~N.}\ \bibnamefont {Hall{\'e}n}},\ and\ \bibinfo {author} {\bibfnamefont {C.~R.}\ \bibnamefont {Laumann}},\ }\bibfield  {title} {\bibinfo {title} {[111]-strained spin ice: Localization of thermodynamically deconfined monopoles},\ }\href@noop {} {\bibfield  {journal} {\bibinfo  {journal} {Physical Review B}\ }\textbf {\bibinfo {volume} {110}},\ \bibinfo {pages} {184421} (\bibinfo {year} {2024})}\BibitemShut {NoStop}%
\bibitem [{\citenamefont {Yu}(2004)}]{yu2004non}%
  \BibitemOpen
  \bibfield  {author} {\bibinfo {author} {\bibfnamefont {T.}~\bibnamefont {Yu}},\ }\bibfield  {title} {\bibinfo {title} {Non-markovian quantum trajectories versus master equations: Finite-temperature heat bath},\ }\href@noop {} {\bibfield  {journal} {\bibinfo  {journal} {Physical Review A—Atomic, Molecular, and Optical Physics}\ }\textbf {\bibinfo {volume} {69}},\ \bibinfo {pages} {062107} (\bibinfo {year} {2004})}\BibitemShut {NoStop}%
\bibitem [{\citenamefont {Tomasello}\ \emph {et~al.}(2019)\citenamefont {Tomasello}, \citenamefont {Castelnovo}, \citenamefont {Moessner},\ and\ \citenamefont {Quintanilla}}]{tomasello2019correlated}%
  \BibitemOpen
  \bibfield  {author} {\bibinfo {author} {\bibfnamefont {B.}~\bibnamefont {Tomasello}}, \bibinfo {author} {\bibfnamefont {C.}~\bibnamefont {Castelnovo}}, \bibinfo {author} {\bibfnamefont {R.}~\bibnamefont {Moessner}},\ and\ \bibinfo {author} {\bibfnamefont {J.}~\bibnamefont {Quintanilla}},\ }\bibfield  {title} {\bibinfo {title} {Correlated quantum tunneling of monopoles in spin ice},\ }\href@noop {} {\bibfield  {journal} {\bibinfo  {journal} {Physical review letters}\ }\textbf {\bibinfo {volume} {123}},\ \bibinfo {pages} {067204} (\bibinfo {year} {2019})}\BibitemShut {NoStop}%
\bibitem [{\citenamefont {Wang}\ \emph {et~al.}(2021)\citenamefont {Wang}, \citenamefont {Reeder}, \citenamefont {Karaki}, \citenamefont {Kindervater}, \citenamefont {Halloran}, \citenamefont {Maliszewskyj}, \citenamefont {Qiu}, \citenamefont {Rodriguez}, \citenamefont {Gladchenko}, \citenamefont {Koohpayeh} \emph {et~al.}}]{wang2021monopolar}%
  \BibitemOpen
  \bibfield  {author} {\bibinfo {author} {\bibfnamefont {Y.}~\bibnamefont {Wang}}, \bibinfo {author} {\bibfnamefont {T.}~\bibnamefont {Reeder}}, \bibinfo {author} {\bibfnamefont {Y.}~\bibnamefont {Karaki}}, \bibinfo {author} {\bibfnamefont {J.}~\bibnamefont {Kindervater}}, \bibinfo {author} {\bibfnamefont {T.}~\bibnamefont {Halloran}}, \bibinfo {author} {\bibfnamefont {N.}~\bibnamefont {Maliszewskyj}}, \bibinfo {author} {\bibfnamefont {Y.}~\bibnamefont {Qiu}}, \bibinfo {author} {\bibfnamefont {J.}~\bibnamefont {Rodriguez}}, \bibinfo {author} {\bibfnamefont {S.}~\bibnamefont {Gladchenko}}, \bibinfo {author} {\bibfnamefont {S.}~\bibnamefont {Koohpayeh}}, \emph {et~al.},\ }\bibfield  {title} {\bibinfo {title} {Monopolar and dipolar relaxation in spin ice ho2ti2o7},\ }\href@noop {} {\bibfield  {journal} {\bibinfo  {journal} {Science Advances}\ }\textbf {\bibinfo {volume} {7}},\ \bibinfo {pages} {eabg0908} (\bibinfo {year} {2021})}\BibitemShut {NoStop}%
\bibitem [{\citenamefont {Radicevic}(2018)}]{radicevic2018spin}%
  \BibitemOpen
  \bibfield  {author} {\bibinfo {author} {\bibfnamefont {D.}~\bibnamefont {Radicevic}},\ }\bibfield  {title} {\bibinfo {title} {Spin structures and exact dualities in low dimensions},\ }\href@noop {} {\bibfield  {journal} {\bibinfo  {journal} {arXiv preprint arXiv:1809.07757}\ } (\bibinfo {year} {2018})}\BibitemShut {NoStop}%
\bibitem [{\citenamefont {Sandvik}(2010)}]{sandvik2010computational}%
  \BibitemOpen
  \bibfield  {author} {\bibinfo {author} {\bibfnamefont {A.~W.}\ \bibnamefont {Sandvik}},\ }\bibfield  {title} {\bibinfo {title} {Computational studies of quantum spin systems},\ }in\ \href@noop {} {\emph {\bibinfo {booktitle} {AIP Conference Proceedings}}},\ Vol.\ \bibinfo {volume} {1297}\ (\bibinfo {organization} {American Institute of Physics},\ \bibinfo {year} {2010})\ pp.\ \bibinfo {pages} {135--338}\BibitemShut {NoStop}%
\bibitem [{\citenamefont {Sandvik}(2019)}]{sandvik2019stochastic}%
  \BibitemOpen
  \bibfield  {author} {\bibinfo {author} {\bibfnamefont {A.~W.}\ \bibnamefont {Sandvik}},\ }\bibfield  {title} {\bibinfo {title} {Stochastic series expansion methods},\ }\href@noop {} {\bibfield  {journal} {\bibinfo  {journal} {arXiv preprint arXiv:1909.10591}\ } (\bibinfo {year} {2019})}\BibitemShut {NoStop}%
\bibitem [{\citenamefont {Lauritsen}\ and\ \citenamefont {Fogedby}(1993)}]{lauritsen1993critical}%
  \BibitemOpen
  \bibfield  {author} {\bibinfo {author} {\bibfnamefont {K.~B.}\ \bibnamefont {Lauritsen}}\ and\ \bibinfo {author} {\bibfnamefont {H.~C.}\ \bibnamefont {Fogedby}},\ }\bibfield  {title} {\bibinfo {title} {Critical exponents from power spectra},\ }\href@noop {} {\bibfield  {journal} {\bibinfo  {journal} {Journal of statistical physics}\ }\textbf {\bibinfo {volume} {72}},\ \bibinfo {pages} {189} (\bibinfo {year} {1993})}\BibitemShut {NoStop}%
\bibitem [{\citenamefont {Lauritsen}\ and\ \citenamefont {Ito}(1994)}]{lauritsen1994spectral}%
  \BibitemOpen
  \bibfield  {author} {\bibinfo {author} {\bibfnamefont {K.~B.}\ \bibnamefont {Lauritsen}}\ and\ \bibinfo {author} {\bibfnamefont {N.}~\bibnamefont {Ito}},\ }\bibfield  {title} {\bibinfo {title} {Spectral method determination of dynamic exponents for ising models},\ }\href@noop {} {\bibfield  {journal} {\bibinfo  {journal} {Physica A: Statistical Mechanics and its Applications}\ }\textbf {\bibinfo {volume} {202}},\ \bibinfo {pages} {224} (\bibinfo {year} {1994})}\BibitemShut {NoStop}%
\bibitem [{\citenamefont {MacIsaac}\ and\ \citenamefont {Jan}(1992)}]{macisaac1992dynamic}%
  \BibitemOpen
  \bibfield  {author} {\bibinfo {author} {\bibfnamefont {K.}~\bibnamefont {MacIsaac}}\ and\ \bibinfo {author} {\bibfnamefont {N.}~\bibnamefont {Jan}},\ }\bibfield  {title} {\bibinfo {title} {On the dynamic exponent of the two-dimensional ising model},\ }\href@noop {} {\bibfield  {journal} {\bibinfo  {journal} {Journal of Physics A: Mathematical and General}\ }\textbf {\bibinfo {volume} {25}},\ \bibinfo {pages} {2139} (\bibinfo {year} {1992})}\BibitemShut {NoStop}%
\bibitem [{\citenamefont {Landau}\ and\ \citenamefont {Lifshitz}(1980)}]{LANDAU1980333}%
  \BibitemOpen
  \bibfield  {author} {\bibinfo {author} {\bibfnamefont {L.}~\bibnamefont {Landau}}\ and\ \bibinfo {author} {\bibfnamefont {E.}~\bibnamefont {Lifshitz}},\ }\bibfield  {title} {\bibinfo {title} {Chapter xii - fluctuations},\ }in\ \href {https://doi.org/https://doi.org/10.1016/B978-0-08-057046-4.50019-1} {\emph {\bibinfo {booktitle} {Statistical Physics (Third Edition)}}},\ \bibinfo {editor} {edited by\ \bibinfo {editor} {\bibfnamefont {L.}~\bibnamefont {Landau}}\ and\ \bibinfo {editor} {\bibfnamefont {E.}~\bibnamefont {Lifshitz}}}\ (\bibinfo  {publisher} {Butterworth-Heinemann},\ \bibinfo {address} {Oxford},\ \bibinfo {year} {1980})\ \bibinfo {edition} {third edition}\ ed.,\ pp.\ \bibinfo {pages} {333--400}\BibitemShut {NoStop}%
\bibitem [{\citenamefont {Clerk}\ \emph {et~al.}(2010)\citenamefont {Clerk}, \citenamefont {Devoret}, \citenamefont {Girvin}, \citenamefont {Marquardt},\ and\ \citenamefont {Schoelkopf}}]{clerk2010introduction}%
  \BibitemOpen
  \bibfield  {author} {\bibinfo {author} {\bibfnamefont {A.~A.}\ \bibnamefont {Clerk}}, \bibinfo {author} {\bibfnamefont {M.~H.}\ \bibnamefont {Devoret}}, \bibinfo {author} {\bibfnamefont {S.~M.}\ \bibnamefont {Girvin}}, \bibinfo {author} {\bibfnamefont {F.}~\bibnamefont {Marquardt}},\ and\ \bibinfo {author} {\bibfnamefont {R.~J.}\ \bibnamefont {Schoelkopf}},\ }\bibfield  {title} {\bibinfo {title} {Introduction to quantum noise, measurement, and amplification},\ }\href@noop {} {\bibfield  {journal} {\bibinfo  {journal} {Reviews of Modern Physics}\ }\textbf {\bibinfo {volume} {82}},\ \bibinfo {pages} {1155} (\bibinfo {year} {2010})}\BibitemShut {NoStop}%
\bibitem [{\citenamefont {Daley}(2014)}]{daley2014quantum}%
  \BibitemOpen
  \bibfield  {author} {\bibinfo {author} {\bibfnamefont {A.~J.}\ \bibnamefont {Daley}},\ }\bibfield  {title} {\bibinfo {title} {Quantum trajectories and open many-body quantum systems},\ }\href@noop {} {\bibfield  {journal} {\bibinfo  {journal} {Advances in Physics}\ }\textbf {\bibinfo {volume} {63}},\ \bibinfo {pages} {77} (\bibinfo {year} {2014})}\BibitemShut {NoStop}%
\bibitem [{\citenamefont {van Beijeren}(1982)}]{van1982transport}%
  \BibitemOpen
  \bibfield  {author} {\bibinfo {author} {\bibfnamefont {H.}~\bibnamefont {van Beijeren}},\ }\bibfield  {title} {\bibinfo {title} {Transport properties of stochastic lorentz models},\ }\href@noop {} {\bibfield  {journal} {\bibinfo  {journal} {Reviews of Modern Physics}\ }\textbf {\bibinfo {volume} {54}},\ \bibinfo {pages} {195} (\bibinfo {year} {1982})}\BibitemShut {NoStop}%
\bibitem [{\citenamefont {Derrida}(1983)}]{derrida1983velocity}%
  \BibitemOpen
  \bibfield  {author} {\bibinfo {author} {\bibfnamefont {B.}~\bibnamefont {Derrida}},\ }\bibfield  {title} {\bibinfo {title} {Velocity and diffusion constant of a periodic one-dimensional hopping model},\ }\href@noop {} {\bibfield  {journal} {\bibinfo  {journal} {Journal of statistical physics}\ }\textbf {\bibinfo {volume} {31}},\ \bibinfo {pages} {433} (\bibinfo {year} {1983})}\BibitemShut {NoStop}%
\bibitem [{wel(2024)}]{welch}%
  \BibitemOpen
  \href {https://docs.scipy.org/doc/scipy/reference/generated/scipy.signal.welch.html} {\bibinfo {title} {Welch -- scipy v1.14.1 manual}} (\bibinfo {year} {2024})\BibitemShut {NoStop}%
\end{thebibliography}%

\end{document}